\documentclass{theoretics}

\usepackage[T5,T1]{fontenc}
\usepackage[utf8]{inputenc}

\DeclareRobustCommand{\tildecircumE}{\~{ê}}

\ThCSauthor[1]{Paul Gallot}{gallot@uni-bremen.de}[0000-0001-6742-4885]
\ThCSauthor[2]{Nathan Lhote}{nathan.lhote@lis-lab.fr}[0000-0003-3303-5368]
\ThCSauthor[2,3]{Lê Thành Dũng (Tito) Nguy{\tildecircumE}n}{nltd@nguyentito.eu}[0000-0002-6900-5577]
\ThCSaffil[1]{Universität Bremen, Germany}
\ThCSaffil[2]{LIS, Aix-Marseille Univ., France}
\ThCSaffil[3]{CNRS, France}
\ThCSshortnames{P.~Gallot, N.~Lhote \& L.~T.~D.~Nguy{\tildecircumE}n}
\ThCSthanks{%
  This collaboration was initiated thanks to Sylvain Salvati's
  intermediation and then highly stimulated by the Dagstuhl Seminar on Regular
  Transformations (where we gave talks about macro tree
  transducers~\cite[Sections~3.11 \&~3.17]{Dagstuhl23202} and MSO set
  interpretations~\cite[\S3.15]{Dagstuhl23202}). Our approach to size-to-height
  increase was inspired by Thomas Colcombet's advice on handling maxima
  using nondeterminism.\\
  We also thank
  Antoine Amarilli, Mikołaj Bojańczyk, Gaëtan Douéneau-Tabot, Emmanuel Filiot, Louis Jachiet, Denis
  Kuperberg, Filip Mazowiecki, Anca Muscholl, Pierre-Alain Reynier, Sylvain Schmitz and Olivier Serre for interesting discussions.
  L.~T.~D.\ Nguy{\~{\^e}}n is especially grateful to Sandra Kiefer and Cécilia
  Pradic for their collaboration on the related work~\cite{Sandra}, and to Charles Paperman for performing some message-passing on IRC.
  Finally, we thank the anonymous reviewers for their detailed feedback and for interesting pointers to the literature, which have significantly shaped the paper.}

\title{The structure of polynomial growth for tree automata / transducers and MSO set queries}
\ThCSshorttitle{The structure of polynomial growth for tree automata ...} 
\ThCSyear{2026}
\ThCSarticlenum{20}
\ThCSreceived{Mar 15, 2025}
\ThCSrevised{Jun 11, 2026}
\ThCSaccepted{Sep 29, 2026}
\ThCSpublished{Nov 27, 2026}
\ThCSdoicreatedtrue
\ThCSkeywords{monadic second-order logic, polynomial ambiguity, polyregular functions, set interpretations, tree transducers, weighted tree automata}

\usepackage{mathtools}
\usepackage{csquotes}
\usepackage{tikz}
\usetikzlibrary{shapes,arrows,positioning,automata,shadows}
\usepackage{xspace}
\usepackage{stmaryrd}
\usepackage{cmll}
\usepackage{lowco2}
\usepackage{lettrine}

\renewcommand{\phi}{\varphi}
\renewcommand{\epsilon}{\varepsilon}

\newcommand{\bbN}{\mathbb{N}}
\newcommand{\bbQ}{\mathbb{Q}}
\newcommand{\bbZ}{\mathbb{Z}}
\newcommand{\bbR}{\mathbb{R}}

\newcommand{\unfold}{\mathsf{unfold}}

\newcommand{\cA}{\mathcal{A}}
\newcommand{\cB}{\mathcal{B}}
\newcommand{\cI}{\mathcal{I}}
\newcommand{\cJ}{\mathcal{J}}
\newcommand{\cM}{\mathcal{M}}
\newcommand{\cT}{\mathcal{T}}

\newcommand{\set}[1]{\left\{ #1 \right\}}
\newcommand{\powerset}{\mathcal{P}}

\newcommand{\components}{\mathrm{comp}}

\newcommand{\marked}{\mathsf{marked}}

\newcommand{\growth}{\mathsf{growth}}
\newcommand{\rank}{\mathsf{rank}}
\newcommand{\height}{\mathsf{height}}
\newcommand{\tree}{\mathsf{T}}
\newcommand{\nodes}{\mathsf{nodes}}
\newcommand{\child}{\mathsf{child}}

\newcommand\dom{\mathrm{dom}}

\newcommand{\pump}{\mathsf{pump}}
\newcommand{\pumpnode}[1]{\langle #1 \rangle}

\newcommand\graph{\mathfrak{G}}
\newcommand\Fpath{F_{\mathrm{path}}}
\newcommand\Gpath{G_{\mathrm{path}}}

\newcommand{\mttarg}[1]{\langle #1 \rangle}
\newcommand{\sem}[1]{\llbracket #1 \rrbracket}
\newcommand{\yield}{\mathsf{yield}}

\newcommand{\Branches}{\mathsf{Branches}}
\newcommand{\RHS}{\mathsf{RHS}}
\newcommand{\rhs}{\mathsf{rhs}}

\newcommand{\entropy}{\mathcal{H}}

\begin{document}
 \maketitle
\begin{abstract}
  Given an $\bbN$-weighted tree automaton, we give a decision procedure for
  exponential vs polynomial growth (with respect to the input size) in quadratic
  time, and an algorithm that computes the exact polynomial degree of growth in
  cubic time. As a special case, they apply to the growth of the ambiguity of a
  nondeterministic tree automaton, i.e.\ the number of distinct accepting runs
  over a given input.

  We deduce analogous decidability results (ignoring complexity) for the
  growth of the number of results of set queries in Monadic Second-Order logic
  (MSO) over ranked trees. In the case of polynomial growth of degree $k$, we
  also prove a reparameterisation theorem for such queries: their results can be
  mapped to $k$-tuples of input nodes in a finite-to-one and MSO-definable
  fashion.

  We then apply these tools to study growth rates and subclass membership problems for tree-to-tree functions. Using new proof strategies, we recover and generalise known results concerning polyregular functions, total deterministic macro tree transducers, and partial nondeterministic top-down tree transducers. In particular, we give a procedure to decide polynomial size-to-height increase for both macro tree transducers and MSO set interpretations, and compute the degree.    
  The paper concludes with a survey of a wide range of related work.
\end{abstract}

\section{Introduction}\label{sec:intro}

This paper is concerned with the asymptotics of various quantities, defined by
automata or logic, depending on a (finite) input tree over a fixed ranked
alphabet. More specifically, for a map $f$ from trees to $\bbN=\{0,1,\dots\}$, we consider the
associated \emph{growth rate function}\footnote{When $f$ is the output length of a string-to-string function $g$, the growth rate of $f$ is called the \emph{(function) length} of $g$ in higher-order computational complexity, where it plays a key role~\cite[Definition~4.1]{Kapron1996}.}
{\let\thefootnote\relax\footnote{\lettrine{\lowcotwo}{}\ \textbf{This is a low-CO\textsubscript{2} research paper: \lowcotwourl[v1]. This research
was developed, written, submitted\\ \phantom{.} and presented without the use of air travel or generative AI.}}}
\addtocounter{footnote}{-1}
\[\growth[f] \colon n \in \bbN \mapsto \displaystyle \max_{|t| \leqslant n} f(t)\]
where the size $|t|$ of the tree $t$ is its number of nodes.
We shall say that:
\begin{itemize}
  \item $f$ has \emph{polynomial growth of degree} $k\in\bbN$ when $\growth[f](n) = \Theta(n^{k})$;
  \item $f$ has \emph{exponential growth} when $\growth[f](n) = 2^{\Theta(n)}$ (sometimes one can be more precise, cf.~Theorem~\ref{thm:precise-exp})
        --- in that case we shall say that the degree of growth of $f$ is $\infty$.
\end{itemize}

\subparagraph{Ambiguity of tree automata (\S\ref{sec:ambiguity}).}

The ambiguity of a non-deterministic automaton over a given input tree is its
number of distinct runs over this input. This notion plays an important role in the
theory of automata over words (cf.\ e.g.~\cite{AmbiguitySurvey}); the definition extends as it is to tree automata,
and it has been occasionally studied in this setting, for instance
in~\cite{MalettiNSU23}.

E.~Paul has shown~\cite[Corollary~8.27]{ErikPaulDiploma} that the ambiguity of a
tree automaton must have a well-defined degree of growth in
$\bbN\cup\{\infty\}$, in accordance with our definition above. In particular,
following standard terminology, every tree automaton is either
\emph{polynomially ambiguous} or \emph{exponentially ambiguous}. We reprove this
result and make it effective:
\begin{theorem}\label{thm:degree-ambiguity}
  Exponential vs polynomial ambiguity is decidable in quadratic time for tree
  automata, and one can compute the degree of growth of the ambiguity in cubic
  time.
\end{theorem}

This was already known in the special case of
strings~\cite{WeberSeidl,AllauzenMR11}, cf.\ Section~\ref{sec:related-work-intro};
furthermore, the time complexity bounds were recently shown to be optimal over
strings, conditionally to standard hypotheses in fine-grained complexity, by
Drabik, Dürr, Frei, Mazowiecki and Węgrzycki~\cite{FineGrainedAmbiguity}. But
surprisingly, we have not found these computability results --- even without
complexity bounds --- explicitly stated for trees in the literature.

\subparagraph{$\bbN$-weighted tree automata (\S\ref{sec:ambiguity}).}

It is well known that the string-to-$\bbN$ functions that describe the ambiguity
of some nondeterministic automaton are exactly the \emph{$\bbN$-rational
  series}, i.e.\ the functions computed by \emph{$\bbN$-weighted
  automata}. Their classical theory is covered for instance in the
books~\cite{Sakarovitch,BerstelReutenauer}, and the case of polynomial growth
remains an active research topic (see e.g.~\cite{Npolyreg}). This correspondence
between $\bbN$-rationality and ambiguity extends to trees; for a book reference,
c.f.~\cite[Theorem~4.3.7]{WTAbook}.

In one direction, the ambiguity of a tree automaton is just the value of the
$\bbN$-weighted automaton obtained by setting all weights to 1. The converse
reduction, however, induces an exponential blow-up (unless the weights are given
in unary). By working directly in the weighted setting, we still manage to
generalise Theorem~\ref{thm:degree-ambiguity} to $\bbN$-weighted tree automata with the
same complexity bounds (Theorem~\ref{thm:degree-weighted}).

\subparagraph{MSO set queries (\S\ref{sec:queries}).}

Let $F(X_1,\dots,X_\ell)$ be a formula of \emph{monadic second-order logic} (MSO) over trees, encoded as relational structures in the usual way. We capitalize the free variables $X_1,\dots,X_\ell$ to indicate that they are second-order variables -- that is, they range over sets of nodes. Such a formula defines an \emph{MSO set query}
\[ \wn F \colon \text{tree}\ t \mapsto \big\{\underbrace{(P_1,\dots,P_\ell)}_{\mathclap{\qquad\qquad\text{each of these tuples is called a \emph{result} of the query}}} \in \{\text{subsets of nodes of}\ t\}^\ell \mid t \models F(P_1,\dots,P_\ell)\big\} \]
The enumeration of the results of such queries has been an active topic of research in database
theory, see for instance~\cite[Sections~3.4~\&~4.3]{AmarilliHDR}. Here, we start with this elementary remark:
\begin{claim}\label{clm:query-vs-ambiguity-intro}
  From an MSO set query over trees (given by a formula), one can effectively compute a tree
  automaton whose ambiguity coincides, over any input tree, with the number of
  results of the query. Thus, the growth rate of this query enjoys a
  polynomial/exponential dichotomy and its degree is computable.
\end{claim}

We can say more about queries of polynomial growth, thanks to structural
properties that arise in our study of polynomially ambiguous tree automata. Let us say that a
family of functions is \emph{finite-to-one} if there is some global finite bound
on the cardinality of every fiber $f^{-1}(\{x\})$, for $f$ in the family and $x$
in the codomain of $f$.
\begin{theorem}\label{thm:reparam-intro}
  Suppose that $F(X_{1},\dots,X_{\ell})$ defines an MSO set query over trees that has polynomial
  growth, with degree $k\in\bbN$. Then there exists a \emph{finite-to-one and
    effectively MSO-definable} family of functions $\wn F(t) \to \{\text{nodes
    of}\ t\}^{k}$, indexed by input trees $t$.
\end{theorem}

We shall make the meaning of \enquote{MSO-definable} in this statement precise
in Section~\ref{sec:queries}; \enquote{effectively} means that the MSO formula defining
this family can be computed from $F$. Basically, the results of the query can be
\emph{reparameterised} in MSO by $k$-tuples of \emph{nodes} instead of $\ell$-tuples of
\emph{sets}, up to bounded imprecision.
\begin{example}\label{ex:paths-intro}
  Let us consider an MSO set query over trees that selects paths from the root --- let us say \emph{rooted paths} for short ---
  under a suitable encoding as tuples of sets of nodes. We give such an encoding
  (for a larger class of graphs) in Example~\ref{ex:paths}. The number of
  rooted paths grows linearly in the size of the tree;
  Theorem~\ref{thm:reparam-intro} then says that they can be mapped in a finite-to-one
  fashion to tree nodes. To do so explicitly, we can map a rooted path
  to its endpoint --- this is in fact one-to-one, as well as MSO-definable
  (Example~\ref{ex:gpath}).
\end{example}

\subparagraph{MSO (set) interpretations (\S\ref{sec:set-interpretations}).}

Theorem~\ref{thm:reparam-intro} bears a close resemblance to Bojańczyk's \enquote{Seed
  Lemma}\footnote{The lemma is stated without proof in~\cite{PolyregSurvey}. All
  the proof ingredients may be found in~\cite[Section~2]{PolyregGrowth}, even
  though the statements given there are formally slightly different. There have been a few later generalisations~\cite{FinerReparam,RabinovichWoLLIC,RabinovichICALP} which we discuss in Remark~\ref{rem:reparam-ext} at the beginning of the conclusion.}~\cite[Lemma~6.2]{PolyregSurvey}. While our conclusions are slightly weaker,
the Seed Lemma as stated in~\cite{PolyregSurvey} only applies to MSO
formulas over strings with first-order free variables.

The original purpose of this lemma was to prove a \emph{dimension minimization}\footnote{When 
	the set of output nodes/positions of an MSO interpretation is parameterised by $k$-tuples 
	of input nodes/positions, $k$ is called the \emph{dimension}, and is an upper-bound 
	of the degree of polynomial growth (relative size of the output w.r.t.\ the input).}
property for \emph{MSO interpretations} on strings, a logical formalism for
defining functions~\cite[Theorem~6.1]{PolyregSurvey}. Our results on MSO queries
therefore allow us to extend this property to trees. And since we can handle
\emph{set} queries, we are even led to prove that the more expressive
\enquote{MSO \emph{set} interpretations} (introduced by Colcombet and
Löding~\cite{ColcombetL07}) enjoy dimension minimization.
\begin{theorem}\label{thm:msosi-min}
  The output size of an \emph{MSO set interpretation} $\cI$ from trees to relational structures grows either polynomially or exponentially in the input size, with a computable degree $k \in \bbN\cup\{\infty\}$.
  Furthermore, when $k<\infty$, one can compute a \emph{$k$-dimensional MSO interpretation} that defines the same function as $\cI$ (up to isomorphism of structures).
\end{theorem}

(Conversely, the growth rate of a $k$-dimensional interpretation has degree at most $k$.)
\begin{example}\label{ex:comb-intro}
  The following function can be defined by an MSO set interpretation:
  \[ \underbrace{S^{n}(\underline{0})}_{\mathclap{\text{size}\ n+1}} \quad \mapsto \quad \underbrace{a(b^{n}(c),a(b^{n-1}(c),\dots a(b(c),c)\dots))}_{\text{size \( n + (n+1)(n+3)/2 = \Theta(n^{2}) \)}} \quad(\text{cf.\ Figure~\ref{fig:msots}}).\]
  The algorithm provided by Theorem~\ref{thm:msosi-min} can determine that this
  tree-to-tree function has quadratic growth, and compute an equivalent
  2-dimensional MSO interpretation.
\end{example}
\begin{figure}
  \centering
    \begin{tikzpicture}
      \node (a) at (0,0) {$S$};
      \node (b) at (0,-2) {$S$};
      \node (c) at (0,-4) {$\underline{0}$};
      \draw (a) -- (b);
      \draw (b) -- (c);

      \node at (2,-2) {$\mapsto$};

      \node (a') at (6,0) {$a$};
      \node (a'') at (4,-0.8) {$b$};
      \node (b') at (8,-2) {$a$};
      \node (b'') at (6,-2.8) {$b$};
      \node (c') at (8,-4) {$c$};
      \draw[->] (a') -- (a'');
      \draw[->] (a') -- (b');
      \draw[->] (b') -- (b'');
      \draw[->] (a'') -- (b'');
      \draw[->] (b') -- (c');
      \draw[->] (b'') -- (c');

      \node at (10,-2) {$\mapsto$};

      \node (a') at (14,0) {$a$};
      \node (a'') at (12,-0.8) {$b$};
      \node (b') at (16,-2) {$a$};
      \node (b'') at (14,-2.8) {$b$};
      \node (bb) at (12,-2.8) {$b$};
      \node (c') at (16,-4) {$c$};
      \node (c'') at (14,-4) {$c$};
      \node (cc) at (12,-4) {$c$};
      \draw (a') -- (a'');
      \draw (a') -- (b');
      \draw (b') -- (b'');
      \draw (a'') -- (bb);
      \draw (bb) -- (cc);
      \draw (b') -- (c');
      \draw (b'') -- (c'');
    \end{tikzpicture}
  \caption{A tree-to-graph MSO transduction (left; cf.\ Example~\ref{ex:fot-dag-comb}) followed by the unfolding operation (right). Their composition is the tree-to-tree function of Example~\ref{ex:comb-intro}. As explained in Section~\ref{sec:transducers-intro}, this entails that this function is definable by an MSO set interpretation.}
  \label{fig:msots}
\end{figure}

\subparagraph{Tree transducers (\S\ref{sec:transducers-intro} \& \S\ref{sec:msot-msots}).}

The dimension minimization theorem has several applications to classes of tree-to-tree functions defined by \emph{transducers}, i.e.\ automata with outputs. In some cases, the applications are direct, when these transducers can be translated to MSO set interpretations; in other cases, they require slightly more work.
As an example, let us discuss a seminal result concerning (total deterministic) \emph{macro tree transducers}, introduced in~\cite{Macro}:
\begin{theorem}[{Engelfriet and Maneth~\cite[Theorem~7.1]{MacroMSOLinear}}]\label{thm:em}
  It is decidable whether the output size of a macro tree transducer \emph{(MTT)} has linear growth. Furthermore, if it does, then the MTT can be translated to a 1-dimensional MSO interpretation, a.k.a.\ an \emph{MSO transduction}, that defines the same function.
\end{theorem}

In general, there exist MTTs not equivalent to any MSO set interpretation.\footnote{The reason is simple: MSO set interpretations have at most exponential growth, whereas MTTs may have doubly exponential growth.} However, we still manage to recover the above result as a corollary of Theorem~\ref{thm:msosi-min}, although we do not know how to extend it to polynomial growth. Our new proof combines some short and conceptual arguments on MSO set interpretations with effective translations, found in the literature, between machine models --- in fact, \emph{we do not even need to recall the definition of an MTT}. We give more details about this, and more results on transducers, in Section~\ref{sec:transducers-intro}.

\subparagraph{Polynomial size-to-height increase (\S\ref{sec:msosi-height} \& \S\ref{sec:mtt-height}).}

More recently, Gallot, Maneth, Nakano and Peyrat have used arguments
similar to the aforementioned~\cite{MacroMSOLinear} to establish the decidability of \emph{linear
  size-to-height increase} for macro tree transducers~\mbox{\cite[Theorem~17(1)]{LSHI}}. This asymptotic property may be
illustrated by the function of Example~\ref{ex:comb-intro}, which maps an input
$S^{n}(\underline{0})$ to an output of height $n+1$. We demonstrate that our
methods are also applicable to such questions of output height (Gallot et al.'s
result can be recovered by testing whether $k\leqslant1$).
\begin{theorem}\label{thm:pshi}
  Let $f$ be a tree-to-tree function defined either by an MSO set interpretation
  or by an MTT. Then $\height \circ f$ has a well-defined and computable degree of growth in $\bbN\cup\{\infty\}$. 
\end{theorem}

Our proof reduces the case of MSO set interpretations to Claim~\ref{clm:query-vs-ambiguity-intro} on MSO queries, and the case of MTTs to the properties of another transducer model that we now discuss.

\subparagraph{Nondeterministic top-down tree transducers (\S\ref{sec:top-down}).}

Of all tree transducer models mentioned in this paper, this is the only one whose definition we recall, in order to directly manipulate it. It is also the only nondeterministic one, which entails that an input tree may be mapped to multiple outputs. Therefore, we need to extend our definition of growth rate from functions to binary relations: for $R \subset \{\text{trees}\} \times \bbN$,
\[ \growth[R] \colon n \in \bbN \mapsto \max \{ m \mid (t,m) \in R,\; |t| \leqslant n\}\quad \text{with the convention}\ \max\varnothing = 0. \]
\begin{theorem}\label{thm:top-down-intro}
  The output size of a \emph{partial nondeterministic top-down tree transducer} has a well-defined degree of growth in $\bbN\cup\{\infty\}$ that can be computed:
  \begin{itemize}
    \item in \emph{exponential time} in general;
    \item in \emph{polynomial time} when the transducer is \emph{total}.
  \end{itemize}
\end{theorem}

For strings as inputs (seen as unary trees), this was proved by Berglund, Drewes and van der Merwe~\cite[Theorems~20 and~24]{BerglundDM18}, by reducing the problem to computing the degree of growth of the ambiguity of a nondeterministic automaton.\footnote{They rely on the work of Allauzen, Mohri and Rastogi~\cite{AllauzenMR11} on nondeterministic automata with $\varepsilon$-transitions (cf.\ Section~\ref{sec:related-work-intro}). This allows them to handle top-down string-to-tree transducers with $\varepsilon$-transitions, which means that our result is not entirely a generalisation of theirs.}
Very similarly, we reduce Theorem~\ref{thm:top-down-intro} to the growth of $\bbN$-weighted tree automata. Unlike for our other results on transducers, we get elementary complexity bounds because the proof does not involve the MSO-based part of our toolbox.
Also, Berglund et al.~\cite{BerglundDM18} consider both the \enquote{full output size} counting all nodes, and the \enquote{yield output size} counting only the leaves. We prove a parameterised version of Theorem~\ref{thm:top-down-intro} that specialises to both.

\begin{remark}\label{rem:total-det-top-down}
  For trees as input, the fact that the degree of growth is well-defined and computable 
  for deterministic top-down tree transducers with regular lookahead:
  \begin{itemize}
    \item already appears (under a different guise) in the work of Aho and Ullman from the late 1960s~\cite{AhoU71}, as we explain in Section~\ref{sec:related-work-trans};
    \item is yet another consequence of Theorem~\ref{thm:msosi-min}, since MSO set interpretations subsume these transducers, as explained in the next subsection.
  \end{itemize}
\end{remark}

\subsection{More on transducers and MSO}
\label{sec:transducers-intro}

We show that MSO set interpretations can express all \emph{MSO transductions with sharing}: tree-to-tree functions of the form
\[ \unfold \circ [\text{tree-to-graph MSO transduction (MSOT)}] \]
where the unfolding operation \enquote{decompresses} a graph representation of a tree that allows isomorphic subtrees to be shared. An example is given in Figure~\ref{fig:msots}. In fact, we prove more (indeed, the identity is an MSO transduction):
\begin{proposition}[proved in~\S\ref{sec:msot-msots}]\label{prop:msot-msots}
  Every function of the form $\text{MSOT} \circ \unfold \circ \text{MSOT}$ between relational structures can also be defined by an MSO set interpretation, effectively.
\end{proposition}

MSO transductions with sharing form a well-established class of tree transformations, that includes some functions with exponential growth --- which means that working with \emph{set} interpretations matters here. While we prefer to work with their MSO-based definition, they can be characterised by several (total deterministic) transducer models:
\begin{itemize}
  \item attributed tree transducers with regular lookaround~\cite{AttributedMSO};
  \item tree-walking tree transducers\footnote{Actually these are very close syntactically to attributed tree transducers. We refer to~\cite[Section~3.2]{tito24} for further discussion of this point.} with regular lookaround (cf.~e.g.~\cite[Chapter~8]{courcellebook}), which syntactically generalise top-down tree transducers, hence Remark~\ref{rem:total-det-top-down};
  \item a suitable restriction of macro tree transducers~\cite{HashimotoM23}.
\end{itemize}
As for general MTTs, they can be translated to invisible pebble tree transducers~\cite[Corollary~42]{InvisiblePebbles}, which in turn are equivalent in expressive power to compositions of two tree-walking tree transducers with lookaround~\cite[Theorem~53]{InvisiblePebbles}. \emph{In this paper, we do not define any of these tree transducer models,} and do not work with them. However, the information that we have just given situates our work in relation to the literature, and suffices to deduce that
\[ \text{MTT} \subset \text{invisible pebbles} = (\unfold\circ\text{MSOT})^2 \subset \unfold \circ [\text{MSO set interpretation}] \]
using Proposition~\ref{prop:msot-msots} for the last inclusion. Theorem~\ref{thm:em} on MTTs is then a consequence of:
\begin{corollary}[of Theorem~\ref{thm:msosi-min}]\label{cor:msosis}
  Given a tree-to-graph function $f$ defined by an MSO set interpretation, it is
  decidable whether $\unfold \circ f$ has linear growth. Furthermore, if that is
  the case, then one can compute an MSO transduction that defines $\unfold \circ f$.
\end{corollary}
\begin{proof}
  We apply Theorem~\ref{thm:msosi-min} to $f$ to decide whether its growth rate
  is at most linear.
  \begin{itemize}
    \item If it is, then one can compute an MSOT that defines $f$.
          Thanks to Proposition~\ref{prop:msot-msots}, we get an MSO set interpretation defining $\unfold\circ f$.
          A second invocation of Theorem~\ref{thm:msosi-min} is enough to finish the job.
    \item Otherwise, $\unfold\circ f$ also grows faster than linearly, because
          $|\unfold(\graph)|\geqslant|\graph|$ for any rooted DAG $\graph$ (each
          vertex of the graph $\graph$ is reachable from the root, and therefore
          has at least one \enquote{copy} in $\unfold(\graph)$). \qedhere
  \end{itemize}
\end{proof}

\noindent
To conclude this subsection, let us discuss \emph{string transducers}. String-to-string MSO interpretations characterise the class
of \emph{polyregular functions}~\cite[Theorem~7(1)]{msoInterpretations}, whose
study has become quite active since the late 2010s
(see~\cite{PolyregSurvey,Kiefer24} for surveys). The recently introduced and larger class
of lexicographic string functions~\cite{Lex} also has, among its equivalent
definitions, a restriction of MSO \emph{set} interpretations.
Finally, the second author of the present paper has shown, together with Colcombet and Ohlmann, that MSO set interpretations can define the same functions as certain machine models~\cite{StringMSOSI}, cf.~\S\ref{sec:related-work-trans}.

For strings encoded as unary trees, the unfolding operation is the identity. Therefore,
\begin{align*}
  \text{invisible pebble string transducer} &\subset \text{MSOT} \circ \unfold \circ \text{MSOT}
  \subset \text{MSO set interpretation}
\end{align*}
By applying Theorem~\ref{thm:msosi-min} we get the following result, which constitutes some modest progress towards a conjecture about these string transducers by Douéneau-Tabot~\cite[Conjecture~4.56]{gaetanPhD}:
\begin{corollary}
  Every invisible pebble string transducer of polynomial growth computes a polyregular function. 
\end{corollary}

\subsection{Some closely related work: ambiguity \& $\bbN$-weighted automata}\label{sec:related-work-intro}

We only discuss a small sample of the related work here, which has inspired our
technical development in Section~\ref{sec:ambiguity}. The rest is covered partially
throughout the paper, and partially in the conclusion
(Sections~\ref{sec:related-work-trans} and~\ref{sec:related-work-misc}).

\subparagraph{Automata on strings.}

Through the correspondence between
ambiguity and $\bbN$-rationality, the structural characterisation of finitely
ambiguous automata --- i.e.\ whose growth rate has degree zero --- by Weber and
Seidl~\cite[Theorem~3.1]{WeberSeidl} is very similar (though not identical) to
Mandel and Simon's earlier decision procedure~\cite[Corollary~5.2]{MandelS77} for the boundedness of
\mbox{$\bbN$-weighted} automata.\footnote{Around the same time (1977), Jacob independently obtained a more general decidability result~\cite{Jacob1977}, which we discuss towards the end of Section~\ref{sec:related-work-misc}.} Weber and Seidl~\cite{WeberSeidl} also
prove\footnote{They are not the first to give a decidable characterisation, though their algorithm is the
  first one that runs in polynomial time. See~\cite[Section~3]{AmbiguitySurvey}
  for a discussion of earlier work. The complexity bounds explicitly stated
  in~\cite{WeberSeidl} are worse than those of~Theorem~\ref{thm:degree-ambiguity}, but in
  fact the actual time complexity of their algorithms match ours; this becomes
  clear in the subsequent work of Allauzen et al.~\cite{AllauzenMR11}.} the
special case of Theorem~\ref{thm:degree-ambiguity} over strings; their algorithms were
extended by Allauzen, Mohri and Rastogi~\cite{AllauzenMR11} to automata with
$\varepsilon$-transitions. The characterisation given
in~\cite{WeberSeidl,AllauzenMR11} of the degree of growth in
$\bbN\cup\{\infty\}$ of the ambiguity of a word automaton admits a
\enquote{natural} extension to \mbox{$\bbN$-weighted} automata, that has been proved by
Jungers, Protasov and Blondel~\cite[Theorem~3]{Jungers2008}\footnote{Strictly speaking, they do not talk about automata; we discuss the connection in Section~\ref{sec:open-problems}.}
and later (independently) by Douéneau-Tabot, Filiot and
Gastin~\cite[Section~5.2]{Marble} --- leading explicitly in both cases to
polynomial-time algorithms.\footnote{While Douéneau-Tabot et al.~\cite{Marble}
  cite Weber and Seidl's results~\cite{WeberSeidl}, Jungers et
  al.~\cite{Jungers2008} seem to have discovered their $\bbN$-weighted version
  independently (though they reuse the work of Mandel and Simon~\cite{MandelS77}). The criteria that appear
  in~\cite{WeberSeidl,AllauzenMR11} for finite ambiguity and exponential
  ambiguity have also been rediscovered by Crespi, Cybenko and
  Jiang~\cite{Crespi2008}, who relate their work on nondeterministic automata to
  the notion of joint spectral radius (cf.\ the discussion following Theorem~\ref{thm:precise-exp} in \S\ref{sec:open-problems}).}
We further extend this characterisation to trees.

The work of Aho and Ullman~\cite{AhoU71}, mentioned earlier in relation to Theorem~\ref{thm:top-down-intro}, exhibits similar structural patterns in its study of a degree of growth. Their patterns are slightly more complicated and involve subsets (hence exponential blowup); modulo this difference, one may compare the \enquote{barbells} of \S\ref{sec:ambiguity} with~\cite[Lemma~4.2]{AhoU71} and the \enquote{heavy cycles} with~\cite[Theorem~4.6]{AhoU71}. In Section~\ref{sec:related-work-trans}, we explain why the existence and computability of the degree of growth of $\bbN$-weighted automata over strings is actually an easy consequence of~\cite{AhoU71} --- a work that predates all other citations in this subsection by several years!

\subparagraph{Trees.}

Extending the ideas of his paper with Weber~\cite{WeberSeidl}, Seidl gave a
decision procedure in cubic time for finite ambiguity of tree
automata~\cite[Theorem~2.5]{Seidl}, based again on a structural criterion.
Caralp, Reynier and Talbot proved an analogous result for
$\bbN$-weighted visibly pushdown automata~\cite[Theorem~1]{VPAmult} and
deduced the case of $\bbN$-weighted tree automata~\cite[\S6]{VPAmult}. The
latter plays a crucial role in our proof of Theorem~\ref{thm:degree-weighted}.

Paul's work~\cite[Chapter~8]{ErikPaulDiploma} on the structure of polynomially ambiguous
tree automata, which we mentioned just before Theorem~\ref{thm:degree-ambiguity}, also leverages
Seidl's theorem. Although Paul does not propose a decision procedure, a
non-effective version of the reparameterisation of MSO set queries
(Theorem~\ref{thm:reparam-intro}) could almost be derived from~\cite[Lemmas~8.11~\&~8.20]{ErikPaulDiploma}. Furthermore, we
reuse his ideas to prove our characterisation of exponential growth.

\subsection{Conventions on trees}

A \emph{ranked alphabet} is a finite set $\Sigma$ given together with a
\emph{rank function} $\rank:\Sigma \rightarrow \bbN$; we use the notation $\Sigma = \{a : \rank(a),\; \dots\}$ to define concrete ranked alphabets. The set $\tree\Sigma$
of \emph{trees} over $\Sigma$ is freely inductively generated by:
\[ \text{for}\ a\in\Sigma\ \text{and}\ \overbrace{t_{1},\dots,t_{\rank(a)}}^{\mathclap{\text{empty family when}\ \rank(a)=0}} \in \tree\Sigma,\quad a(t_{1},\dots,t_{\rank(a)}) \in \tree\Sigma ~\text{.}\]
This inductive definition means that trees are subject to structural equality:
\[ a(t_{1},\dots,t_{k}) = b(t'_{1},\dots,t'_{\ell}) \iff a = b\ (\text{implying}\ k=\ell)\ \text{and}\ \forall i,\; t_{i} = t'_{i} ~\text{.}\]
A tree $t\in\tree\Sigma$ has a set of \emph{nodes}, with the usual meaning. We shall use the identification
\[ \nodes(a(t_{1},\dots,t_{k})) = \{\text{root of}\ t\} \cup \bigcup_{i=0}^{k} \nodes(t_{i}),\quad \text{the union being disjoint.} \]
Even though it is formally incompatible with structural equality (in $a(t,t)$,
do the two copies of $t$ have the same nodes?), this slight abuse will
not cause any issue for us in this paper. (See the \enquote{tree structures} of
Section~\ref{sec:queries} for a rigorous treatment.)
Each node has a \emph{label} in $\Sigma$, and the nodes whose label have rank 0
are the \emph{leaves}. To relate nodes in a tree, we use standard terminology:
\emph{parent}, $i$-th \emph{child}, \emph{ancestor/descendant}, etc.

A \emph{(one-hole) context} over $\Sigma$ is a tree over
$\Sigma\cup\{\square : 0\}$ with exactly one leaf labeled by $\square$, 
the \emph{hole}. A context $C$ can be applied to a tree by
\enquote{plugging the root of $t$ into the hole of $C$}, producing a tree
$C[t]$; for example, $a(b(\square),c)[a(c,c)] = a(b(a(c,c)),c)$. Contexts can be
composed so that $(C\circ C')[t] = C[C'[t]]$, and $\square$ is the neutral element 
for this composition.

\section{Growth of $\bbN$-weighted tree automata (and ambiguity)}\label{sec:ambiguity}

In this section, we prove the following result, which implies Theorem~\ref{thm:degree-ambiguity}:
\begin{theorem}\label{thm:degree-weighted}
  Every $\bbN$-weighted tree automaton has a well-defined (in
  $\bbN\cup\{\infty\}$) degree of growth. Exponential vs polynomial growth can be decided in
  quadratic time, and the degree can be computed in cubic time.
\end{theorem}

\noindent
Let us fix a ranked alphabet $\Sigma$. As usual, a \emph{tree automaton} over $\Sigma$ consists of:
\begin{itemize}
  \item a finite set of \emph{states}, with a subset of \emph{accepting states};
  \item a set of \emph{transitions} --- each transition is a tuple
        $(q_{1},\dots,q_{\rank(a)},a,q)$ where $a\in\Sigma$ and
        $q_{1},\dots,q_{\rank(a)},q$ are states. 

\end{itemize}
An \emph{$\bbN$-weighted} tree automaton additionally comes with:
\begin{itemize}
  \item a \emph{weight} in $\bbN\setminus\{0\}$ for each transition, and for each accepting state.
\end{itemize}

\noindent
A \emph{run} of the automaton over an input tree is a map from its nodes to
states where for each node labeled by $a\in\Sigma$, its state and its children's
states are allowed to be $q$ and $q_{1},\dots,q_{\rank(a)}$ (in order) respectively only if
$(q_{1},\dots,q_{\rank(a)},a,q)$ belongs to the set of transitions --- we then
call it the \emph{transition taken at this node}. A run is \emph{accepting} if
the state at the root is an accepting state.

On a given input, the
\emph{ambiguity} of a tree automaton is the number of accepting runs.
We have already mentioned the notions of finite/polynomial/exponential
ambiguity. More generally a tree automaton is called \emph{ambiguous} if its
ambiguity is at least two on some input, and it is \emph{unambiguous} otherwise.

In the $\bbN$-weighted case, the \emph{weight of a run} is the product of the
weights of the transitions taken at each node. The \emph{value} of an
$\bbN$-weighted tree automaton over a given input is the sum, over all accepting
runs, of $[\text{weight of the run}]\times[\text{weight of the state at the root}]$.
Thus, the ambiguity is a lower bound for the value, and they coincide when all transitions and accepting states have weight 1.

\begin{remark}
The way we write the transitions suggests a bottom-up point of view on tree
automata; the top-down version would be $(q,a,q_{1},\dots,q_{m})$. However,
we insist that for nondeterministic tree automata, this is only a matter of
 perspective: the bottom-up and top-down versions are syntactically
isomorphic. Therefore, we can freely apply results stated for top-down automata,
such as those of~\cite{Seidl} (where accepting states are said to be \enquote{initial}).
\end{remark}

A run of a tree automaton over a context $C$ is like a run over an input tree (and its weight is defined similarly),
except that the hole $\square$ may be assigned any state consistent with the transition at its parent node. 
\begin{definition}\label{def:acc-rel}
  For $q,q'$ two states and $C$ a one-hole context, we denote by $q' \to_C q$
  the existence of a run on $C$ with $q'$ at the hole and $q$ at the root. When
  we drop the subscript, $q' \to q$ means that there exists a context $C$ such
  that $q' \to_C q$.
\end{definition}

A state $q$ is \emph{accessible} if there exists a tree $t$ and a run on $t$ with $q$ at the root. We say that $q$ is \emph{co-accessible} if $q \to q_a$ where $q_a$ is an accepting state.

\subsection{A preliminary step: trimming in linear time (and reusable lemmas)}\label{sec:trim}

A tree automaton is \emph{trim} if all its states are accessible and co-accessible. Tree automata can be trimmed by keeping only the states that are both accessible and co-accessible, and the transitions that do not involve some removed state. This transformation preserves the set of accepting runs over each input tree; therefore, it preserves the ambiguity, as well as the value in the $\bbN$-weighted case.

It is often stated without proof that \enquote{trimming ranked tree automata is standard (and can be performed in linear time)} (quoting Caralp et al.~\cite[\S1]{VPAtrim}); for another occurrence of this claim, see e.g.\ Seidl's~\cite[Proposition~1.1]{Seidl}, which he uses to get a quadratic-time algorithm to decide unambiguity~\cite[Proposition~1.2]{Seidl}. For the sake of completeness, we sketch the folklore linear-time trimming algorithm.

First, to remove inaccessible states, we use:
\begin{lemma}\label{lem:acc-lin}
  The set of accessible states of a tree automaton is computable in linear time.
\end{lemma}
\begin{proof}
  While there is an intuitive graph-traversal-like algorithm that runs in polynomial time (cf.~e.g.\ \textit{Tree Automata Techniques and Applications}~\cite[Theorem~1.1.8]{TATA} or the \textit{Handbook of Automata Theory}~\cite[Section~2.3]{HandbookAutomataTrees}), it is not so trivial to achieve a linear running time. In the appendix (\S\ref{app:acc-lin}), we explain a folklore solution based on a reduction to \emph{Horn satisfiability}, and relate it to the graph traversal intuition.
\end{proof}

Let us say that a one-hole context is \emph{shallow} if its hole is a child of
the root.
\begin{definition}
  The \emph{shallow digraph} of a tree automaton has all the states as vertices,
  and it has an edge from $q'$ to $q$ if and only if there is some shallow
  context $C$ such that $q' \to_{C} q$.
\end{definition}

Since a context whose hole is at depth $h$ can be (uniquely) written as a composition of $h$
shallow contexts, the reachability relation in this digraph is equal to
the relation `$\to$' on states.
\begin{lemma}\label{lem:shallow-digraph}
  The shallow digraph can be computed in linear time.
\end{lemma}
\begin{proof}
  After precomputing the accessible states (Lemma~\ref{lem:acc-lin}),
  we build the shallow digraph by successively adding edges: for each
  transition $(q_{1},\dots,q_{m},a,q)$,
  \begin{itemize}
    \item if $q_{1},\dots,q_{m}$ are all accessible, we add all the edges $(q_{1},q),\dots,(q_{m},q)$;
    \item if they are all accessible except for a single $q_{i}$, we add only $(q_{i},q)$;
    \item otherwise we do not add any edge for this transition. \qedhere
  \end{itemize}
\end{proof}

\noindent
The trimming algorithm first removes the inaccessible states using
Lemma~\ref{lem:acc-lin}, then performs a linear-time backwards traversal of the
shallow digraph starting from the accepting states to determine the remaining
co-accessible states.

\subsection{Deciding exponential vs polynomial growth}\label{sec:exp-vs-poly}

\begin{definition}
  The \emph{value of a context $C$ from a state $q'$ to a state $q$} is the sum
  of the weights of all runs on $C$ with $q'$ at the hole and $q$ at the root.

  We write $q' \twoheadrightarrow_C q$ when this value is at least two. As in
  Definition~\ref{def:acc-rel}, we drop the subscript to existentially quantify on $C$.
  A \emph{heavy cycle} consists of a state $q$ and a context $C$ such that
  $q \twoheadrightarrow_{C} q$.
\end{definition}

Heavy cycles generalise Weber and Seidl's condition (EDA) characterising
exponentially ambiguous word automata~\cite[Theorem~4.1]{WeberSeidl}.\footnote{See also
  Ibarra and Ravikumar's~\cite[Theorem~3.1]{IbarraR86}. Weber and Seidl also
  cite Reutenauer's 1977 PhD thesis for this condition, but we have not been
  able to access it.}
\begin{theorem}\label{thm:heavy-cycle-exp}
  If a \emph{trim} $\bbN$-weighted tree automaton admits a heavy cycle, then it has
  exponential growth; otherwise, its growth is bounded by a
  polynomial.\footnote{We do not yet say \enquote{polynomial growth} because at
    this point, we have not yet proved that the growth rate is $\Theta(n^{k})$
    for some $k\in\bbN$. This could be quickly deduced from the case of
    ambiguity of tree automata~\cite[Corollary~8.27]{ErikPaulDiploma}, but we
    want to (re)prove it on the way to making it effective.}
  This condition can be tested in quadratic time.
\end{theorem}

This subsection is dedicated to proving the above theorem.
We fix a trim $\bbN$-weighted tree automaton $\cA$ over a ranked alphabet $\Sigma$.

\subparagraph{Heavy cycle $\Rightarrow$ exponential growth.}

To get an exponential upper bound, observe that the number of runs on an input tree of size $n$ is at most $(\text{number of states})^{n}$, and the weight of each run is a product of linearly many bounded numbers. 
The lower bound is established by a simple pumping argument, that already appears
in~\cite[proof of Proposition~2.1]{Seidl}. Let $q \twoheadrightarrow_{C} q$; let
$q \to_{C'} q_{\mathsf{a}}$ where $q_{\mathsf{a}}$ is accepting (by trimness,
$q$ is co-accessible) and $t$ be a tree that admits a run where the root has
state $q$ (states are accessible). Then $t_{n} = C'[C^{n}[t]]$ has size $O(n)$ and value at least $2^{n}$ --- this can be seen from runs of the form
$q \to_{C} \dots \to_{C} q \to_{C'} q_{\mathsf{a}}$.

\subparagraph{No heavy cycle $\Rightarrow$ polynomial bound.}

Our argument is heavily inspired by~\cite[\S8.1--8.2]{ErikPaulDiploma}.

Suppose also, without loss of generality, that the $\bbN$-weighted tree automaton $\cA$ has a
single accepting state $q_{\mathsf{a}}$ which moreover has weight~1, and that
$\cA$ has value $\geqslant2$ on some input (otherwise the result is trivial).
Let us call a state $q$ \emph{heavy} if for some input, the sum of weights of
runs with $q$ at the root is at least two.

\begin{claim}\label{clm:subtree}
  The states $q$ such that $q_{\mathsf{a}}\to q$ are heavy. Moreover, in any run, the set of nodes labeled by such states is
  closed under taking ancestors.
\end{claim}
\begin{proof}
  Our assumptions on $\cA$ entail that $q_{\mathsf{a}}$ is heavy. To deduce the
  first half of the claim, note that if $q_{1} \to_{C} q_{2}$ and $q_{1}$ is
  heavy (with witness $t$), then so is $q_{2}$ (with witness $C[t]$).

  Let us fix an input tree and a run. Consider a node in the above-defined set,
  with state $q$, and one of its ancestors with state $q'$. They can be
  described as the respective roots of $t$ and $C[t]$ for some decomposition of
  the input as $C'[C[t]]$. We then have $q \to_{C} q'$
  as part of our fixed run. From the assumption that $q_{\mathsf{a}} \to q$, we then get $q_{\mathsf{a}} \to q'$.
\end{proof}

Call a transition $(q_1,\dots,q_m,a,q)$ a \emph{final bridge}\footnote{Inspired by Paul's notion of \enquote{bridge} from~\cite[Definition~8.5]{ErikPaulDiploma}; and the claim below is similar to~\cite[Proposition~8.13]{ErikPaulDiploma}.} when
$q_{\mathsf{a}} \to q$ but $q_{\mathsf{a}} \not\to q_i$ for $i=1,\dots,m$. Note
that the second half of the definition is vacuously true for $m=0$.
  \begin{claim}
    Each accepting run contains exactly one occurrence of a final bridge.
  \end{claim}
  \begin{proof}
    Let us fix an input tree and an accepting run on that input. Let $S$ be the set of nodes considered in Claim~\ref{clm:subtree}. The final bridges are exactly the transitions for the maximal nodes in $S$ (for the descendant ordering). Since $S\neq\varnothing$ (it contains the root), it suffices to show that it has only one maximal element.

    Suppose for the sake of contradiction that $S$ has at least two distinct maximal nodes. They are descendants of the $i$-th and $j$-th children, respectively, of their lowest common ancestor, for some $i\neq j$. Let $(q_1,\dots,q_m,a,q)$ be the transition taken at this common ancestor; then $q$, $q_{i}$ and $q_{j}$ are states labeling nodes in $S$. By Claim~\ref{clm:subtree} again, $q_j$ is heavy; this is witnessed by some tree $t$. Then $q_i \twoheadrightarrow_{C} q$ where $C = a(t_{1},\dots,t_{i-1},\square,t_{i+1},\dots,t_{m})$, with $t_{\ell}$ chosen so that has a run with $q_{\ell}$ at the root for each $\ell$, and $t_{j} = t$. Furthermore, $q_{\mathsf{a}}\to q_{i}$ by definition of $S$, and since the automaton is trim, $q \to q_{\mathsf{a}}$. Thus, we get a heavy cycle $q \twoheadrightarrow_{C'\circ C} q$.
  \end{proof}

We now prove that any trim $\bbN$-weighted tree automaton $\cA$ without a heavy cycle 
has its growth bounded by a polynomial, by induction over the number of states of $\cA$. 

Let $(q_{1},\dots,q_{m},a,q)$ be a final bridge and $t = C[a(t_{1},\dots,t_{m})]$
an input tree. Decomposing the tree in such a way corresponds to choosing a
distinguished node (with label $a$). Consider the accepting runs where this
final bridge is taken at this distinguished node. The sum of their weights can
be factored as the product of:
\begin{description}
	\item[Above the final bridge:] The value of $C$ from $q$ to $q_{\mathsf{a}}$. Since $q_{\mathsf{a}}\to q$, this value must be 1; otherwise, $q \twoheadrightarrow_{C} q_{\mathsf{a}}$ would imply $q \twoheadrightarrow q$, a heavy cycle.
	\item[Below the final bridge:] For each $i\in\{1,\dots,m\}$, the value over $t_{i}$ of the $\bbN$-weighted tree automaton $\cA_{q_{i}}$ obtained from $\cA$ by setting $q_{i}$ to be the new unique accepting state, with weight 1, then trimming. This last step removes at least the state $q_{\mathsf{a}}$, because $q_{\mathsf{a}} \not\to q_{i}$. Furthermore, $\cA_{q_{i}}$ still does not have heavy cycles.

	\item[In-between:] The weight of the final bridge.
\end{description}
The value of $\cA$ on $t$ is the sum, over all possible 
final bridges and all possible nodes where the final bridge may be taken, of the 
product of the above three items. Since $|t_i| \leqslant |t|$, we have:
\[\growth[\cA](n) ~\leqslant~ n \times \sum_{\mathclap{\text{final bridge } b}} \text{weight}(b) \prod_{i=1}^{R} \growth[\cA_{q_i}](n) ~\leqslant~ nW\times \big(\max_{i\leq m} \growth[\cA_{q_i}](n)\big)^R \]
where $R = \max(\rank(\Sigma))$ and $W$ is the sum of weights of all transitions in $\cA$. 

We can perform an induction on the number $N$ of states because the automata $\cA_{q_i}$ have at least one less state than $\cA$, and the sum of weights of their transitions are still bounded by $k$. In the end, we get that $\growth[\cA](n) \leqslant (nW)^{1 + R + \dots + R^{N-1}} = n^{O(1)}$.

\subparagraph{Quadratic-time detection of heavy cycles.}

We distinguish three cases (which are not mutually exclusive): a heavy cycle $q \twoheadrightarrow_{C} q$ is called
\begin{itemize}
  \item a \emph{scalar-heavy cycle} if there is some run $q \to_{C} q$ with weight at least two;
  \item a \emph{center-ambiguous cycle} if there are two runs $q \to_{C} q$ which
        differ on some node on the path from the hole of $C$ to the root;
  \item a \emph{side-ambiguous cycle} if there are two runs $q \to_{C} q$ which
        differ on some node outside the aforementioned path.
\end{itemize}
If there is only one run $q \to_{C} q$, then $q \twoheadrightarrow_{C} q$ must
be scalar-heavy; otherwise, it must be either center-ambiguous or
side-ambiguous. We provide decision procedures below for the existence of each
of these three kinds of heavy cycles. In the following, we assume that the input
automata are trim.

\begin{proposition}
  Scalar-heavy cycles can be detected in linear time.
\end{proposition}
\begin{proof}
  We begin by building the shallow digraph (cf.~Lemma~\ref{lem:shallow-digraph})
  of the input automaton. By performing a linear-time graph traversal, starting
  from the states $q'$ for which there is some transition
  $(q'_{1},\dots,q'_{\ell},a,q')$ of weight at least two, we can determine all
  the states that appear at the root of some run of weight $\geqslant2$ ---
  let us call them \enquote{scalar-heavy states}.

  Then we compute the \emph{strongly connected components} of
  the shallow digraph; there are classic linear-time algorithms for this task (see
  e.g.~\cite[\S6.8]{DurrVie}). Finally, there exists a scalar-heavy cycle if and
  only if, for some transition $(q_{1},\dots,q_{m},a,q)$ and
  $i\in\{1,\dots,m\}$:
  \begin{itemize}
    \item $q_{i}$ and $q$ belong to the same strongly connected component;
    \item either this transition has weight $\geqslant2$ or some $q_{j}$ with $j\neq i$ is scalar-heavy (basically, these two alternatives are respectively \enquote{center-heavy} and \enquote{side-heavy}). \qedhere
  \end{itemize}
\end{proof}

\noindent
Next, recall that for two tree automata $\cA$ and $\cB$ on the same input alphabet, the
states of the \emph{product automaton} $\cA\times\cB$ are the pairs $(\text{state of}\ \cA,\; \text{state of}\ \cB)$, and its transitions are
\[ ((q_{1},p_{1}),\dots,(q_{m},p_{m}),a,(q,p))\ \text{for}\ (q_{1},\dots,q_{m},a,q)\ \text{in}\ \cA\ \text{and}\ (p_{1},\dots,p_{m},a,p)\ \text{in}\ \cB. \]
There are multiple canonical choices for the accepting states of $\cA\times\cB$, but they are not relevant to our algorithmic applications.
We did not mention its weights either, because center-ambiguous and side-ambiguous cycles (and later, barbells) do not depend on weights.

\begin{proposition}\label{prop:center-ambiguous-cycles}
  Center-ambiguous cycles can be detected in quadratic time.
\end{proposition}
\begin{proof}
  This is analogous to Weber and Seidl's decision procedure for their condition
  (EDA) over word automata~\cite[p.~338]{WeberSeidl}; Allauzen et al.\ give more
  details in their extension of this algorithm to handle
  $\varepsilon$-transitions~\cite[Theorem~11]{AllauzenMR11}.

  Given a tree automaton $\cA$, we form the product automaton $\cA\times\cA$,
  and build its shallow digraph. By analogy with~\cite[Lemma~10]{AllauzenMR11},
  one can see that $\cA$ admits a center-ambiguous cycle if and only if the
  graph contains two vertices of the form $(q,q)$ and $(q_{1},q_{2})$ with
  $q_{1}\neq q_{2}$ in the same strongly connected component. See \S\ref{app:center-ambiguous-cycles} of the appendix for more details. 
\end{proof}

\begin{proposition}\label{prop:side-amb}
  Side-ambiguous cycles can be detected in quadratic time.
\end{proposition}
\begin{proof}
  Consider two runs $q \to_{C} q$ that differ at some node $u$ outside the
  hole-to-root path. Let $v$ be the lowest ancestor of $u$ that lies on the
  hole-to-root path.
  \begin{itemize}
    \item Suppose that the transitions taken at $v$ are
          $(q'_{1},\dots,q'_{m},a,q') \neq (q''_{1},\dots,q''_{m},a,q'')$ in our
          two runs. They must satisfy some compatibility conditions: for each
          $i$ there exists a tree that admits at least one run with $q'_{i}$ at
          the root and at least one with $q''_{i}$.
    \item Otherwise, the two runs take the same transition
          $(q_{1},\dots,q_{m},a,q')$ at $v$. Then if $u$ is a descendant of the
          $i$-th child of $v$, then $q_{i}$ is an \emph{ambiguous state}: there
          is some tree --- namely, the subtree below this $i$-th child --- that
          has two different runs reaching $q_{i}$.
  \end{itemize}
  Those two situations correspond respectively to Seidl's conditions (T1.1) and
  (T1.2), whose disjunction is called (T1), from~\cite[Proposition~2.1]{Seidl}.
  Conversely, (T1) implies the existence of a side-ambiguous cycle,
  as illustrated in~\cite[Fig.~2]{Seidl}.

  Finally, Seidl gives a quadratic-time decision procedure for (T1)~\cite[proof of Prop.~2.1]{Seidl}.
  In his description, the step that marks all ambiguous states
is claimed to be in quadratic time without detailed justification. For the sake of completeness, we give an explanation of this step in the appendix (\S\ref{app:side-amb}).
\end{proof}

\subsection{A known characterisation of boundedness}

Our next goal is to study the polynomial degree of growth. To do so, we first
need to recall the degree 0 case.
(The terms \enquote{heavy cycle} and \enquote{barbell} are taken from Douéneau-Tabot's PhD thesis~\cite[\S4.4.2]{gaetanPhD} which is the main inspiration for the arguments in Section~\ref{sec:chara-poly}.)

\begin{definition}\label{def:barbell}
  When $q_1 \neq q_2$ and, for a \emph{same} context $C$ (cf.\ Figure~\ref{fig:barbell}), $q_1 \to_C q_1$ and $q_1 \to_C q_2$ and
  $q_2 \to_C q_2$, we write $q_1 \Rrightarrow_C q_2$ and we call
  this configuration a \emph{barbell}.
\end{definition}
\begin{figure}
  \centering
\begin{tikzpicture}
  \node[circle, draw,thick] (q') at (0,0) {\footnotesize $q$};
  \draw[thick,>=stealth] (q') edge[loop above] node[above] {$C$ (weight $\geqslant2$)} (q');

  \node[circle, draw,thick] (q) at (3,0) {\footnotesize $q$};

  \draw[thick,>=stealth] (q) edge[loop left] node[left] {$C$} (q);
  \draw[thick,>=stealth] (q) edge[loop right] node[right] {$C$} (q);

  \node[circle, draw,thick] (q1) at (6,0) {\footnotesize $q_1$};
  
  \node[circle, draw,thick] (q2) at (8,0) {\footnotesize $q_2$};

  \draw[thick,>=stealth,->] (q1) edge[bend left] node[above] {$C$} (q2);
  \draw[thick,>=stealth] (q1) edge[loop above] node[above] {$C$} (q1);
  \draw[thick,>=stealth] (q2) edge[loop above] node[above] {$C$} (q2);

\end{tikzpicture}

\caption{A scalar-heavy cycle (left), a center- or side-ambiguous cycle (middle) and a barbell (right), with the assumption that the $C$-runs are distinct in the middle and $q_1\neq q_2$ on the right.}\label{fig:barbell}
\end{figure}

Note that barbells do not take weights into account.

\begin{theorem}[{Caralp, Reynier \& Talbot~\cite[\S6]{VPAmult}\protect\footnote{See Caralp's dissertation~\cite[Théorème~4.4.1]{Caralp} for the proof (in French).}}]\label{thm:finamb}
  A trim $\bbN$-weighted tree automaton is bounded if and only if it does not contain
  any heavy cycle, nor any barbell.
\end{theorem}

A variant of this theorem was proved earlier by Seidl~\cite[Section~2]{Seidl}.
More precisely, for trim tree automata, he shows that unbounded ambiguity is
equivalent to the disjunction of two conditions (T1) and (T2). While these
conditions are stated in a local fashion using a \enquote{branch automaton}
(cf.~Remark~\ref{rem:branch-automata}), whereas Caralp et al.~\cite{VPAmult} speak of
one-hole contexts, the proofs in~\cite[\S2]{Seidl} make it clear that:
\begin{itemize}
  \item (T1) corresponds precisely to side-ambiguous cycles, as we saw while proving Prop.~\ref{prop:side-amb};
  \item (T2) corresponds precisely to barbells (cf.~\cite[Fig.~3]{Seidl}).
\end{itemize}
Thanks to this last item we can also reuse the following result from~\cite{Seidl}:
\begin{lemma}[Seidl~{\cite[proof of Proposition~2.2]{Seidl}}]\label{lem:all-pairs-barbell}
  One can compute in cubic time all the pairs of states between which there is a
  barbell, i.e.\ the set $\{(q_1,q_{2}) \mid q_{1} \Rrightarrow q_{2}\}$.
\end{lemma}

Let us mention and rephrase the key idea of Seidl's algorithm, which is substantially
the same as in the case of word automata (see for instance~\cite[proof of
Theorem~13]{AllauzenMR11}) modulo the fact that a one-hole context does not consist only of its central branch:
\begin{claim}
  Consider the shallow digraph of the product $\cA\times\cA\times\cA$. For each
  pair of states $q,q'$, let us add an edge from $(q,q',q')$ to $(q,q,q')$.
  Then $q_{1} \Rrightarrow q_{2}$ in $\cA$ if and only if:
  \begin{itemize}
    \item the vertices $(q_{1},q_{1},q_{2})$ and $(q_{1},q_{2},q_{2})$ are
          distinct, but belong to the same strongly connected component of this
          extended graph;
    \item this component contains some edge that was already in the shallow digraph of $\cA\times\cA\times\cA$.
  \end{itemize}
\end{claim}
\begin{remark}\label{rem:branch-automata}
  The presentation using \enquote{branch automata} in~\cite{Seidl} seems to be
  chosen in order to make algorithmic testing convenient. For us, this purpose
  is fulfilled by the linear-time construction of the shallow digraph
  (Lemma~\ref{lem:shallow-digraph}) which encapsulates the algorithmics needed to
  work effectively with conditions stated in terms of contexts.
\end{remark}

\subsection{Characterising the polynomial degree of growth}\label{sec:chara-poly}

In this section, we consider the polynomially bounded case. In other words,
\textbf{we assume that our trim $\bbN$-weighted tree automaton $\cA$ has no heavy cycle}.

However, it may have barbells. Those play an important role in determining the
degree of growth, but the situation is a bit subtle (at least more so over trees
than over strings). This is already true in the unweighted case, when looking at ambiguity.

\begin{example}\label{ex:exp-degree}
  Consider, for $N\in\bbN$, the tree automaton whose transitions are:
  \[ \{(c,q'),\; (q',b,q'),\; (q',b,q_{N}),\; (q_{N},b,q_{N}),\; (q_{N},q_{N},a,q_{N-1}),\; \dots,\; (q_{1},q_{1},a,q_{0})\} \]
  (where $\rank(a)=2$, $\rank(b)=1$ and $\rank(c)=0$), with accepting state $q_{0}$.

  It has a single pair of states forming a barbell, namely
  $q' \Rrightarrow q_{N}$ (this can be witnessed by the context $b(\square)$, or
  by $b(b(\square))$, etc.). 
  So the ambiguity on the tree $b^n(c)$ would be $n+1$ (if $N = 0$). 
  For $N=1$, the tree $a(b^n(c),b^n(c))$ has ambiguity $(n+1)^2$ because we have $n+1$ possible runs for each $b^n(c)$. 
  For $N=2$, the tree $a(a(b^n(c),b^n(c)),a(b^n(c),b^n(c)))$ has ambiguity $(n+1)^4$ and so on. 
  In the end, the degree of growth of its ambiguity is ${2^{N}}$. 
\end{example}

We now introduce patterns involving barbells that
witness lower bounds for the degree of growth. They are finite ranked trees over
an infinite alphabet that consists of the original ranked alphabet $\Sigma$
plus, for each one-hole context $C$, a rank 1 node $\pumpnode{C}$.
\begin{definition}\label{def:pumping}
  We define inductively the \emph{pumping patterns} for each state of $\cA$:
  \begin{itemize}
    \item if $\cA$ contains a transition $(q_1,\dots,q_k,a,q)$ and
          $\Pi_{i}$ is a pumping pattern for $q_{i}$, for each $i\in\{1,\dots,k\}$,
          then $a(\Pi_{1},\dots,\Pi_{k})$ is a pumping pattern for $q$;
    \item if $q' \Rrightarrow_{C} q$ and $\Pi'$ is a pumping pattern for $q'$ then
          $\pumpnode{C}(\Pi')$ is a pumping pattern for $q$.
  \end{itemize}
  The \emph{degree} $\deg(\Pi)$ of a pattern $\Pi$ is its number of
  $\pumpnode{-}$ nodes. For $n\in\bbN$, by structural induction on a pumping
  pattern, we set:
  \begin{align*}
    \pump(n,a(\Pi_{1},\dots,\Pi_{k})) &= a(\pump(n,\Pi_{1}),\dots,\pump(n,\Pi_{k}))\\
    \pump(n,\pumpnode{C}(\Pi)) &= C^{n}[\pump(n,\Pi)]
  \end{align*}
\end{definition}
In particular, the base case is that if $\cA$ contains a transition $(c,q)$ with $\rank(c)=0$, then $c$ is a pumping pattern for $q$ and $\pump(n,c)=c$.
\begin{remark}\label{rem:zero-degree-pattern}
  If $t\in\tree\Sigma$ admits a run ending in state $q$ at the root, then $t$
  can be seen as a pumping pattern of degree 0 for $q$. In particular, since
  the tree automaton $\cA$ is assumed to be trim, for any state there is at
  least one pumping pattern.
\end{remark}

As a non-degenerate example, the automaton of Example~\ref{ex:exp-degree}
(parameterised by $N\in\bbN$) admits a pumping pattern $\Pi_{N}$ of degree
$2^{N}$, defined by $\Pi_{0} = \pumpnode{b(\square)}(c)$ and
$\Pi_{i+1} = a(\Pi_{i},\Pi_{i})$, for $i\in \set{0,\ldots,N-1}$.

\begin{claim}\label{clm:pump-lower-bound}
  Let $\Pi$ be a pumping pattern for a state $q$ and $q \to_{C} q_{\mathsf{a}}$ where
  $q_{\mathsf{a}}$ is accepting. Then $|C[\pump(n,\Pi)]|=O(n)$ while the ambiguity over
  $C[\pump(n,\Pi)]$ is at least $n^{\deg(\Pi)}$.
\end{claim}
\begin{proof}
  We only detail here the proof for the lower bound on the ambiguity (the proof of $O(n)$ is in \S\ref{app:pump-lower-bound} of the appendix). It suffices to
  show that there are at least $n^{\deg(\Pi)}$ runs on $\pump(n,\Pi)$ with $q$
  at the root. By structural induction:
  \begin{description}
    \item[case $\Pi=a(\Pi_{1},\dots,\Pi_{k})$:] by definition, there is a
          transition $(q_{1},\dots,q_{k},a,q)$ such that each $\Pi_{i}$ is a
          pumping pattern for $q_{i}$. By induction hypothesis, there are at
          least $n^{\deg(\Pi_{i})}$ runs on $\pump(n,\Pi_{i})$ with $q_{i}$ at
          the root. These runs plus the aforementioned transition can be
          assembled to yield
          $n^{\deg(\Pi_{1})} \times \dots \times n^{\deg(\Pi_{k})} = n^{\deg(\Pi)}$
          distinct runs on $\pump(n,\Pi)$.
    \item[case $\Pi = \pumpnode{C}(\Pi')$:] by definition, $\Pi'$ is a pumping
          pattern for some state $q'$ such that $q' \Rrightarrow_{C} q$. The key
          observation (from~\cite[proof of Proposition~2.2]{Seidl}) is that
          there are at least $n$ runs over the context $C^{n}$ with $q'$ at the
          hole and $q$ at the root, namely:
          \[ \overbrace{q' \to_{C} \dots \to_{C} q'\phantom{'}}^{\text{\(m-1\) times \(C\)}} \to_{C} \overbrace{\phantom{'}q \to_{C} \dots \to_{C} q}^{\text{\(n-m\) times \(C\)}} \]
		  There are $n^{\deg(\Pi')}$ runs over $\pump(n,\Pi')$ with $q'$ at the root, each such run can be combined with each of the $n$ runs over $C^n$ to obtain a distinct run over the tree $C^{n}[\pump(n,\Pi')]$. In total we get $n\times n^{\deg(\Pi')}=n^{\deg(\Pi)}$ runs on tree $\pump(n,\Pi) = C^{n}[\pump(n,\Pi')]$.
		  \qedhere
  \end{description}
\end{proof}

\noindent
The ambiguity of $\cA$ on an input is at most its value on the same input, and
we have assumed in this subsection that $\cA$ is polynomially bounded. Together
with Claim~\ref{clm:pump-lower-bound}, this entails that there must be a global bound
on the degrees of pumping patterns --- which ensures that the degree of a state,
introduced below, is a well-defined natural number. (It is an adaptation to
tree automata of the notion of \enquote{height of a state}
from~\cite[\S4.4.2]{gaetanPhD}.)
\begin{definition}
  The \emph{degree} $\deg(q)$ is the maximum degree of a pumping pattern for $q$.
\end{definition}

\begin{claim}\label{clm:deg-ineq}
  These inequalities follow from the definition of pumping patterns:
    \begin{itemize}
    \item if $\cA$ contains a transition $(q_1,\dots,q_k,a,q)$ then
          $\deg(q_1) + \cdots + \deg(q_k) \leqslant \deg(q)$ -- in particular,
          $\deg$ is monotone with respect to $\to$;
    \item if $q' \Rrightarrow q$ then $\deg(q') + 1 \leqslant \deg(q)$.
  \end{itemize}
\end{claim}

\noindent
The degrees of the states are lower
bounds on the polynomial degree of growth of $\cA$. Next, we prove a
matching upper bound. To do so, we start by distinguishing some transitions.
\begin{definition}\label{def:critical}
  A transition $(q_{1},\dots,q_{m},a,q)$ is \emph{critical} when
  \[\deg(q) > \deg(q_{1}) + \cdots + \deg(q_{m}) \qquad [=0\ \text{when}\ m=0].\]
  A \emph{critical node} for a run over some tree is a node where the transition
  taken is critical.
\end{definition}

We may draw an analogy with flows in directed graphs: non-critical nodes satisfy
the \enquote{conservation of flow} equation
$\deg(q) = \deg(q_{1}) + \cdots + \deg(q_{m})$ while critical nodes serve as
\enquote{sources} for the flow of degrees. Since the root is the only
\enquote{sink}, we have:
\begin{claim}\label{clm:crit-degree}
  The number of critical nodes in a (not necessarily accepting) run is bounded
  by the degree of the state at the root.
\end{claim}
\begin{proof}
  For each transition $(q_{1},\dots,q_{m},a,q)$, we have
  $\deg(q) - \deg(q_{1}) - \cdots - \deg(q_{m}) \in \bbN$; for critical transitions this \enquote{entering flow} is strictly positive. Therefore, the sum of entering flows over all nodes bounds the number of critical nodes. In the expression for this big sum, for each non-root node, the degree of its state appears once negatively (from its parent's flow) and once positively (from its own flow). Thus, the only term that does not get canceled out is the degree of the state at the root.
\end{proof}

Furthermore, the set of critical nodes of an accepting run in our $\bbN$-weighted tree automaton $\cA$ almost determines
this run, and the weight of this run is bounded:
\begin{lemma}\label{lem:crit-bounded}
  The sum of weights of accepting runs in $\cA$ over any given input tree whose set of critical
  nodes equals any prescribed value is bounded by a computable function of $\cA$.
\end{lemma}
\begin{proof}
  We build an $\bbN$-weighted tree automaton $\cB$ with the same states as $\cA$, but with input alphabet $\Sigma\times\{0,1\}$ (the ranks are inherited from $\Sigma$).
  For each critical (resp.\ non-critical) transition $(q_{1},\dots,q_{m},a,q)$ in $\cA$, we add a transition $(q_{1},\dots,q_{m},(a,[\text{1 (resp.\ 0)}]), q)$ to $\cB$, with the same weight. This way, the runs in $\cA$ over an input $t\in\tree\Sigma$ whose set of critical nodes is $S\subseteq\nodes(t)$ correspond precisely to the runs in $\cB$ over the tree $t_{S}$ obtained by replacing the label $a$ of each node $\nu$ by $(a,1)$ if $\nu\in S$ or by $(a,0)$ if $\nu\notin S$.

  Therefore, it suffices to show that $\cB$ is bounded. To do so, we apply Theorem~\ref{thm:finamb}. Let us check its assumptions:
  \begin{itemize}
    \item $\cA$ is trim and has no heavy cycles (by assumption), therefore $\cB$ too.
    \item Suppose for the sake of contradiction that $q_{1} \Rrightarrow_{C} q_{2}$ in $\cB$. Then $q_{1} \Rrightarrow_{C'} q_{2}$ in $\cA$, where $C'$ is the context obtained from $C$ by projecting on the first component; thus, $\deg(q_{1}) < \deg(q_{2})$ (we only speak of degrees in $\cA$). Furthermore, by definition of $\Rrightarrow$, we have $q_{1} \to_{C} q_{1}$ and $q_{1} \to_{C} q_{2}$ in $\cB$. These are jointly impossible because, by a \enquote{conservation of flow} argument (cf.\ above), if $q' \to_{C} q''$, then $\deg(q') < \deg(q'')$ if and only if $C$ contains a node with label in $\Sigma\times\{1\}$.
  \end{itemize}
  Finally, now that we know that $\cB$ is bounded, its maximum value is computable because:
  \begin{itemize}
    \item $\cB$ can be translated into an $\bbN$-weighted visibly pushdown automaton ($\bbN$-VPA) with the same range~\cite[\S2.3.3]{Caralp};
    \item the maximum value of a bounded $\bbN$-VPA is computable~\cite[Theorem~4]{VPAmult}.
  \end{itemize}
  (Note that this final observation does not require knowing what an $\bbN$-VPA is!)
\end{proof}

\noindent
We are now in a position to bound the value of $\cA$ over an input tree of size $n$:
\begin{itemize}
  \item there are $O(n^{\max(\deg(\text{states}))})$ possible sets of critical
        nodes for the runs by Claim~\ref{clm:crit-degree};
  \item for each of these possibilities, the contribution of the corresponding
        runs to the total value is $O(1)$ by the above lemma.
\end{itemize}
Combined with the matching lower bound obtained in Claim~\ref{clm:pump-lower-bound}, we see that, using the definition of growth rate from the beginning of the introduction:
\begin{claim}\label{clm:growth-tight-bound}
  $\growth[\cA](n) = \Theta(n^{\max(\deg(\text{states}))})$.
\end{claim}
Thus, to prove Theorem~\ref{thm:degree-weighted}, we just have to efficiently compute the degrees.

\subsection{Computing the degrees of all states in cubic time}

To compute the degree map, we first characterise it more abstractly as the
\emph{smallest} map that satisfies the inequalities of Claim~\ref{clm:deg-ineq}. In
other words:
\begin{lemma}\label{lem:deg-abstract}
  The function that maps each state to its degree is the least fixed point of
  \[ f \mapsto \left[ q \mapsto \max\left(\bigl\{f(q)\bigr\}\cup\left\{\sum_{i=1}^{k} f(q_i) \middle| (q_1,\dots,q_k,a,q)\ \text{trans.}\right\} \cup \Bigl\{f\bigl(q'\bigl)+1 \,\Big|\, q' \Rrightarrow q\Bigr\}\right) \right]. \]
\end{lemma}
\begin{proof} We sketch the high-level idea here and give details in \S\ref{app:deg-abstract} of the appendix.
  
  Let $S(q) = \{\deg(\Pi) \mid \Pi\ \text{is a pumping pattern for}\ q\}$. From
  Definition~\ref{def:pumping} we can extract a direct inductive definition of $S$, that
  can be expressed by general considerations as a least fixed point in a lattice
  of set-valued functions. To translate this into an expression for the degree
  map, observe that $S(q) = \{0,\dots,\deg(q)\}$ (indeed, $S(q)$ is
  downwards-closed because $\pumpnode{C}(\Pi)$ can be replaced by $C[\Pi]$ to
  decrement the degree of a pattern).
\end{proof}

  To compute the degree map, we can iterate the above operator (starting from
  $q\mapsto0$) until we reach a fixed point. This naive algorithm terminates
  because it builds a bounded increasing sequence of $\bbN$-valued functions
  with finite domain: such a sequence must be finite.
  Slightly less naively, we can precompute
  $\{(q_{1},q_{2}) \mid q_{1}\Rrightarrow q_{2}\}$ in cubic time
  (cf.~Lemma~\ref{lem:all-pairs-barbell}) before iterating. Subsequently, each
  iteration can be carried out in quadratic time, by examining each of the
  $O(|\cA|)$ transitions once and each of the $O(|\cA|^{2})$ barbells once.

  To get a cubic bound on the total running time, it now suffices to show that \emph{the fixed point is reached after a linear number of iterations}.
  We denote by $f_\tau$ the map from states to $\bbN$ obtained after $\tau$
  iterations of the aforementioned operator, and $Q(\tau)$ the set of states that are mapped to the correct value after $\tau$ iterations: $Q(\tau) = \{q \mid f_\tau(q)=\deg(q)\}$.
  We call a set of states $S$ a \emph{trigger} for a state $q$ if:
  \begin{itemize}
    \item either $S=\set{q'}$ and $q'\Rrightarrow q$ and $\deg(q)=\deg(q')+1$,
    \item or $S=\set{q_1,\ldots,q_k}$ 
    and $(q_1,\dots,q_k,a,q)$ is a transition
    and $\deg(q)=\displaystyle\sum_{\mathclap{1\leqslant i \leqslant k}}\deg(q_i)$. 
  \end{itemize}

\begin{claim}\label{clm:trigger}
   $Q(\tau+1)=Q(\tau) \cup \set{q \mid q\text{ has a trigger included in }Q(\tau)}$.
\end{claim}
\begin{proof}
  The expression in Lemma~\ref{lem:deg-abstract} gives us $f_{\tau+1}(q) = \max(\{f_\tau(q)\} \cup D_\tau(q))$ where
  \[  D_\tau(q) = \left\{\sum_{i=1}^{k} f_\tau(q_i) \middle| (q_1,\dots,q_k,a,q)\ \text{trans.}\right\} \cup\{f_\tau(q')+1 \mid q' \Rrightarrow q\} \]
  If a state $q$ has a trigger included in $Q(\tau)$, then this trigger witnesses that $\deg(q) \in D_\tau(q)$, 
  hence $f_{\tau+1}(q) \geqslant \deg(q)$. Since $f_{\tau+1}$ is part of an increasing sequence that eventually reaches $\deg$, we have the converse inequality, so $q \in Q(\tau+1)$. We also have $Q(\tau) \subseteq Q(\tau+1)$, again because the iterations are increasing.

  Conversely, if $q \in Q(\tau+1) \setminus Q(\tau)$, then $\deg(q) = f_{\tau+1}(q) = \max(\{f_\tau(q)\} \cup D_\tau(q))$.
  \begin{itemize}
    \item The case $\deg(q) = f_{\tau}(q)$ is impossible because $q \notin Q(\tau)$.
    \item If $\deg(q) = f_\tau(q_1) + \dots + f_\tau(q_k)$ for some transition $(q_1,\dots,q_k,a,q)$, we also know that:
    \begin{itemize}
      \item $f_\tau(q_i) \leqslant \deg(q_i)$ for all $i$, and therefore $\deg(q) \leqslant \deg(q_1) + \dots + \deg(q_k)$; 
      \item $\deg(q) \geqslant \deg(q_1) + \dots + \deg(q_k)$ by Claim~\ref{clm:deg-ineq}.
    \end{itemize}
    This forces all these inequalities to be equalities, which means that $\{q_1,\dots,q_k\}$ is a trigger for $q$ and is included in $Q(\tau)$ (by definition).
    \item If $\deg(q) = f_\tau(q') + 1$ for some $q' \Rrightarrow q$, then Claim~\ref{clm:deg-ineq} again forces $f_\tau(q') = \deg(q')$, so $q' \in Q(\tau)$ and $\{q'\}$ is a trigger for $q$. \qedhere
  \end{itemize}
\end{proof}
\begin{claim}
  If $Q(\tau+1)=Q(\tau)$ then a fixed point is reached in at most $\tau$ iterations.
\end{claim}
\begin{proof}
  As a consequence of the previous claim, whenever $Q(\tau+1)=Q(\tau)$, for any $N\in\bbN$ we have $Q(\tau)=Q(\tau+N)$. This means that the fixed point has been reached: indeed, by definition, the algorithm halts at the first time $\tau'$ when $Q(\tau')$ is the full set of states.
\end{proof}

Thus, $Q(\tau)$ has to be strictly increasing before it stabilises. Therefore, the number of iterations is bounded by the number of states. This concludes the proof of Theorem~\ref{thm:degree-weighted}.

\begin{remark}
  In the case of word automata, the computation of the degrees can be seen as a
  longest-path problem in a weighted graph without positive cycles: take the
  transitions of the automaton as weight 0 edges, and barbells as weight 1
  edges. This problem can be solved by the classical Bellman--Ford algorithm
  (cf.~\cite[\S8.4]{DurrVie}), whose iterative relaxations are very similar to
  our own procedure.
\end{remark}

\section{Reparameterisation of MSO queries}
\label{sec:queries}

Let us now deduce Theorem~\ref{thm:reparam-intro} (stated in more detail as
Theorem~\ref{thm:reparam-details}) from the properties of critical transitions that we
just saw in the previous section.

We use the standard definition of monadic second-order (MSO) logic and its
semantics over relational structures; see for instance~\cite[Chapter~5]{courcellebook}. As
usual, uppercase letters $X,Y,\ldots$ denote second-order variables (ranging
over subsets of the domain) whereas lowercase letters $x,y,\ldots$ denote
first-order variables (ranging over elements of the domain). A formula is closed
when it has no free variables.

\subparagraph{Trees and DAGs as relational structures.}

To a ranked alphabet $\Sigma$, we associate a relational signature that consists of:
\begin{itemize}
  \item a unary symbol $a$ for each $a\in \Sigma$;
  \item a binary symbol $\child_i$ for each $i\in\set{1,\ldots,\max(\rank(\Sigma))}$;
  \item a binary symbol $\leqslant$.
\end{itemize}
\begin{definition}
We define a \emph{rooted DAG (directed acyclic graph)} over $\Sigma$ to be a finite structure $\graph$ over the above signature, such that the following hold:
\begin{itemize}
  \item $\graph$ interprets the formula $\child_1(x_1,x_2)\lor\dots\lor\child_{\max(\rank(\Sigma))}(x_1,x_2)$
    as an \emph{irreflexive and acyclic} binary relation over its domain, whose reflexive transitive closure is equal to $\leqslant$;
  \item the partial order $\leqslant$ has a unique minimum element, called the \emph{root};
  \item for each vertex (i.e.\ domain element) $u\in\dom(\graph)$:
    \begin{itemize}
    \item there is a unique $a\in\Sigma$ such that $\graph\models a(u)$;
    \item for $i\in\set{1,\ldots,\rank(a)}$, there is a unique $v \in \dom(\graph)$ such that $\graph\models \child_i(u,v)$;
    \item for $i > \rank(a)$, there is no $v\in \dom(\graph)$ such that $\graph\models \child_i(u,v)$.
    \end{itemize}
\end{itemize}
When $\graph\models u \leqslant v$, we say that $u$ is an \emph{ancestor} of $v$ and that $v$ is a \emph{descendant} of $u$.
\end{definition}
\begin{remark}\label{rem:multigraph}
  Strictly speaking, our rooted DAGs may be multigraphs, i.e.\ contain parallel edges: when $\graph\models \child_i(u,v) \land \child_j(u,v)$, it does not necessarily imply that $i=j$.
\end{remark}
\begin{remark}
  All these axioms are definable in MSO.
\end{remark}

Let us illustrate the power of MSO for querying DAGs encoded as relational structures.
\begin{example}\label{ex:paths}
  A \emph{rooted path} in $\graph$ is a finite sequence $(u_{0},i_{1},u_{1},\dots,i_{n},u_{n})$ such that:
  \begin{itemize}
    \item $u_{0}$ is the root;
    \item $\graph\models \child_{i_{k}}(u_{k-1},u_{k})$ for $k\in\{1,\dots,n\}$.
  \end{itemize}
  Note that with this definition, because of Remark~\ref{rem:multigraph}, a path is \emph{not} uniquely identified by its sequence of vertices $(u_{0},u_{1},\dots,u_{n})$.
  This is why we are going to encode such a path not as a single set of vertices, but as an $r$-tuple of sets, for $r=\max(\rank(\Sigma))$:
  \[ (P_{1},\dots,P_{r}) \in \powerset(\dom(\graph))^{r} \quad\text{where}\quad P_{i} = \{u_{k-1} \mid k \in \{1,\dots,n\},\; i_{k} = i\}\]
  (in particular, the path with only the root ($k=0$) is encoded as $(\varnothing,\dots,\varnothing)$).

  Then $\vec{P}$ is an encoding of a rooted path if and only if $\graph\models \Fpath(\vec{P})$
  for some MSO formula $\Fpath(X_{1},\dots,X_{r})$. We can then say that this formula defines an MSO set query whose results are the rooted paths.
  Explicitly, $\Fpath$ may be defined as the conjunction of:
  \begin{itemize}
    \item $\forall x.\; X_{i}(x) \Rightarrow \displaystyle\bigvee_{\mathclap{a\in\Sigma,i\leqslant\rank(a)}} a(x)$ (take one such subformula for each $i\in\{1,\dots,r\}$);
    \item $\forall x.\; \displaystyle\bigvee_{\mathclap{1\leqslant i \leqslant r}} X_{i}(x) \Rightarrow\Big( (\forall y.\; x \leqslant y) \lor \bigvee_{\mathclap{1\leqslant j\leqslant r}} \exists y.\; X_{j}(y) \land \child_{j}(y,x)\Big)$;
    \item the predicates $X_{i}$ define disjoint subsets (this can be expressed in MSO). \qedhere
  \end{itemize}

\end{example}
\begin{remark}
  We work with structures whose interpretation of $\leqslant$ is fully
  determined by their interpretation of the other symbols. But from the point of
  view of expressive power, it is well known that having direct access to the
  descendant order makes a difference in first-order logic (see e.g.\ the
  introduction to~\cite{BenediktS09}) --- this is important for the
  decomposition of Lemma~\ref{lem:mso-via-fo}. It also matters for the structures
  outputted by MSO interpretations (\S\ref{sec:set-interpretations}), as
  explained in~\cite[\S2.2]{msoInterpretations}.
\end{remark}

Any tree over $\Sigma$ can be turned into a rooted DAG whose domain is the set
of nodes of the tree, with the intuitive interpretations of the relation
symbols. This defines an injection from trees to DAGs, whose range can be
characterised up to isomorphism of structures:
\begin{definition}
  If every non-root vertex of a rooted DAG $\graph$ has a unique parent \emph{and in a unique way} --- in other words, for every non-root $v \in \dom(\graph)$ there are a unique $u \in \dom(\graph)$ and a unique $i$ such that $\graph \models \child_{i}(u,v)$ --- then $\graph$ is called a \emph{tree structure}.
\end{definition}

For $t \in \tree\Sigma$ and $F$ a formula, we abbreviate \enquote{$[\text{the
      tree structure for}\ t] \models F$} as $t \models F$. This abuse of
notation extends to formulas with first-order (resp.\ second-order) variables
instantiated by nodes (resp.\ subsets of nodes) of $t$ --- as used, for
instance, in the introduction when defining the set query $\wn F$ specified
by an MSO formula $F(X_1,\dots,X_\ell)$.

We recall a basic and useful syntactic property.
\begin{proposition}\label{prop:mso-marked}
  From an MSO formula $F(X_1,\dots,X_{\ell},Y_1,\dots,Y_m)$ over $\tree\Sigma$ one can build an MSO formula $G(Y_1,\dots,Y_m)$ over $\tree(\Sigma\times\{0,1\}^\ell)$ such that    
  \[ \forall t \in \tree\Sigma,\; \forall \bigl(\vec{P},\vec{S}\bigr) \in \powerset(\nodes(t))^{\ell+m},\quad  t \models F\bigl(\vec{P},\vec{S}\bigr) \iff \marked\bigl(t,\vec{P}\bigr) \models G\bigl(\vec{S}\bigr)\]
  where $\marked(t,P_1,\dots,P_\ell)$ is the tree such that:
  \begin{itemize}
    \item by projecting all labels to their first coordinate, one recovers the tree $t$ (for this to make sense, we define the ranks on $\Sigma\times\{0,1\}^{\ell}$ to be inherited from the ranks on $\Sigma$);
    \item the $(i+1)$-th coordinate of the label of a node is 1 if and only if the corresponding node in $t$ belongs to $P_{i}$.
  \end{itemize}
  Conversely, from $G\bigl(\vec{Y}\bigr)$ one can build $F\bigl(\vec{X},\vec{Y}\bigr)$ such that the above equivalence holds. 
\end{proposition}
\begin{proof}
  From $F$, build $G$ by replacing, for all $a\in\Sigma$, $i \in\{1,\dots,\ell\}$, first-order variables $x$:
  \begin{itemize}
    \item each occurrence of $a(x)$ by $\bigvee_{\vec{b} \in \{0,1\}^\ell} (a,\vec{b})(x)$;
    \item each occurrence of $X_i(x)$ by the disjunction of $(a',\vec{b})(x)$ ranging over all $a'\in\Sigma$ and all $\vec{b} \in \{0,1\}^\ell$ whose $i$-th components equal 1. 
  \end{itemize}
  From $G$, build $F$ by replacing $(a,(b_1,\dots,b_\ell))(x)$ by $\displaystyle a(x) \land \bigwedge_{b_i = 0} \lnot X_i(x) \land \bigwedge_{b_i = 1} X_i(x)$. 
\end{proof}

\subparagraph{Logic-automata correspondence.}

The fundamental theorem of MSO on trees is that the sets of trees defined by
closed MSO formulas are precisely the regular tree languages, that is, those recognised
by tree automata; cf.\ e.g.~{\cite[Theorem~2.7]{HandbookAutomataTrees}}. (The language recognised by an automaton is the set of inputs with an accepting run.) Furthermore, this
equivalence is effective, i.e.\ witnessed by computable translations.
Thanks to Proposition~\ref{prop:mso-marked}, one can also state a correspondence for formulas with free variables defining
queries, as follows:
\begin{theorem}\label{thm:mso-vs-tree-automata}
  Let $\ell\in\bbN$ and, for each $t\in\tree\Sigma$, let $\Xi_t \subseteq \powerset(\nodes(t))^\ell$.
  The following are effectively equivalent:
  \begin{itemize}
    \item there is an MSO formula $F(X_1,\dots,X_\ell)$ such that $\Xi_t = \wn F(t)$ for all $t\in\tree\Sigma$;
    \item the \enquote{language of marked trees} $\{\marked(t,\vec{P}) \mid t\in\tree\Sigma,\, \vec{P} \in \Xi_t\}$ is regular.
  \end{itemize}
\end{theorem}
\begin{remark}
  Actually, one classical translation from MSO to tree automata --- which is the one given in~\cite[proof of Theorem~2.7]{HandbookAutomataTrees} --- proceeds by induction over the syntax; the induction hypothesis needs to apply to subformulas with free variables, so it ends up being the statement of Theorem~\ref{thm:mso-vs-tree-automata} that is proved.
\end{remark}

\subparagraph{Number of query results vs ambiguity.}

Let $F(X_{1},\dots,X_{\ell})$ be an MSO formula on trees over the ranked alphabet $\Sigma$.
By Theorem~\ref{thm:mso-vs-tree-automata}, its language of marked trees is regular.
Therefore, it is recognised by some \emph{deterministic bottom-up} tree automaton $\cA_{F}$ over $\Sigma\times\{0,1\}^\ell$ (see for instance~\cite[Theorem~2.1]{HandbookAutomataTrees} or~\cite[Theorem~1.1.9]{TATA}); what is important for us is that determinism implies \emph{unambiguity}. 
We build a tree automaton $\cB_{F}$ over $\Sigma$ as follows:
\begin{itemize}
  \item its states are of the form $(q,\vec{b})$ where $q$ is a state of $\cA_{F}$ and $\vec{b} \in\{0,1\}^{\ell}$;
  \item the accepting states in $\cB_{F}$ are those whose first component is accepting in $\cA_{F}$;
  \item its transitions are $((q_{1},\vec{b}_{1}),\dots,(q_{m},\vec{b}_{m}),a,(q,\vec{b}))$ for each $\vec{b}_{1},\dots,\vec{b}_{m},\vec{b}\in\{0,1\}^{\ell}$ and $a\in\Sigma$ such that $\cA_{F}$ contains the transition $(q_{1},\dots,q_{m},(a,\vec{b}),q)$.
\end{itemize}
\begin{claim}[Explicit version of~Claim~\ref{clm:query-vs-ambiguity-intro}]\label{clm:query-vs-ambiguity}
  The results of the query $\wn F$ are in bijection with the set of accepting runs of $\cB_{F}$ over every input tree in $\tree\Sigma$.
\end{claim}
\begin{proof}
  By construction, the accepting runs of $\cB_{F}$ over an input $t\in\tree\Sigma$
  are in bijection with
  \[ \{ (t',\; \underbrace{\text{some accepting run of}\ \cA_{F}\ \text{over}\ t'}_{\mathclap{\text{unique because}\ \cA_{F}\ \text{is unambiguous}}}) \mid [\text{project node labels on}\ \Sigma](t') = t \} \]
  By definition of
  the language of marked trees recognised by $\cA_{F}$, the trees $t'$ appearing in the above set are
  precisely those of the form $\marked(t,\vec{P})$ for $\vec{P} \in \wn F(t)$.
\end{proof}

As stated in the introduction, the properties that we have established
concerning the ambiguity of tree automata can therefore be transferred to MSO set
queries. Since the correspondence in Theorem~\ref{thm:mso-vs-tree-automata} is
effective, what we get is:
\begin{corollary}\label{cor:set-query-growth}
  The number of results of an MSO set query grows either polynomially or
  exponentially in the input size, with a computable degree in $\bbN\cup\{\infty\}$.
\end{corollary}

\subparagraph{The reparameterisation theorem.}

In the case of polynomial growth, we have:

\begin{theorem}[precise statement of {Theorem~\ref{thm:reparam-intro}}]\label{thm:reparam-details}
  Given an MSO formula $F(X_1,\dots,X_\ell)$ over trees on $\Sigma$ whose
  number of results has a growth rate of degree $k < \infty$, one can compute
  an MSO formula $G(X_1,\dots,X_\ell,z_1,\dots,z_k)$ and $B\in\bbN$ such that,
  for every $t \in \tree\Sigma$:
  \begin{itemize}
    \item $\wn G(t) \subset \powerset(\nodes(t))^{\ell} \times \nodes(t)^{k}$ is the graph of some function $g_{t} \colon \wn F(t) \to \nodes(t)^{k}$;
    \item this function satisfies $|g_{t}^{-1}(\{u\})| \leqslant B$ for every $u \in \nodes(t)$.
  \end{itemize}
  Thus, $(g_{t})_{t\in\tree\Sigma}$ is a finite-to-one family of functions with computable bound.
\end{theorem}
\begin{example}[detailed version of {Example~\ref{ex:paths-intro}}]\label{ex:gpath}
  Consider the formula $\Fpath(X_{1},\dots,X_{r})$ where
  $r=\max(\rank(\Sigma))$ from Example~\ref{ex:paths}, which selects the encodings of
  rooted paths in a DAG over $\Sigma$. Over \emph{trees}, the query $\wn\Fpath$ has linear
  growth. The above theorem then says that one can compute an MSO formula $\Gpath(\vec{X},z)$ defining a reparameterisation for $\Fpath$.

  An explicit reparameterisation is given by the family of functions that maps
  each rooted path in a tree to its endpoint (i.e.\ the last vertex in the
  path). Indeed, it is injective (again, over trees), so its fibers are of cardinality at most
  $B=1$, and it is MSO-definable by:
  \[ \Gpath(\vec{X},z) = \Fpath(\vec{X}) \land \forall z'.\; z \leqslant z' \Leftrightarrow \overbrace{\bigwedge_{\mathclap{1\leqslant i \leqslant r}} \forall x.\,\forall y.\, X_{i}(x) \land \child_{i}(x,y) \Rightarrow y \leqslant z'}^{\text{\enquote{\(z'\) is an upper bound of all vertices in the path}}} \]
  This MSO formula selects the endpoint of a rooted path in any rooted DAG --- a level of generality that will be convenient for later reuse. (Though this is not finite-to-one over DAGs; in fact, there
  can be up to exponentially many rooted paths to a given endpoint.)
\end{example}
\begin{proof}[Proof of {Theorem~\ref{thm:reparam-details}}]
  Let $\cB'_{F}$ be obtained by trimming the tree automaton $\cB_{F}$ defined earlier.
  By the assumption $k <\infty$ and Claim~\ref{clm:query-vs-ambiguity}, $\cB'_{F}$ is polynomially ambiguous: it satisfies the assumptions of Section~\ref{sec:chara-poly}. This allows us to speak of the critical nodes (Definition~\ref{def:critical}) for the runs of $\cB'_{F}$, which are the same as the runs of $\cB_F$.
  Using the bijective correspondence of Claim~\ref{clm:query-vs-ambiguity}, we can define a family of functions indexed by $t\in\tree\Sigma$ as follows:
  \[ \vec{P} \in \wn F(t) \mapsto \{\text{critical nodes for the accepting run of}\ \cB_{F}\ \text{over}\ t\ \text{corresponding to}\ \vec{P}\} \]
  By Lemma~\ref{lem:crit-bounded}, this family is finite-to-one with computable bound. Furthermore, it can be defined in MSO by some formula $H(X_{1},\ldots,X_{\ell},Y)$ using standard techniques for representing runs of tree automata in MSO (cf.\ Appendix~\ref{app:reparam-details} for more details). Note also that the range of any function in this family only contains sets of nodes of cardinality at most $k$, according to Claims~\ref{clm:crit-degree} and~\ref{clm:growth-tight-bound}.

  The next step is to devise an MSO formula $J(Y,z_{1}\dots,z_{k})$ that defines a finite-to-one family of functions
  $\{ \text{subsets of}\ \nodes(t)\ \text{of size} \leqslant k \} \to \nodes(t)^{k}$ for $t\in\tree\Sigma$.
  One possible solution is to use a prefix of the output tuple to list the elements of the input set, in a fixed MSO-definable order, and pad the remaining output coordinates with the root node.

  Finally, we take $G(\vec{X},\vec{z}) = \exists Y.\, H(\vec{X},Y) \land J(Y,\vec{z})$. This works because the composition of two finite-to-one families of functions is itself finite-to-one.
\end{proof}

\section{Around MSO set interpretations}
\label{sec:set-interpretations}

We now apply the newly established Theorem~\ref{thm:reparam-details} to the study of functions from trees to trees (or sometimes to more general relational structures).

Let us first recall the notion of MSO set interpretation. The basic idea is that
the domain of an output structure contains the results of an MSO set query over
the input structure.
\begin{definition}
  \label{def:mso-set-interp}
  An \emph{MSO set interpretation} $\cI$ from a family of finite relational
  structures to another, over fixed input and output signatures, consists of a
  positive integer\footnote{Strictly speaking, Colcombet and Löding's original
    definition~\cite[\S2.3]{ColcombetL07} corresponds to taking $\ell=1$. The
    case $\ell>1$ has been considered for example by Rabinovich and
    Rubin~\cite[§II.C]{RabinovichR12} and by Filiot, Lhote and
    Reynier~\cite{Lex}; as far as we know, it does not pose additional
    difficulties.} $\ell$ and some MSO formulas over the
  input signature with \emph{second-order} free variables:
  \[  \cI_\dom(X_1,\dots,X_\ell) \qquad\qquad \underbrace{\cI_R(X_{1,1},\dots,X_{1,\ell},X_{2,1},\dots,X_{m,\ell})}_{\mathclap{\text{for each $m$-ary relation symbol $R$ in the output signature}}}\]
  It defines the function $f$ such that $\dom(f(S)) = \wn\cI_{\dom}(S)  \subseteq \powerset(\dom(S))^{\ell}$ and
  \[ f(S) \models R(\langle P_{1,1},\dots,P_{1,\ell}\rangle,\dots, \langle P_{m,1},\dots,P_{m,\ell}\rangle) \quad \iff \quad S\models\cI_R(P_{1,1},\dots,P_{m,\ell})\]
  for every input structure $S$ and output symbol $R$.
\end{definition}

An MSO set interpretation maps isomorphic inputs to isomorphic outputs. We may therefore identify structures up to isomorphism; this allows us to treat trees (processed by automata and transducers) and tree structures (handled by interpretations) interchangeably without losing relevant information.

\begin{example}
  \label{ex:unfold}
  We define the \emph{unfolding} $\unfold(\graph)$ of a rooted DAG $\graph$, illustrated by the right-hand side of Figure~\ref{fig:msots} in \S\ref{sec:intro}, as a tree structure (over the same ranked alphabet $\Sigma$):
  \begin{itemize}
    \item the domain elements (nodes) are the rooted paths in $\graph$, cf.\ Example~\ref{ex:paths};
    \item each path $\pi\in\dom(\unfold(\graph))$ inherits the label of its endpoint: $\unfold(\graph) \models a(\pi)$ if and only if $\pi = (\dots,u)$ with $\graph\models a(u)$;
    \item the descendant order $\leqslant$ is the prefix ordering on paths;
    \item $\unfold(\graph) \models \child_{i}(\pi,\pi')$ for
          $\pi,\pi' \in \dom(\unfold(\graph))$ if and only if $\pi$ extended by
          going to the $i$-th child of its endpoint equals $\pi'$ --- in other
          words, the two paths are of the form $(u_{0},i_{1},u_{1},\dots,u_{n})$ and
          $(u_{0},i_{1},u_{1},\dots,u_{n},i,u_{n+1})$ respectively.
  \end{itemize}
  We can define $\unfold$ (up to isomorphism) as an MSO set interpretation, by working over the encodings of rooted paths introduced in Example~\ref{ex:paths}:
  \begin{itemize}
    \item $\cI_\dom(X_{1},\dots,X_{\ell}) = \Fpath(X_{1},\dots,X_{\ell})$ with $\ell=\max(\rank(\Sigma))$;
    \item $\cI_{a}(\vec{X}) = \exists z.\, \Gpath(\vec{X},z) \land a(z)$ for $a\in\Sigma$, using the formula $\Gpath$ from Example~\ref{ex:gpath} which selects the endpoint of a rooted path;
    \item $\cI_\leqslant(X_{1},\dots,X_{\ell},Y_{1},\dots,Y_{\ell}) = X_{1} \subseteq Y_{1} \land \dots \land X_{\ell} \subseteq Y_{\ell}$;
    \item $\cI_{\child_{i}}(\vec{X},\vec{Y}) = \displaystyle\bigwedge_{j \neq i} X_{j} = Y_{j} \land \exists z.\; \Gpath(\vec{X},z) \land \underbrace{(\forall z'.\; Y_{i}(z') \Leftrightarrow X_{i}(z') \lor z' = z)}_{Y_{i} = X_{i}\cup\{z\}}$. \qedhere
  \end{itemize}
\end{example}

An \emph{MSO interpretation} is mostly like an MSO set interpretation, except that the output domain contains tuples of input domain elements, rather than tuples of sets. Another difference, in the definition that we take, is to allow a bounded amount of \enquote{copying} thanks to a finite set of \emph{components}; we discuss this feature in Remark~\ref{rem:components}.
\begin{definition}
  \label{def:mso-interp}
  For $k\in\bbN$, a \emph{$k$-dimensional MSO interpretation} $\cI$ from a family of finite relational structures to another consists of:
  \begin{itemize}
    \item a finite set of \emph{components} $\cI_\components$;
    \item some MSO formulas over the input signature with \emph{first-order} free variables, namely
          \[ \underbrace{\cI_\dom^\alpha(x_1,\dots,x_k)}_{\mathclap{\text{for each $\alpha\in\cI_\components$}}} \qquad \underbrace{\cI_R^{\alpha_1,\dots,\alpha_m}(x_{1,1},\dots,x_{1,k},x_{2,1},\dots,x_{m,k})}_{\mathclap{\text{for each output symbol $R$ and $\alpha_{i} \in \cI_{\components}$}}} \]
  \end{itemize}
  It defines the function that maps an input structure $S$ to the output structure:
  \begin{itemize}
    \item with domain $\{ \langle \alpha, u_1,\dots,u_k \rangle \in \cI_\components\times\dom(S)^k \mid S \models \cI_\dom^\alpha(u_1,\dots,u_k) \}$;
    \item where $R(\langle \alpha_1, u_{1,1},\dots,u_{1,k}\rangle,\dots, \langle \alpha_m, u_{m,1},\dots,u_{m,k}\rangle)$ iff $S\models\cI_R^{\alpha_1,\dots,\alpha_m}(u_{1,1},\dots,u_{m,k})$.
  \end{itemize}
  In the case $k=1$, a 1-dimensional MSO interpretation is also called an \emph{MSO transduction}.
\end{definition}
\begin{example}\label{ex:fot-dag-comb}
  The following MSO transduction $\cT$ defines a function from trees over the alphabet $\{S : 1,\; \underline{0} : 0\}$ to rooted DAGs over $\{a : 2,\; b : 1,\; c : 0\}$. This map is drawn on the left-hand side of Figure~\ref{fig:msots} in \S\ref{sec:intro}.
  \begin{itemize}
    \item $\cT_{\components} = \{\alpha,\beta\}$ (the names of the components are just formal symbols);
    \item The $\alpha$ component contains the $a$-nodes, one for each input $S$-node, forming a $\searrow$-path:
    \[ \cT_{\dom}^{\alpha}(x) = \cT_{a}^{\alpha}(x) = S(x) \qquad \cT_{\leqslant}^{\alpha,\alpha}(x,y) = (x \leqslant y) \qquad \cT_{\child_{2}}^{\alpha,\alpha}(x,y) = \child_{1}(x,y)\]
    \item The $\beta$ component has one copy of each input node ($\cT_{\dom}^{\beta}(x) = \mathtt{true}$) and forms a path:
    \[ \cT_{b}^{\beta}(x) = S(x)\qquad \cT_{c}^{\beta}(x) = \underline{0}(x) \qquad \cT_{\leqslant}^{\beta,\beta}(x,y) = (x \leqslant y) \qquad \cT_{\child_{1}}^{\beta,\beta}(x,y) = \child_{1}(x,y)\]
    \item To add the remaining cross-component edges we take:
    \[ \cT_{\leqslant}^{\alpha,\beta}(x,y) = (x \leqslant y) \qquad \underbrace{\cT_{\child_{1}}^{\alpha,\beta}(x,y) = (x=y)}_{\text{case}\ a\;\to\;b} \qquad \underbrace{\cT_{\child_{2}}^{\alpha,\beta}(x,y) = \underline{0}(y) \land \child_{1}(x,y)}_{\text{case}\ a\;\to\;c} \]
    \item Every formula in $\cT$ that has not been defined above (e.g.\ $\cT^{\beta,\alpha}_{\leqslant}$) is set to $\mathtt{false}$. \qedhere
  \end{itemize}
\end{example}
\begin{proposition}
  Any function defined by an MSO interpretation can also be defined by an MSO set interpretation (up to isomorphism of output structures).
\end{proposition}
\begin{proof}
  The basic idea is to replace each first-order free variable $x_{i}$ by a second-order variable $X_{i}$ with the constraint $X_{i} = \{x_{i}\}$ (definable by $\forall y.\; X_{i}(y) \Leftrightarrow y = x_{i}$). We also take care of the components by using extra second-order variables, whose values are restricted to be either the full set or the empty set, as binary \enquote{flags}; this works because our input structures are nonempty (they contain at least the root). Thus, we translate a $k$-dimensional MSO interpretation $\cI$ whose components are $\{1,\dots,N\}$ (without loss of generality) to an MSO set interpretation $\cJ$ with a domain formula $\cJ_{\dom}(X_{1},\dots,X_{k+N})$ defined as follows:
  \[ \exists x_{1}\dots x_{k}.\; \bigwedge_{i=1}^{k} X_{i} = \{x_{i}\} \land \bigvee_{\alpha=1}^{N} \cI_{\dom}^{\alpha}(x_{1},\dots,x_{k}) \land \overbrace{\forall y.\; X_{k+\alpha}(y) \land \bigwedge_{\mathclap{\beta\neq\alpha}} \lnot X_{k+\beta}(y)}^{\mathclap{\text{detects that we are in the component}\ \alpha}} \]
  The definition of the formulas in $\cJ$ for the output relations follows a similar scheme.
\end{proof}

\subsection{Dimension minimization (proof of Theorem~\ref{thm:msosi-min})}

Let us now discuss the growth rate of MSO (set) interpretations. It follows
directly from the definitions that if a function $f$ between relational structures is defined by a
$k$-dimensional MSO interpretation, then
$|f(\graph)| = O(|\graph|^k)$, where the size $|\graph|$ of the structure
$\graph$ is the cardinality of its domain (thus, for trees, it coincides with the number of nodes).
The dimension minimization property announced in the introduction (Theorem~\ref{thm:msosi-min}) is a sort of converse, over trees.
\begin{proof}[Proof of {Theorem~\ref{thm:msosi-min}}]
  Let $\cI$ be an MSO set interpretation with trees as inputs. We want to compute its degree of growth, and if it is finite (i.e.\ the growth is polynomial), to compute an equivalent MSO interpretation of optimal dimension.

  First, by definition, $\dom(\cI(t)) = \wn\cI_{\dom}(t)$ for every input tree $t$. Thus, the output size is the number of results of an MSO set query. By Corollary~\ref{cor:set-query-growth}, its degree of growth $k\in\bbN\cup\{\infty\}$ is well-defined and can be computed from $\cI_{\dom}$.

  Suppose now that $k < \infty$. By Theorem~\ref{thm:reparam-details}, one can compute $B\in\bbN$ and an MSO formula $G(\vec{X},z_{1},\dots,z_{k})$ that defines a family of functions $\dom(\cI(t)) \to \nodes(t)^{k}$, indexed by input trees $t$, which is finite-to-one with bound $B$.
  For $\alpha\in\{1,\dots,B\}$, there is an MSO formula $H_\alpha(\vec{X},\vec{z})$ whose meaning is: there are at least $\alpha$ tuples $\vec{Y}$ such that $G(\vec{Y},\vec{z})$, and $\vec{X}$ is the $\alpha$-th such tuple for some fixed total order, e.g.\ lexicographic with respect to some tree traversal. The following $k$-dimensional interpretation $\cJ$ then defines the same function as $\cI$:
  \begin{itemize}
    \item $\cJ_\components = \{1,\dots,B\}$
    \item $\cJ_\dom^{\alpha}(z_1,\dots,z_k) = \exists \vec{X}.\; H_\alpha(\vec{X},z_1,\dots,z_k)$ (this implies $\cI_\dom(\vec{X})$)
    \item $\cJ_R^{\alpha_1,\dots,\alpha_m}(\overrightarrow{z_1},\dots,\overrightarrow{z_m}) = \exists\overrightarrow{X_1}.\; \dots \exists\overrightarrow{X_m}.\; \displaystyle \bigwedge_{\mathclap{1 \leqslant i \leqslant m}} H_\alpha(\overrightarrow{X_i},\overrightarrow{z_i}) \land \cI_R(\overrightarrow{X_1},\dots,\overrightarrow{X_m})$
  \end{itemize}
  This proof has been directly adapted from Bojańczyk's~{\cite[proof of Theorem~6.1]{PolyregSurvey}}.
\end{proof}
\begin{remark}\label{rem:components}
  The use of components is crucial to make Theorem~\ref{thm:msosi-min} work, since they
  allow the constant factor in $|f(t)|=O(|t|^{k})$ to be larger than 1. That is
  why, even though the original definition of string-to-string MSO
  interpretations~\cite{msoInterpretations} did not involve components,
  Bojańczyk added them in~\cite{PolyregSurvey,PolyregGrowth} when dealing with
  dimension minimization.

  In fact, components, also called \enquote{copy indices}, are standard in the
  1-dimensional case of \emph{MSO transductions} (MSOT) --- cf.\
  e.g.~\cite[Chapter~7]{courcellebook}.
\end{remark}
\begin{remark}
  MSO transductions predate MSO (set) interpretations. Indeed, they were
  introduced in the late
  1980s~\cite{DBLP:conf/icalp/ArnborgLS88,DBLP:conf/gg/Engelfriet90,DBLP:journals/tcs/Courcelle91a}
  and have gone on to acquire a central importance in automata theory and
  adjacent fields (such as parts of graph theory, cf.\
  e.g.~\cite{courcellebook,categoryMSO}). String-to-string MSO transductions
  capture the well-established class of \emph{regular functions},
  surveyed in~\cite{MuschollPuppis}.
\end{remark}

\subsection{On functions defined using unfoldings (proof of Proposition~\ref{prop:msot-msots})}
\label{sec:msot-msots}

We now prove the following inclusion for maps between relational structures:
  \[ \text{MSOT} \circ \unfold \circ \text{MSOT} \subset \text{MSO set interpretations} \]
(all such inclusions / equalities between function classes stated in this subsection are witnessed by \emph{effective} translations).
As explained in Section~\ref{sec:transducers-intro}, this result enables several applications of Theorem~\ref{thm:msosi-min} to tree transducers.

First, we recall some cases in which MSO (set) interpretations can be
  composed by simple, yet powerful, syntactic substitution: replacing relation symbols by
  formulas defining them. (The issues that arise in other cases are discussed
  in~\cite[Sections~1~and~2.1]{msoInterpretations}.) 
\begin{lemma}[{\cite[Proposition~2.4]{ColcombetL07}}]\label{lem:msosi-closure-prop}
  A composition of functions of the form
  \[ [\text{FO interpretation}] \circ [\text{MSO set interpretation}] \circ [\text{MSO transduction}] \]
  (between arbitrary relational structures) can be defined by a single MSO set interpretation.
\end{lemma}
Here, an \emph{FO interpretation} is an MSO interpretation defined using
formulas in \emph{first-order logic}, i.e.\ without using the second-order
quantifiers $\exists X / \forall X$. For instance, Example~\ref{ex:fot-dag-comb} is
actually a \emph{1-dimensional} FO interpretation, also known as an \emph{FO
  transduction}.

Our next lemmas also involve the class of \emph{MSO relabelings}: tree-to-tree MSO transductions that change only the labels of the nodes. 
\begin{lemma}[Colcombet~{\cite[Theorem~2]{DetFF}}]\label{lem:mso-via-fo}
  Every MSO transduction \emph{from trees} can be expressed as an MSO relabeling
  followed by an FO transduction.
\end{lemma}
\begin{remark}\label{rem:detff}
  This decomposition relies on a deep result~\cite[Theorem~1]{DetFF} on semigroup
  theory, which is therefore a dependency in our proof
  of Proposition~\ref{prop:msot-msots}.
\end{remark}
\begin{lemma}\label{lem:commutation}
  $\text{MSO relabeling} \circ \unfold \subset \unfold \circ \text{MSOT}$. 
\end{lemma}
\begin{proof}[Explanation]
  Colcombet and Löding have shown~\cite[Theorem~1]{ColcombetL04}\footnote{The theorem actually applies to rooted digraphs that are not necessarily acyclic, whose unfoldings are possibly infinite trees. Its proof may be found
  in Colcombet's PhD thesis (in French)~\cite[\S3.4.4]{ColcombetPhD}.} that
  \[ [\text{deterministic top-down tree transducer with
      lookahead}] \circ \unfold \subset \unfold \circ \text{MSOT} \]
  According to Bloem and Engelfriet~\cite{AttributedMSO}, the above variant of top-down tree transducers (whose definition we do not need here) contains a class called ATT-REL~\cite[p.~46]{AttributedMSO} which is equal to the class of MSO relabelings~\cite[Theorem~10]{AttributedMSO}.
\end{proof}

With these lemmas in place, we can now conclude.

\begin{proof}[Proof of {Proposition~\ref{prop:msot-msots}}]
  By the following reasoning:
  \begin{align*}
    \text{MSOT} \circ \unfold \circ \text{MSOT} &= \text{FOT} \circ \text{relabeling} \circ \unfold \circ \text{MSOT} & \text{(by Lemma~\ref{lem:mso-via-fo})}\\
    &= \text{FOT} \circ \unfold \circ \text{MSOT} \circ \text{MSOT} & \text{(by Lemma~\ref{lem:commutation})}\\
    &\subset \text{MSO set interpretation} \circ \text{MSOT} & (*)\\
    &= \text{MSO set interpretation} &\text{(by Lemma~\ref{lem:msosi-closure-prop})}
  \end{align*}
  where the step $(*)$ combines Lemma~\ref{lem:msosi-closure-prop} with the fact that $\unfold$ is defined by an MSO set interpretation (Example~\ref{ex:unfold}).
\end{proof}

\begin{remark}
  While we only use FO transductions as a convenient tool, they have been
  intensively studied for their own sake in the string-to-string case, see
  e.g.~\cite[p.~2:9--2:11]{MuschollPuppis}. Bojańczyk and Doumane have also investigated
  tree-to-tree FO transductions~\cite{FOTree}. In higher dimensions, FO interpretations
  of strings correspond to a well-behaved subclass of polyregular
  functions~\cite[Theorem~7(2)]{msoInterpretations}. Finally, let us remark that FO
  transductions of graphs\footnote{Confusingly, the literature on structural graph theory often uses \enquote{FO interpretations} to refer to a special case of FO transductions.} play a key role in recent progress
  in algorithmic model theory, cf.\
  e.g.~\cite[Section~5]{siebertz_et_al:LIPIcs.FSTTCS.2024.2}.
\end{remark}

\subsection{Output height of tree-to-tree MSO set interpretations}%
\label{sec:msosi-height}

Our last result on MSO set interpretations is half of Theorem~\ref{thm:pshi}. We reduce it to Claim~\ref{clm:query-vs-ambiguity-intro} on the growth rate of MSO queries:
\begin{theorem}\label{thm:msosi-height}
  Given an MSO set interpretation defining a tree-to-tree function $f$,
  one can compute an MSO formula $F(\vec{X})$ over a \emph{different input alphabet} than $f$ such that 
  \[ \growth[\height \circ f] = \growth[t' \mapsto |\wn F(t')|] \]
\end{theorem}
The following lemma gives more details regarding the properties of the construction:
\begin{lemma}\label{lem:msosi-height}
  Given an MSO set interpretation defining a function $f \colon \tree\Sigma \to \tree\Gamma$,
  one can compute a ranked alphabet $\Sigma'$ endowed with a rank-preserving surjection $\pi\colon\Sigma'\to\Sigma$, as well as an MSO formula $F(\vec{X})$ over $\tree\Sigma'$, such that 
  \[ \forall t \in \tree\Sigma, \quad \{ |\wn F(t')| \mid t \in \tree\Sigma',\; \pi_{*}(t') = t \} = \{0,\dots,\height(f(t))\} \]
  where $\pi_{*} \colon \tree\Sigma' \to \tree\Sigma$ is the tree relabeling which applies $\pi$ to all node labels.
\end{lemma}
\begin{proof}[Proof of the lemma]
Let $f$ be defined by the set interpretation $\cI$ with domain formula
$\cI_{\dom}(X_{1},\dots,X_{\ell})$. Consider the following MSO formula on $\tree\Sigma$:
\[ G(X_{1},\dots,X_{\ell}, Y_{1},\dots,Y_{\ell}) = \cI_\dom(\vec{X}) \land \cI_\dom(\vec{Y}) \land \vec{X} \neq \vec{Y} \land \cI_\leqslant(\vec{Y},\vec{X}) \]
It defines a set query whose results on $t\in\tree\Sigma$ are tuples $(P_1,\dots,P_\ell,S_1,\dots,S_\ell)$ such that:
\begin{itemize}
  \item $t \models \cI_\dom(\vec{P})$ and $t \models \cI_\dom(\vec{S})$ --- thus, $\vec{P}$ and $\vec{S}$ identify two nodes $p$ and $s$ of $f(t)$;
  \item $s$ is a strict ancestor of $p$ in the tree $f(t)$.
\end{itemize}
Let $\Sigma' = \Sigma \times \{0,1\}^{\ell}$. By Proposition~\ref{prop:mso-marked}, we can build $F(X_1,\dots,X_\ell)$ such that
\[\marked\bigl(t,\vec{P}\bigr) \models F\bigl(\vec{S}\bigr) \iff t \models G\bigl(\vec{P},\vec{S}\bigr)\]
for a certain bijection $\marked \colon \{(t,\vec{P}) \mid t \in \tree\Sigma,\; \vec{P} \in \powerset(\nodes(t))^{\ell}\} \to \tree\Sigma'$.
Therefore,
\[ \wn F\bigl(\marked\bigl(t,\vec{P}\bigr)\bigr) =
  \begin{cases}
    \{\text{strict ancestors of}\ p\} & \text{if $\vec{P}$ identifies some node $p$ in $f(t)$}\\
    \varnothing & \text{otherwise}
  \end{cases}
\]
Every output node is identified by some $\vec{P}$ since $\nodes(f(t)) = \wn\cI_{\dom}(t)$. The number of strict ancestors of a node is its depth, and the depths of the nodes in a tree range from 0 to its height. Hence:
\[ \forall t \in \tree\Sigma,\quad \bigl\{ \bigl|\wn F\bigl(\marked\bigl(t,\vec{P}\bigr)\bigr)\bigr| \;\big|\; \vec{P} \in \powerset(\nodes(t))^{\ell} \bigr\} = \{0,\dots,\height(t)\}  \]
To conclude, note that by definition, $\pi_*(\marked(t,\vec{P})) = t$ where $\pi \colon \Sigma' \to \Sigma$ is the first projection of the product $\Sigma \times \{0,1\}^{\ell}$.
\end{proof}

\begin{proof}[Proof of {Theorem~\ref{thm:msosi-height}}]
  For any $n\in\bbN$, we have:
  \[
    \{t' \in \tree\Sigma' \mid |t'| \leqslant n \} = \bigcup_{|t| \leqslant n} \pi_{*}^{-1}(\{t\})
  \]
  because the tree relabeling $\pi_{*}$ is size-preserving. Therefore, thanks to Lemma~\ref{lem:msosi-height},
  \[ \max_{|t'|\leqslant n} |\wn F(t')| = \max_{|t|\leqslant n} \max_{\pi_{*}(t')=t} |\wn F(t')|
    = \max_{|t|\leqslant n} \max(\{0,\dots,\height(f(t))\}) = \max_{|t|\leqslant n} \height(f(t)) \]
  which is what we want by definition of the growth rate function.
\end{proof}

\section{From tree transducers to $\bbN$-weighted tree automata, directly}%
\label{sec:directly}

In this section, which can be read independently of \S\ref{sec:queries}--\S\ref{sec:set-interpretations}, we prove some results on transducers by reduction to Theorem~\ref{thm:degree-weighted}, without invoking MSO.
\begin{itemize}
  \item Section~\ref{sec:top-down} is devoted to the proof of Theorem~\ref{thm:top-down-intro} on top-down tree transducers. In fact, we state a parameterised version (Theorem~\ref{thm:top-down} below) which, as promised in the introduction, can be used to count (\textit{inter alia}) all nodes or just the leaves.
  \item We then apply this result in Section~\ref{sec:mtt-height} in order to prove the part of Theorem~\ref{thm:pshi} (on size-to-height increase) that concerns macro tree transducers.
\end{itemize}

\subsection{Output size of nondeterministic top-down tree transducers}%
\label{sec:top-down}

\begin{definition}\label{def:top-down}
  A \emph{partial nondeterministic top-down tree transducer} $\cT$ from an input
  alphabet $\Sigma$ to an output alphabet $\Gamma$ (both ranked) consists of:
  \begin{itemize}
    \item a finite set $Q$ of states, with an initial state $q_{0}$;
    \item for each state $q\in Q$ and each input letter $a\in\Sigma$, a finite set of trees
    \[ \RHS_{\cT}(q,a) \subset \tree(\Gamma \cup \underbrace{(Q \times \{x_{1},\dots,x_{\rank(a)}\})}_{\mathclap{\text{we write \(q\mttarg{x_{i}}\) for \((q,x_{i})\) and its rank is}\ 0}}) \]
    which should be understood as the collection of right-hand sides of \enquote{rules}
    \[ q\mttarg{a(x_{1},\dots,x_{\rank(a)})} \rightsquigarrow \rho \quad\text{for}\ \rho \in \RHS_{\cT}(q,a) \]
  \end{itemize}
  The transducer $\cT$ defines a set of values
  $q\mttarg{t} \subset \tree(\Gamma)$ for every state $q$ and input tree $t \in \tree\Sigma$, by structural induction: 
  $q\mttarg{a(t_{1},\dots,t_{\rank(a)})}$ is the set of trees that can be obtained by
  \begin{itemize}
    \item first choosing a rule for $q,a$ i.e.\ a tree $\rho \in \RHS_{\cT}(q,a)$,
    \item then, for each occurrence of a leaf with label $q'\mttarg{x_{i}}$, replacing it by some choice of tree in the set $q'\mttarg{t_{i}}$ (one may perform different choices for different leaves with the same label).
  \end{itemize}
  The semantics of $\cT$ is the binary relation $\sem{\cT} = \{(t,s) \in \tree\Sigma \times \tree\Gamma \mid s \in q_{0}\mttarg{t}\}$.
\end{definition}
\begin{definition}
  A partial nondeterministic top-down tree transducer $\cT$ is said to be:
  \begin{itemize}
    \item \emph{total} when $|\RHS_{\cT}(q,a)| \geqslant 1$ for all $q,a$; this entails that $|q\mttarg{t}| \geqslant 1$ for all $q,t$;
    \item \emph{deterministic} when $|\RHS_{\cT}(q,a)| \leqslant 1$ for all $q,a$; this entails that $|q\mttarg{t}| \leqslant 1$ for all $q,t$.
  \end{itemize}
  In the deterministic case, we write $\rhs_\cT(q,a)$ for the unique element of $\RHS_{\cT}(q,a)$ whenever it exists (if $\RHS_{\cT}(q,a) = \varnothing$ then $\rhs_\cT(q,a)$ is undefined).
  
  If $\cT$ is deterministic, then $\sem{\cT}$ is the graph of a partial function $\tree\Sigma \rightharpoonup \tree\Gamma$ (and it is total if $\cT$ is total). Abusing notation, we also denote this partial function by $\sem{\cT}$.
\end{definition}

\begin{example}
  Let $\Sigma = \{S : 1,\; \underline{0} : 0\}$ and $\Gamma = \{a : 2,\; b : 0,\; c : 0\}$.
  \begin{itemize}
    \item The function $S^{n}(\underline{0}) \in \tree\Sigma \mapsto a(b^{n}(c),a(b^{n-1}(c),\dots a(b(c),c)\dots)) \in \tree\Gamma$ of Example~\ref{ex:comb-intro} is computed by a total deterministic transducer with two states $q_0,q_1$ and the transitions
    \begin{align*}
      q_0\mttarg{S(x_1)} &\rightsquigarrow a(q_1\mttarg{x_1},q_0\mttarg{x_1}) & q_0\mttarg{\underline{0}} &\rightsquigarrow c\\
      q_1\mttarg{S(x_1)} &\rightsquigarrow b(q_1\mttarg{x_1}) & q_1\mttarg{\underline{0}} &\rightsquigarrow b(c)
    \end{align*}
    \item The following total nondeterministic example relates each $S^n(\underline{0})$ to all unary-binary trees whose leaves are all at depth $n$: it has a single state $q$ and its transitions are
    \[ q\mttarg{S(x_1)} \rightsquigarrow a(q\mttarg{x_1},q\mttarg{x_1}) \qquad q\mttarg{S(x_1)} \rightsquigarrow b(q\mttarg{x_1}) \qquad q\mttarg{\underline{0}} \rightsquigarrow c\]
    \item Let $\lambda_c,\lambda_d \in \bbN$. Consider the partial nondeterministic top-down tree transducer with input $\tree\{a : 2,\; c : 0,\; d : 0\}$ and output $\tree\Sigma$ whose states are $q_0,q_c,q_d$ and whose rules are:
          \begin{align*}
            q_0\mttarg{a(x_1,x_2)} &\rightsquigarrow S^{\lambda_z}(q_z\mttarg{x_i}) \quad \text{for}\ z \in \{c,d\},\, i \in \{1,2\}\quad (\text{so}\ |\RHS(q_0,a)| = 4)\\
            q_z\mttarg{a(x_1,x_2)} &\rightsquigarrow S^{\lambda_z}(q_z\mttarg{x_i}) \quad \text{for}\ z \in \{c,d\},\, i \in \{1,2\}\quad (\text{so}\ |\RHS(q_z,a)| = 2)\\
            q_0\mttarg{z} &\rightsquigarrow \underline{0} \quad\text{and}\quad q_z\mttarg{z} \rightsquigarrow \underline{0} \quad \text{for}\ z \in \{c,d\}
          \end{align*}
          Its semantics is the following relation, which is total (every tree has at least one leaf): 
          \[ \bigl\{\bigl(t,S^{\lambda_z \times n}(\underline{0})\bigr) \mid t\ \text{contains a}\ z\in\{c,d\}\ \text{at depth}\ n\in\bbN \bigr\} \]
          and yet the transducer is not total ($\RHS(q_c,d) = \varnothing$). \qedhere
  \end{itemize}
\end{example}

\begin{definition}
  For a ranked alphabet $\Gamma$, a subset $\Pi \subset \Gamma$, and a tree $t \in \tree\Gamma$, we write $|t|_{\Pi}$ for the number of nodes of $t$ whose labels belong in $\Pi$. We abbreviate $|t|_{c} = |t|_{\{c\}}$ for $c\in\Gamma$. 
\end{definition}
\begin{theorem}[{Theorem~\ref{thm:top-down-intro} parameterised by $\Pi$}]\label{thm:top-down}
  For any partial nondeterministic top-down tree transducer $\cT$, the binary relation $\{(t,|s|_\Pi) \mid (t,s) \in \sem{\mathcal{T}}\}$ has a well-defined degree of growth in $\bbN\cup\{\infty\}$ that can be computed from $\cT$:
  \begin{itemize}
    \item in \emph{exponential time} in general;
    \item in \emph{polynomial time} when $\cT$ is \emph{total}.
  \end{itemize}
\end{theorem}

\noindent
There already exists a reduction from the nondeterministic case to the deterministic case:
\begin{lemma}[Drewes~{\cite[Lemma~3.2]{Drewes01}}]
  For every top-down tree transducer $\cT$, one can construct a deterministic top-down tree transducer $\cT'$ such that, for some constant $K \in \bbN$,\footnote{Where $R \circ S = \{(x,z) \mid \exists y : (x,y) \in S\ \text{and}\ (y,z) \in R\}$ is the usual composition of binary relations, extending function composition (a function is identified with its graph).}
  \[ \growth[|\cdot|_\Pi \circ \sem{\cT}](n) \leqslant \growth[|\cdot|_\Pi \circ \sem{\cT'}](Kn) \leqslant \growth[|\cdot|_\Pi \circ \sem{\cT}](Kn) \]
  for every $n\in\bbN$ and every set $\Pi$ of output letters ($\cT$ and $\cT'$ have the same output alphabet). In particular, the degree of growth of $|\cdot|_\Pi \circ \sem{\cT}$ is equal to that of $|\cdot|_\Pi \circ \sem{\cT'}$ (as we shall see below, the latter degree is well-defined).

  The construction preserves totality and can be carried out in polynomial time.
\end{lemma}
\begin{remark}
  Drewes originally stated this result without the parameter $\Pi$. Berglund, Drewes and van der Merwe later claimed that the construction also works for the number of output leaves, cf.~\cite[proof of Theorem~20]{BerglundDM18}. Likewise, we claim that it works for any $|\cdot|_\Pi$.
\end{remark}

There remains to prove the deterministic case. We start by reducing the \emph{total} deterministic case to Theorem~\ref{thm:degree-weighted} on $\bbN$-weighted tree automata. Our construction extends Berglund et al.'s treatment of the string-to-tree case~\cite[Lemma~18]{BerglundDM18}.

\begin{lemma}\label{lem:total-det-top-down}
  Let $\cT$ be a total deterministic top-down tree transducer with $\sem{\cT} \colon \tree\Sigma \to \tree\Gamma$ and let $\Pi \subseteq \Gamma$. One can build in time $O(|\cT| \times \max(\rank(\Sigma)))$ an $\bbN$-weighted tree automaton that computes $t \mapsto |\sem{\cT}(t)|_{\Pi}$.
\end{lemma}
\begin{proof}
  Let $Q$ be the set of states of $\cT$ and $q_0$ be its initial state. We build an $\bbN$-weighted tree automaton $\cA$ with the set of states $Q \cup \{\mathbf{1}\}$, the single accepting state $q_0$ with weight 1, and the following transitions:
  \begin{itemize}
    \item $(\mathbf{1},\dots,\mathbf{1},a,\mathbf{1})$ with weight 1 for each $a \in \Sigma$;
    \item $(\mathbf{1},\dots,\mathbf{1},a,q)$ with weight $|\rhs(q,a)|_{\Pi}$ for each $a \in \Sigma$ such that $\rhs(q,a)$ contains at least one node with label in $\Pi$;
    \item $(\mathbf{1},\dots,\mathbf{1},q',\mathbf{1},\dots,\mathbf{1},a,q)$ with $q'$ appearing in the $i$-th component of the tuple and with weight $|\rhs(q,a)|_{q'\mttarg{x_i}}$ for each $a \in \Sigma$ and $i \leqslant \rank(a)$ such that $\rhs(q,a)$ contains at least one leaf with label $q'\mttarg{x_i}$.
  \end{itemize}
  For each $q,a$ and for each letter that labels at least one node in $\rhs(q,a)$, we have a new transition of size $\rank(a)$, hence the $O(|\cT| \times \max(\rank(\Sigma)))$ time bound for building $\cA$.
  
  For $p \in Q \cup \{\mathbf{1}\}$, let $\cA_p$ be obtained from $\cA$ by setting $p$ to be the single accepting state, with weight 1. Since $(\mathbf{1},\dots,\mathbf{1},a,\mathbf{1})$ is the only transition whose last component is $\mathbf{1}$, the value of $\cA_{\mathbf{1}}$ is 1 on every input tree.
  Let us prove by induction that for every $t \in \tree\Sigma$ and $q \in Q$, we have $|q\mttarg{t}|_{\Pi} = \cA_q(t)$ (the case $q_0$ gives us the lemma statement):

  \begin{align*}
    |q\mttarg{a(t_1,\dots,t_{\rank(a)})}|_{\Pi}
    &= |\rhs_{\cT}(q,a)[q'\mttarg{x_{i}} \leftarrow q'\mttarg{t_{i}}\ \text{for all}\ q',\, x_{i}]|_\Pi \\
    &= |\rhs_{\cT}(q,a)|_\Pi + \sum_{q',i} |\rhs_{\cT}(q,a)|_{q'\mttarg{x_{i}}} \times |q'\mttarg{t_{i}}|_\Pi\\
    &= |\rhs_{\cT}(q,a)|_\Pi + \sum_{q',i} |\rhs_{\cT}(q,a)|_{q'\mttarg{x_{i}}} \times \cA_{q'}(t_i)\quad \text{by ind.\ hyp.}
  \end{align*}
  Using the fact that $\cA_{\mathbf{1}}(\tree\Sigma) = \{1\}$, the last expression is equal to
  \[ |\rhs_{\cT}(q,a)|_\Pi \times \prod_j \cA_{\mathbf{1}}(t_j) + \sum_{q',i} |\rhs_{\cT}(q,a)|_{q'\mttarg{x_{i}}}  \times \cA_{q'}(t_i) \times \prod_{j\neq i} \cA_{\mathbf{1}}(t_j) \]
  which is the sum, over all transitions $(p_1,\dots,p_{\rank(a)},a,q)$ in $\cA_q$, of their weight times the product of the $\cA_{p_i}(t_i)$. Thus, it is equal to $\cA_q(a(t_1,\dots,t_{\rank(a)}))$.
\end{proof}
\begin{remark}\label{rem:weighted-to-top-down}
  In the case of strings (unary trees) as input treated by Berglund et al.~\cite{BerglundDM18}, this construction almost admits a right inverse. Indeed, let $\cA$ be an $\bbN$-weighted automaton over a ranked alphabet $\Sigma$ with $\max(\rank(\Sigma)) = 1$, assuming without loss of generality that $\cA$ has a single accepting state $q_0$ of weight 1. One can build a total deterministic top-down tree transducer $\tree\Sigma \to \tree\Gamma$ where $\Gamma=\{\underline{+} : 2,\; \underline{0} : 0,\; \underline{1} : 0\}$ such that:
  \begin{itemize}
    \item its states are the same as $\cA$;
    \item its transitions for each state $q$ satisfy:
    \begin{itemize}
      \item for $\rank(c)=0$, $|\rhs(q,c)|_{\underline{1}} = \text{weight of the transition}\ (c,q)$ in $\cA$ (take the syntax tree of an expression that adds the right number of copies of $\underline{1}$; we say that the weight of a non-existing transition is 0 and in this case we take $\rhs(q,c)=\underline{0}$);
      \item for $\rank(a)=1$, $|\rhs(q,a)|_\Gamma = 0$ and $|\rhs(q,a)|_{q'\mttarg{x_1}} = \text{weight of}\ (q',a,q)$ for all $q'$. 
    \end{itemize}
  \end{itemize}
  Applying the construction of Lemma~\ref{lem:total-det-top-down} for $\Pi = \{\underline{1}\} \subset \Gamma$ to this transducer, we get a weighted automaton $\cA'$ which is almost the same as $\cA$, the only differences being a new state $\mathbf{1}$ and new transitions that are all of the form $(\dots,a,\mathbf{1})$.
  In particular, $\cA'$ computes the same function as $\cA$ (indeed, $\mathbf{1}$ is not co-accessible in this specific case).

  We may conclude that \emph{every function computed by an $\bbN$-weighted automaton on strings is the yield output size of some total deterministic top-down string-to-tree transducer} (taking $\yield_{\Gamma,\{\underline{0}\}}$ as defined in Definition~\ref{def:branches-yield} below).
\end{remark}
\begin{remark}
  There is no such converse over non-unary trees. Indeed:
\begin{itemize}
  \item top-down tree transducers have at most linear height increase (cf.~e.g.~\cite[\S1]{LSHI}) and therefore at most exponential height-to-size increase;
  \item if $\Sigma$ contains a letter of rank $k \geqslant 2$, then $t\in\tree\Sigma \mapsto 2^{|t|}$ can be doubly exponential in the input height (consider perfect $k$-ary trees), and this function is the ambiguity of a tree automaton with two states (both accepting) and all possible transitions. 
\end{itemize}
\end{remark}

\noindent
Next, Berglund, Drewes and van der Merwe translate partial deterministic top-down string-to-tree transducers into total ones with the same growth rate~\cite[Theorem~22]{BerglundDM18}. We prefer to take another approach, deducing the partial deterministic case by manipulations on $\bbN$-weighted tree automata. The following lemma completes the proof of Theorem~\ref{thm:top-down}.
\begin{lemma}
  Let $\cT$ be a partial deterministic top-down tree transducer with $\sem{\cT} \colon \tree\Sigma \rightharpoonup \tree\Gamma$ and let $\Pi \subseteq \Gamma$. One can build in exponential time an $\bbN$-weighted tree automaton that computes
  \[ t \in \tree\Sigma \mapsto
    \begin{cases}
      |\sem{\cT}(t)|_{\Pi} &\text{if}\ \sem{\cT}(t)\ \text{is defined}\\
      0 & \text{otherwise}
    \end{cases}
  \]
\end{lemma}
\begin{proof}
  First, we can complete $\cT$ arbitrarily into a total deterministic transducer $\cT'$ such that $\sem{\cT}$ and $\sem{\cT'}$ coincide whenever the former is defined. We can apply the polynomial-time algorithm from Lemma~\ref{lem:total-det-top-down} to $\cT'$ and get an $\bbN$-weighted automaton $\cA$ that computes $t \mapsto |\cT'(t)|_\Pi$. But we do not control the value of $\cA$ outside the domain of $\sem{\cT}$.

  One can build a tree automaton $\cB$ that recognises this domain, using a powerset-based construction that can be carried out from $\cT$ in exponential time~\cite[Theorem~3.1]{Engelfriet77}.\footnote{This paper by Engelfriet, which is half a century old, has had an erratum published nearly 40 years later~\cite{Engelfriet77Erratum}. Fortunately the construction that we cite is not affected by the error.} Furthermore, this construction guarantees that $\cB$ is \emph{deterministic top-down}: for every state $q$ and letter $a$ there is at most one transition of the form $(\dots,a,q)$. This property implies that $\cB$ is \emph{unambiguous}. As a consequence, the $\bbN$-weighted tree automaton $\cB'$ obtained from $\cB$ by setting all weights to 1 maps the domain of $\cT$ to 1 and its complement to 0.

  To conclude, note that the function that we want in the end is $t \mapsto \cA(t) \times \cB'(t)$. It is computed by the product of the weighted automata $\cA$ and $\cB'$, cf.\ e.g.~\cite[Lemma~3.8]{Borchardt04}. This product construction extends the usual product of tree automata (that we described in Section~\ref{sec:exp-vs-poly}) by multiplying weights; it can be done in time $O(|\cA| \times |\cB'|)$.
\end{proof}

\begin{remark}
  Berglund et al.\ actually give a PSPACE algorithm that computes the degree of growth of a partial nondeterministic top-down string-to-tree transducer~\cite[Theorem~24]{BerglundDM18}. However, in the tree-to-tree case, Drewes has shown that deciding polynomial vs exponential growth is EXPTIME-complete, even for partial deterministic transducers~\cite[Theorem~5.1]{Drewes01}. The same theorem also states that for total top-down tree transducers, deciding exponential growth is NL-complete (both in the deterministic and nondeterministic cases). 
\end{remark}

\subsection{Output height of (total deterministic) macro tree transducers}%
\label{sec:mtt-height}

We can now reduce the part of Theorem~\ref{thm:pshi} that concerns MTTs to Theorem~\ref{thm:top-down} proved in the previous subsection:

\begin{lemma}\label{lem:mtt-height}
  From a total deterministic macro tree transducer $\cM$, one can compute a partial nondeterministic top-down tree transducer $\cT$ over the same input alphabet $\Sigma$, and a subset $\Pi$ of the output alphabet of $\cT$, such that
  \[ \forall t \in \tree\Sigma,\quad \height(\sem{\cM}(t)) = \max\{|s|_\Pi \mid (t,s) \in \sem{\cT}\}\]
  and therefore $\growth[\height \circ \sem{\cM}] = \growth[|\cdot|_\Pi \circ \sem{\cT}]$.
\end{lemma}

Following the same methodology as Section~\ref{sec:transducers-intro}, we do not define MTTs; instead we draw on composition properties found in the literature, relating them to top-down tree transducers. We start by recalling two standard operations on trees.
\begin{definition}\label{def:branches-yield}
  Let $\Gamma$ be a ranked alphabet.

  The binary relation $\Branches_\Gamma \subset \tree\Gamma \times \Gamma^*$ contains the pairs $(t,w)$ where the word $w$ lists the node labels on some path from the root of $t$ to a leaf. Inductively:
  \[ (a(t_1,\dots,t_n), w) \in \Branches_\Gamma \iff (w = a \land n=0) \lor (w = aw' \land \exists i : (t_i,w') \in \Branches_\Gamma)\]
  Let $e \in \Gamma^{(0)}$ where $\Gamma^{(0)} \subset \Gamma$ is the set of rank 0 letters. The map $\yield_{\Gamma,e} \colon \tree\Gamma \to (\Gamma^{(0)}\setminus\{e\})^*$ maps a tree to the labels of non-$e$ leaves read from left to right, i.e.,
  \begin{align*}
    \text{for}\ a \in \Gamma^{(0)}\setminus\{e\}, \quad \yield_{\Gamma,e}(a) &= a \\
    \yield_{\Gamma,e}(e) &= \varepsilon\ (\text{the empty string})\\
    \text{for}\ n \geqslant 1,\quad \yield_{\Gamma,e}(a(t_1,\dots,t_{n})) &= \yield(t_1) \cdot \ldots \cdot \yield(t_{n}) 
  \end{align*}
\end{definition}
\begin{proof}[Proof of {Lemma~\ref{lem:mtt-height}}]
  We use two results by Engelfriet and Vogler~\cite{Macro,EngelfrietHighLevel} from the 1980s which involve the notion of nondeterministic macro tree transducer with outside-in (OI) derivations, or OI-MTT for short. This is yet another opaque keyword that we use merely to access the literature.
  \begin{itemize}
    \item A tree-to-string relation can be expressed as $\yield \circ \sem{\mathcal{T}}$ for some top-down tree transducer $\cT$ if and only if can be defined by an OI-MTT that outputs encodings of strings as unary trees, via $\Sigma^* \cong \tree(\{a : 1 \mid a \in \Sigma\} \cup \{\bullet : 0\})$~\cite[Theorem~8.6(a)]{EngelfrietHighLevel}.\footnote{It is the case $n=0$ of this theorem: as explained in~\cite[Fact~4.6]{EngelfrietHighLevel}, the level~0 tree transducers considered in that paper correspond to nondeterministic top-down tree transducers while level~1 transducers correspond to OI-MTTs.}
    \item A tree-to-tree relation can be defined by an OI-MTT if and only if it can be expressed as $\sem{\mathcal{T}'} \circ \sem{\mathcal{M}}$ where $\mathcal{T}'$ is a partial nondeterministic top-down tree transducer and $\mathcal{M}$ is a total deterministic MTT~\cite[Corollary~6.12]{Macro}.\footnote{Beware: in the paper~\cite{Macro}, the notation for composition is in the reverse order.}
  \end{itemize}
  In both cases, the equivalences are witnessed by effective conversions. Let us apply the \enquote{if} directions by taking $\sem{\mathcal{T}'} = \Branches$. (Concretely, we can build such a top-down tree transducer $\mathcal{T}'$ with a single state $q$ and the rules $q\mttarg{a(x_1,\dots,x_n)} \rightsquigarrow a(x_i)$ for every input letter $a$ of rank $n \geqslant 1$ and $i \in \{1,\dots,n\}$ ($a$ is also an output letter of rank 1) and $q\mttarg{c} \rightsquigarrow c(\bullet)$ for $\rank(c) = 0$.) We then get $\cT$ satisfying
  \[ \Branches \circ \sem{\cM} = \yield \circ \sem{\cT} \]
  This implies the desired result because for all $t \in \tree\Gamma$, we have $|\yield_{\Gamma,e}(t)| = |t|_{\Gamma^{(0)}\setminus\{e\}}$ and $\height(t) = \max\{|w| \mid (t,w) \in \Branches_\Gamma\}$.
\end{proof}

One drawback of this abstract argument is that it does not come with complexity bounds. It might be interesting to revisit Engelfriet and Vogler's aforementioned equivalence results to see whether they could be witnessed by polynomial-time translations. We also believe that a direct construction could lead to the following result:
\begin{conjecture}
  From a (total deterministic) macro tree transducer $\cM$, one can compute \emph{in polynomial time} a partial nondeterministic top-down tree transducer $\cT$ over the same input alphabet, such that $\Branches \circ \sem{\cM} = \yield \circ \sem{\cT}$.

  Furthermore, if $\cM$ is \emph{nonerasing~\cite[Definition~3.9]{MacroMSOLinear}}, then $\cT$ is \emph{total}. 
\end{conjecture}
To motivate the second half of this statement, recall from Theorem~\ref{thm:top-down} that computing the degree of growth of a nondeterministic top-down tree transducer can be done more efficiently in the total case. Let us also mention that MTTs are equivalent in expressive power with nonerasing MTTs \emph{with regular lookahead}~\cite[Lemma~3.10]{MacroMSOLinear}. One could hope to get a polynomial-time algorithm that computes the degree of size-to-height increase of these devices, by adapting the lemmas on top-down transducers in the previous subsection to incorporate regular lookahead (see also Remark~\ref{rem:total-det-top-down}).

\section{Conclusion}\label{sec:conclusion}

In this paper, we followed a direct thread:
\begin{itemize}
  \item from effective characterisations for the growth rate of $\bbN$-weighted tree automata,
  \item to reparameterisation of MSO set queries of polynomial growth,
  \item and then to dimension minimization for MSO set interpretations,
\end{itemize}
with the arguments for these last two items being of elementary character. On the contrary,
Bojańczyk's original proofs~\cite{PolyregGrowth} for $k$-ary MSO queries and interpretations on strings are based on the factorization forest
theorem, and does no seem to lend themselves to an extension to set queries/interpretations.
\begin{remark}\label{rem:reparam-ext}
  However, semigroup-theoretic techniques seem to be required for now to prove other variants of reparameterisation:
  \begin{itemize}
    \item a refinement on strings by the third author and Parys~\cite{FinerReparam};
    \item Rabinovich's recent extension to countable labelled chains~~\cite{RabinovichWoLLIC,RabinovichICALP}.
  \end{itemize}
\end{remark}

To demonstrate the wide applicability of our theorems on $\bbN$-weighted tree
automata and MSO set interpretations, we then used them to derive various
results on the asymptotics of tree transducers. These applications sometimes
required a bit of work, but again, no new difficult combinatorics; the only
place where pumping arguments occur is Section~\ref{sec:ambiguity}. Our methodology was
to work as little as possible with machine models for tree-to-tree functions.
The proof of dimension minimization (Theorem~\ref{thm:msosi-min}) leverages the
flexibility of the syntax of MSO (set) interpretations to reduce the problem to
reparameterising the domain query: we do not have to worry about how the order
is defined on the output structure. Meanwhile, combinatorial arguments on macro
tree transducers, such as Engelfriet and Maneth's proof of Theorem~\ref{thm:em}, do not
decouple the size of the output from its shape.

\subsection{Some open problems}\label{sec:open-problems}

\subparagraph{Ambiguity vs height.}

One limitation of our work is that it investigates the growth of some output
parameter depending on the \emph{size} of an input tree. It would also make
sense to study the dependency on the input \emph{height}: the suitable
definition of growth rate would then be
\[ h\in\bbN \mapsto \max\{f(t) \mid \height(t)\leqslant h\} \qquad\text{for}\ f\colon\tree\Sigma\to\bbN. \]
For example, Maletti, Nasz, Stier and Ulbricht~\cite{MalettiNSU23} study tree
automata that are \enquote{polynomially ambiguous in height}, by which they mean
the existence of a polynomial upper bound. It is natural to ask: is this
property decidable, and does it entail that the growth rate of the ambiguity is
$\Theta(h^{k})$ for some computable $k\in\bbN$? A positive answer would also
allow us to compute the polynomial degree of
height increase of an MSO set interpretation (indeed, in Lemma~\ref{lem:msosi-height}, the tree relabeling $\pi_*$ is height-preserving).
For now, we know the case $k=1$:
\begin{theorem}[Gallot, Maneth, Nakano \& Peyrat~{\cite[Theorem~17(2)]{LSHI}}]\label{thm:lhi}
  It is decidable whether a given macro tree transducer is of linear height increase.
\end{theorem}
\begin{corollary}
  It is decidable whether the ambiguity of a given tree automaton is at most linear in the input height.
\end{corollary}
\begin{proof}
  There is (effectively) some MSO query $F$ describing the runs of the tree automaton, such that the latter's ambiguity over an input $t$ is equal to $|\wn F(t)|$ (cf.\ Claim~\ref{clm:runs-in-mso} in the appendix). From this one can derive an MSO set interpretation that defines $g \colon t \mapsto S^{|\wn F(t)|}(\underline{0})$. One can decide by Theorem~\ref{thm:msosi-min} whether $|g(t)| = O(|t|)$. If it does not, then we cannot have $|\wn F(t)| = O(\height(t))$. If it does, then $g$ is effectively definable by some MSO transduction, and therefore by some macro tree transducer~\cite[Lemma~4.9]{MacroMSOLinear}. We can conclude by invoking the above theorem since $|\wn F(t)| = \height(g(t)) = |g(t)|-1$.
\end{proof}

\subparagraph{A finer understanding of exponential growth.}

The question is: does the following theorem, which assumes that the input are
strings, extend to trees?
\begin{theorem}\label{thm:precise-exp}
  If an $\bbN$-weighted automaton $\cA$ on words has exponential growth ---
  which we defined as $\growth[\cA](n) = 2^{\Theta(n)}$ --- then in fact
  $\growth{[\cA]}(n) = {\rho}^{n+o(n)}$ for some $\rho > 1$.
\end{theorem}
This result does not seem to have been explicitly stated in the literature, but see Theorem~\ref{thm:entropy} for a precursor.
Let us explain how it follows directly from the context of the work~\cite{Jungers2008}
mentioned in Section~\ref{sec:related-work-intro}. That work by Jungers et al.\ is
actually phrased in terms of products of matrices rather than
automata, providing decision procedures that are originally about the
asymptotics of
$F_{S} \colon n\in\bbN\mapsto \max\bigl\{\|A_{1} \times \cdots \times A_{n}\| \;\big|\; A_{1},\dots,A_{n}\in S \bigr\}$ for a finite $S\subseteq\bbN^{d\times d}$.
\begin{claim}[proved in Appendix~\ref{app:joint-spectral}]\label{clm:joint-spectral}
If the elements of $S$ are the identity matrix and the transition matrices (one
for each input letter) of an $\bbN$-weighted automaton $\cA$ on strings with $d$
states, then, assuming that $\cA$ is \emph{trim} (without loss of generality,
cf.\ Section~\ref{sec:trim}),
\[ \growth[\cA](n) =  \Theta\bigl(F_{S}(n)\bigr) \quad\text{as}\ n \to \infty. \]
\end{claim}
This connects the asymptotics of $\cA$ to the \emph{joint spectral radius}
$\rho(S)$, defined as the limit of a sequence that, importantly, always
converges~\cite[p.~379]{Rota1960}:
\[ \rho(S) = \lim_{\mathclap{n\to\infty}} \sqrt[n]{F_{S}(n)} \quad\text{and therefore}\quad \frac{1}{n} \log(\growth{[\cA]}(n)) = \log(\rho(S)) + o(1) \]
(indeed, here $\rho(S)\geqslant1$ because we have put the identity matrix in $S$). It follows that:
\begin{itemize}
  \item $\cA$ has exponential growth if and only if $\rho(S)>1$;\footnote{This
        had been observed by Middeldorp, Moser, Neurauter, Waldmann and
        Zankl~\cite[Section~5]{MiddeldorpMNWZ11} but their proof goes through
        \enquote{heavy cycles} (\S\ref{sec:exp-vs-poly}) rather than directly
        relating $\growth[\cA]$ with $F_{S}$ as we do.}
  \item if that is the case, then Theorem~\ref{thm:precise-exp} holds, taking
        $\rho=\rho(S)$.
\end{itemize}
See also~\cite{JointSpectralRadius} for a book reference on the joint spectral
radius, and~\cite{Blondel2008} for historical remarks on this notion.

\subsection{Related work on transducers}\label{sec:related-work-trans}

\subparagraph{Aho and Ullman's~\cite{AhoU71}.}

The \emph{generalised syntax-directed translations} (GSDTs) defined in this work are a class of string-to-string relations. They can be rephrased, for elementary syntactic reasons (cf.~\cite[Section~3]{ERS}), as the relations obtained from a context-free grammar $G$ and a deterministic total top-down tree transducer $f$ as
\[ \{(w,\yield(f(t))) \mid t\ \text{is a parse tree of}\ G\ \text{for}\ w \} \]
They show that $\yield\circ f$, restricted to the set of parse trees of $G$, has a well-defined and computable degree of growth in $\bbN\cup\{+\infty\}$~\cite[Theorem~5.1]{AhoU71} --- hence the connection to Theorem~\ref{thm:top-down-intro}. As a corollary, this also applies to GSDTs defined using unambiguous grammars~\cite[Theorem~5.2]{AhoU71}.

A consequence of Remark~\ref{rem:weighted-to-top-down} is that for any $\bbN$-weighted automaton $\cA$ on words, the string-to-string function $w \mapsto \underline{1}^{\cA(w)}$ is an unambigous GSDT, via a very simple construction. Therefore, Theorem~\ref{thm:degree-weighted} for strings and without complexity bounds could have been derived directly from~\cite[Theorem~5.2]{AhoU71} (though history did not follow this course, cf.~\S\ref{sec:related-work-intro}).

Finally, Aho and Ullman also give a formalism that is expressively equivalent to GSDTs whose output length is linear in the size of the parse tree~\cite[Theorem~7.3]{AhoU71}. It involves the first appearance of tree-walking transducers in the literature (in the tree-to-string case). This result is also a direct inspiration for Engelfriet and Maneth's proof of~Theorem~\ref{thm:em} on MTTs, as discussed in~\cite[Section~1]{MacroMSOLinear}.

\subparagraph{Streaming string transducers.}

The following (total deterministic) transducers all compute the same string-to-string functions, as explained in e.g.~\cite[\S2.5]{tito24} for the first 3 items:
\begin{itemize}
  \item macro tree transducers whose inputs and outputs are encodings of strings as unary trees;
  \item $\yield \circ [\text{top-down tree transducers with unary input trees}]$;    
  \item right-to-left \emph{(copyful) streaming string transducers}\footnote{The \emph{copyless} version was
  introduced earlier by Alur and Černý~\cite{SST} and characterises MSO
  transductions; see also~\cite{MuschollPuppis}. Although we mostly focus on
  deterministic transducers, let us mention a recent result about nondeterminism
  by Filiot, Jecker, Löding, Muscholl, Puppis and
  Winter~\cite{theoretics:13747}: the \emph{finite-valued} string-to-string
  relations computed by nondeterministic copyless SSTs are exactly the relations
  computed by \emph{finitely ambiguous} copyless
  SSTs~\cite[Prop.~3.2]{theoretics:13747} ---
  see~\cite[Theorem~1.2]{theoretics:13747} for a stronger statement. The
  introduction of~\cite{theoretics:13747} also discusses the history of this
  kind of result about finitely ambiguous transducers. A final remark: compare
  the \enquote{barbells} in this paper (Definition~\ref{def:barbell}) with~\cite[Definition~4.7]{theoretics:13747}.} (SSTs)~\cite{CopyfulSST};
  \item HDT0L systems~\cite[Theorem~3.1]{CopyfulSST};\footnote{The closely related EDT0L systems have been known to be connected to top-down tree transducers since the 1970s, see~\cite[Section~3]{ERS} or~\cite[Corollary~4.7]{Engelfriet77} (which point to older references); but the connection was merely at the level of output languages. EDT0L languages are still an active research topic, for instance in relation to finitely generated groups~\cite{CiobanuLevine,bishop2025groupsedt0lwordproblem}.}
  \item some other characterisations studied in~\cite{FerteMarinSenizergues}.
\end{itemize}
Most of these equivalences take advantage of the fact that these formalisms are syntactically very close to each other, even if their usual presentations may evoke different intuitions (top-down states $\approx$ bottom-up registers $\approx$ non-terminals).

Filiot and Reynier~\cite[Theorem~5.2]{CopyfulSST}
have given a polynomial-time procedure that decides whether an SST has linear growth and, if so,
translates it to an MSO transduction.
Without the complexity bound, this would be a special case of Theorem~\ref{thm:em} on MTTs.

Douéneau-Tabot, Filiot and Gastin later showed~\cite[\S5.2]{Marble} that the
polynomial degree of growth of an SST is also computable in polynomial
time, by reduction to their work on $\bbN$-weighted automata described in
Section~\ref{sec:related-work-intro}. Via the above-mentioned equivalences, this result alternatively follows from the earlier work of Berglund, Drewes and van der Merwe~\cite{BerglundDM18} on top-down string-to-tree transducers (which our Theorem~\ref{thm:top-down-intro} extends). That said, Douéneau-Tabot et al.\ also obtain a structure theorem~\cite[\S5.3]{Marble} which implies, in particular, that every function computed by a
(copyful) SST of polynomial growth is polyregular (the inclusion is strict, cf.\
e.g.~\cite[Theorem~8.1]{NguyenNP21}).
It might be possible to get an analogous structure theorem on total deterministic top-down tree transducers of polynomial growth from the properties of the degrees of states seen in Section~\ref{sec:chara-poly}. (Note that we can already apply dimension minimization of MSO set interpretations, cf.\ Remark~\ref{rem:total-det-top-down}.)

Let us also mention that SSTs $\Sigma^*\to\{a\}^*$ and \emph{cost register automata (CRA) with affine updates} $\Sigma^{*} \to \bbN$ are related by the isomorphism of monoids $|\cdot| \colon \{a\}^* \to \bbN$ --- the syntax of these machine models is the same modulo the fact that the SSTs do not \enquote{know} that $\{a\}^*$ is commutative. Affine $\bbN$-CRA are expressively equivalent to $\bbN$-weighted automata, cf.~\cite{Benalioua2024}. (The original paper on cost register automata~\cite{AlurDDRY13} shows that weighted automata are equivalent to CRA with \emph{linear} updates.)

\subparagraph{Tree transducers.}

The theorem stated below is one of the most powerful available results
concerning the growth rate of transducers (on strings or trees). It concerns
linear growth, and we consider that extending it to polynomial growth is a major
open problem.
\begin{theorem}[{\cite{Maneth03,EngelfrietIM21}}]\label{thm:comp-mtt}
  It is decidable whether the composition of a sequence of functions, each
  defined by a macro tree transducer, has linear size increase; and if it does, then it
  is effectively equivalent to a tree-to-tree MSO transduction.
\end{theorem}

This was originally stated and proved by
Maneth~\cite[Theorems~1~and~2]{Maneth03}, by induction over the length of the
composition. The case of a single MTT, namely Theorem~\ref{thm:em}, plays a key
role in the inductive step --- not just in the base case. Engelfriet, Inaba and
Maneth~\cite{EngelfrietIM21} later reproved Theorem~\ref{thm:comp-mtt} by a similar
induction based on tree-walking tree transducers / MSO transductions with sharing (MSOTS, i.e.\ $\unfold \circ \text{MSOT}$) instead, using the fact that
\[ \text{MSOTS}^{k} \subset \text{MTT}^{k} \subset \text{MSOTS}^{k+1} \qquad\text{\cite[Corollary~25]{EngelfrietIM21}.} \]
Their inductive step thus depends on the case of a single MSOTS.\@ We are not
aware of any previous proof of this case that does not go through Theorem~\ref{thm:em} on MTTs, but we can give a new proof as a direct application of Theorem~\ref{thm:msosi-min}. Indeed, the inclusion of MSOTS in MSO set interpretations is not only a special case of Proposition~\ref{prop:msot-msots}, but a direct consequence of Lemma~\ref{lem:msosi-closure-prop} --- which just composes logical transformations by simple syntactic substitution.

It also seems likely that the \enquote{pruning} arguments
in~\cite{EngelfrietIM21}, used in the inductive step, can also be simplified by working with the logical
definition of MSOTS.

\subparagraph{Tree-to-tree Hennie machines.}

As a sort of sequel to the work of Gallot, Maneth, Nakano and Peyrat~\cite{LSHI} (mentioned in the introduction and in Theorem~\ref{thm:lhi}), a robust class of functions of tree-to-tree functions of \emph{linear size-to-height increase} has been studied by Dartois, Nguy{\~{\^e}}n and Peyrat~\cite{TreeHennie}. It subsumes all linear size-to-height MTTs, and is closed under pre-composition (resp.\ post-composition) by MTTs of linear size (resp.\ height) increase, while still being included in MSO set interpretations.

In a closely related work, Colcombet, Lhote and Ohlmann~\cite{StringMSOSI} prove an equality between classes of string-to-string functions that can be rephrased as:
\[ \text{MSO set interpretations} = \yield \circ [\text{string-to-tree Hennie machines}] \]

\subparagraph{Polyregular functions.}

There have been several works on the
degree of growth of polyregular functions, such as Bojańczyk's dimension
minimization theorem for string-to-string MSO interpretations~\cite[Theorem~2.3]{PolyregGrowth} (which Theorem~\ref{thm:msosi-min} generalises, as mentioned in the introduction).

Among the many equivalent formalisms that define polyregular functions
(cf.~\cite{PolyregSurvey}), the earliest one is the \emph{pebble transducer},
originating in the early 2000s~\cite{Pebble,PebbleString}: a machine that
manipulates a bounded stack of \enquote{pebbles} (pointers to input positions)
whose height trivially bounds the degree of growth. Analogously to the case of
MSO interpretations, one can ask the converse question: is a polyregular
function $f$ of growth $O(n^{k})$ always computable by some $k$-pebble
transducer? Douéneau-Tabot~\cite[Chapter~3]{gaetanPhD} showed\footnote{One of
  his results~\cite[Theorem~3.12]{gaetanPhD} is an effective
  version of Nguy{\~{\^e}}n, Noûs \&
  Pradic's earlier~\cite[Theorem~7.1]{NguyenNP21}.} that for a few restricted classes of
pebble transducers, given a device in this class that computes~$f$, one can
compute an equivalent $k$-pebble transducer in the same class. However, this
\enquote{pebble minimization} fails in the general case: Bojańczyk proved that
there is no bounded number of pebbles that suffices to compute all polyregular
functions of quadratic growth~\cite[Theorem~3.3]{PolyregGrowth}. Note that since
1-pebble (a.k.a.\ two-way) transducers are equivalent to MSO
transductions~\cite{EngelfrietHoogeboom}, pebble minimization holds for linear
growth as a consequence of (Bojańczyk's special case of) Theorem~\ref{thm:msosi-min},
cf.~\cite[\S1]{PolyregGrowth}.

This linear growth case can also be derived from Theorem~\ref{thm:comp-mtt} because
compositions of $\ell$ macro tree transducers can simulate
$\ell$-pebble\footnote{Note that the numbering of the pebble transducer
  hierarchy is off-by-one in~\cite{EngelfrietPebbleMacro} compared to more
  recent papers such as~\cite{PolyregSurvey}. We use the latter convention
  here.} transducers~\cite{EngelfrietPebbleMacro}. This connection with the MTT
composition hierarchy has also been used by Kiefer, Nguy{\~{\^e}}n and
Pradic~\cite{Sandra} to give an alternative refutation of pebble minimization.
They show a strenghtening of Bojańczyk's result: there is no $\ell$ such that
compositions of $\ell$ macro tree transducers include all polyregular functions
of quadratic growth. The proof is a direct application of Engelfriet and
Maneth's \enquote{bridge theorem} on the output languages of compositions of
MTTs~\cite{OutputMacro}.

The philosophy of this work by Kiefer et al.~\cite{Sandra} is somehow dual to
ours here: they give a short reduction from a problem on polyregular functions
to older results on macro tree transducers, whereas we take inspiration from
recent work on polyregular functions to simplify proofs concerning tree
transducers such as MTTs.

\subparagraph{Transducer growth vs regex matching.}

Sugiyama and Minamide~\cite{SugiyamaM14} show how to compute, from a regular expression $E$, a top-down string-to-tree transducer with regular lookahead whose output size on an input $w$ is proportional to the running time on $(E,w)$ of a backtracking regex matching algorithm. They then appeal to Engelfriet and Maneth's Theorem~\ref{thm:em} to decide whether, for a fixed $E$, the worst-case running time is linear in $|w|$. Using our results, we can also determine whether it is quadratic, cubic, \dots{} or exponential.

The results of Berglund, Drewes and van der Merwe that we extended in Theorem~\ref{thm:top-down} are also motivated by regex matching, as they discuss in~\cite[Section~3]{BerglundDM18}. They model backtracking regex matching using \enquote{prioritized NFA}~\cite{Berglund2017}, and provide a translation from prioritized NFA to nondeterministic top-down string-to-tree transducers, which allows them to apply their results on transducer growth to the worst-case running time of a matching algorithm.

These results apply to naive backtracking algorithms which are known to have suboptimal performance. Indeed, regex matching can be done in linear time, even with capturing groups, cf.~\cite[Theorem~4]{Berglund2017}. However, these algorithms have the advantage of being extensible to support features such as backreferences, which partly explains their popularity in real-world implementations. See the discussion of~\cite[Remark~III.3]{Terauchi25}, from a recent work of Terauchi on the running time of matching with backreferences (and its relation to a finite ambiguity condition, cf.~\cite[Theorem~III.6]{Terauchi25}).

\subsection{Further related work}\label{sec:related-work-misc}

\subparagraph{The growth rate of a language $L \subseteq \Sigma^{*}$.}
The counting function $f_{L}\colon n\in\bbN \mapsto
\mathrm{card}(L \cap \Sigma^{n})$ was already considered in the 1950s~\cite{Mandelbrot,ChomskyMiller58},
and there have been many works concerning $\growth[f_{L}]$.
  Languages of polynomial growth are called
\enquote{sparse}\footnote{For instance in~\cite{IbarraR86}; it is also the definition used in computational complexity, cf.\ e.g.~\cite{Mahaney1982}. Unfortunately the name \enquote{sparse language} is overloaded: it also refers to languages of asymptotic density zero, i.e.\ $f_{L}(n)/|\Sigma^{n}| \longrightarrow 0$ as $n \to \infty$, which have been investigated in recent years~\cite{ZeroOne,eickmeyer2025decidingsparsenessregularlanguages}. Polynomial growth implies asymptotic density zero, but the converse is not true. (The notion of asymptotic density of a language was introduced by Berstel at the first ICALP~\cite{Berstel72}.)} or \enquote{poly-slender}~\cite{Raz1997}.

From an unambiguous finite automaton that recognises a language $L$, one can derive an automaton over $\{a\}^{*}$ whose
ambiguity on $a^{n}$ is equal to $f_{L}(n)$ for all
$n\in\bbN$~\cite[Lemma~3.2]{IbarraR86}. Therefore, Theorem~\ref{thm:degree-ambiguity}
can be applied to study the growth of $L$.
The results thus obtained for regular languages are true more generally for
\emph{context-free} languages:
\begin{itemize}
  \item They have either polynomial or exponential growth. Furthermore, a
        context-free language has polynomial growth if and only if it is a
        \emph{bounded language}\footnote{Beware: this notion is very different from the boundedness of a function $\Sigma^{*}\to\bbN$ recognised by a weighted automaton.} --- this property \enquote{has been independently discovered at least six times}~\cite[\S1]{GawrychowskiKRS10} and has a recent far-reaching generalisation to languages recognised by \enquote{amalgamation systems}\footnote{A wide class of models which includes pushdown automata, vector addition systems with states (VASS), pushdown VASS, etc.  Anand, Schmitz, Schütze and Zetzsche also show that language boundedness is decidable for amalgamation systems satisfying some mild assumptions~\cite[Main Theorem~A]{AnandSSZ24}. For a recent survey on language boundedness, see~\cite{zetzsche:LIPIcs.FSTTCS.2025.2}.} \cite[Section~4.2]{AnandSSZ24}.
\end{itemize}
        \begin{definition}\label{def:bounded-language}
          A language $L \subseteq \Sigma^{*}$ is \emph{bounded} when
          \[ L \subseteq (w_{1})^{*} \dots (w_{\ell})^{*} \quad \text{for some}\ \ell\in\bbN\ \text{and}\ w_{1},\dots,w_{\ell}\in\Sigma^{*}. \]
        \end{definition}
\begin{itemize}
  \item For any fixed exponent $k\in\bbN$, the context-free languages of growth rate $O(n^k)$ have been characterised in the 2000s by Ilie, Rozenberg and Salomaa~\cite[Theorem~8]{IlieRS00} and by Ito, Martín-Vide and Mitrana~\cite[Theorem~3.2]{ItoMVM04}. In the regular case, Szilard, Yu, Zhang and Shallit had shown earlier~\cite[Theorem~3]{SzilardYZS92} that a regular language has growth $O(n^k)$ if and only if it is a finite union of languages of the form
        \[ u_0 (v_1)^* u_1 \dots (v_\ell)^* u_\ell\qquad \text{for}\ 0 \leqslant \ell \leqslant k+1.\]
  \item Ginsburg and Spanier showed in the 1960s that the boundedness of a
        context-free language is decidable~\cite[Theorem~5.2]{ALGOLlike} (see
        also~\cite[Chapter~5]{Ginsburg}). In the bounded case, Ilie gave an algorithm that computes the polynomial degree of growth~\cite[Theorem~23]{Ilie}, based on the characterisation of~\cite{IlieRS00}. Gawrychowski, Krieger, Rampersad and Shallit later showed that both problems are solvable in polynomial time~\cite[Theorems~19 and~25]{GawrychowskiKRS10}.
\end{itemize}
Although this does not seem to have been explicitly noted, these decidability
and complexity results transfer to regular tree languages:
\begin{theorem}
  Any regular tree language has a well-defined degree of growth in
  $\bbN\cup\{\infty\}$, that can be computed in polynomial time.
\end{theorem}
\begin{proof}[Proof idea]
  From a tree automaton, one can compute in polynomial time a context-free
  grammar that generates the encodings in (e.g.) reverse Polish notation of the
  trees accepted by the automaton. The encoding is injective and
  size-preserving.
\end{proof}

However, over trees, the above theorem cannot be deduced from our results on the
ambiguity of tree automata. The difference with the case of strings is that the
size of a tree does not uniquely determine its shape (the information kept after
forgetting the node labels).

\subparagraph{Entropy of a language.}

The counting function $f_{L}$ of $L \subseteq \Sigma^{*}$ is also related to its \emph{entropy}
\[ \entropy(L) =\limsup_{n\rightarrow \infty}\frac{1}{n}\log_2(f_{L}(n)) \in \{-\infty\} \cup \bbR_{\geqslant0} \]
This formula first appears in the work of Chomsky and Miller~\cite[Equation~(7)]{ChomskyMiller58}, with a limit instead of a limsup --- but the limit does not always exist (e.g.\ $(aa)^{*} \subset a^{*}$).
They were inspired by the definition of channel capacity in information theory. Intuitively, each letter in a word from $L$ provides, asymptotically, $\entropy(L)$ new bits of information. Languages with sub-exponential growth thus have zero entropy, reflecting the fact that the amount of new information per letter gets vanishingly small.

Chomsky and Miller~\cite{ChomskyMiller58} focus on the entropy of a regular language and relate it to the dominant eigenvalue of a nonnegative matrix, i.e.\ its \emph{spectral radius}. Their observation would later be made rigorous by Parker, Yancey \& Yancey~\cite{parker_et_al:LIPIcs.MFCS.2017.3}, who also explain the connection with the notion of topological entropy of a subshift in symbolic dynamics.
\begin{theorem}[{\cite[Section~4]{parker_et_al:LIPIcs.MFCS.2017.3}}]\label{thm:entropy}
  Let $\cA$ be a trim deterministic finite automaton that recognises the language $L$. The spectral radius of the adjacency matrix of $\cA$ (seen as a directed multigraph with self-loops by forgetting the labels) is equal to $\rho = 2^{\entropy(L)}$. Furthermore, if $L$ is infinite,
  \[ \lim_{n\to\infty} \frac{1}{n} \log_2 \sum_{k=0}^n f_L(k) = \entropy(L) \]
\end{theorem}
\begin{corollary}[immediate\protect\footnotemark]\label{cor:entropy-growth}
\footnotetext{It suffices to apply $\log_{\rho}(-)/n$ and then the squeeze theorem to the comparison:
\[ \forall n \in \bbN, \quad \frac{1}{n+1} \sum_{k=0}^n f_L(k) \leqslant \growth[f_{L}](n) = \max_{0\leqslant k\leqslant n} f_L(k) \leqslant \sum_{k=0}^n f_L(k) \]}
  $\growth[f_L](n) = \rho^{n + o(n)}$ when $\entropy(L) > 0$.
\end{corollary}
Via the reduction of $f_{L}$ to automata ambiguity, this can be seen as the special case for a unary alphabet of Theorem~\ref{thm:precise-exp} (which involves a \emph{joint} spectral radius).

Parker et al.\ also survey other works on the entropy of regular languages~\cite[Section~2.1]{parker_et_al:LIPIcs.MFCS.2017.3}. Let us also mention that Kuich~\cite{Kuich1970} has investigated the context-free case.

Finally, the concept of entropy has also been extended to timed automata by Asarin, Basset and Degorre~\cite{AsarinBD15}.
Words of length $n$ in a timed language are elements of $\Sigma^n\times \mathbb R^n$, and their definition of entropy uses the above limsup formula, where the cardinality $f_{L}(n)$ is replaced by the volume of $L\cap \Sigma^n\times \bbR^n$.
They also obtain:
\begin{itemize}
  \item a dichotomy between \emph{thin} and \emph{thick} automata~\cite[\S3.5]{AsarinBD15}, characterising $-\infty$ and $>-\infty$ entropy, respectively, somewhat analogously to our characterisation of exponential growth via heavy cycles (cf.\ Section~\ref{sec:exp-vs-poly});
  \item a connection between the entropy and a spectral radius~\cite[\S4.2.3]{AsarinBD15}.
\end{itemize}

\subparagraph{Infinite trees \& automatic structures.}

A theorem of Niwiński states that a regular language of infinite
trees is either at most countable or of cardinality
$2^{\aleph_{0}}$~\cite[Theorem~2.2]{Niwinski91}. His proof uses a pumping
argument quite similar to Seidl's~\cite[proof of Proposition~2.1]{Seidl} (which
we discussed in~\S\ref{sec:exp-vs-poly}); indeed, this sort of
\enquote{continuum hypothesis} can be seen as the infinite counterpart to the
polynomial/exponential dichotomy in our growth rates, since
$(\aleph_{0})^{k} = \aleph_{0}$. Furthermore, this cardinality in
$\bbN\cup\{\aleph_{0},2^{\aleph_{0}}\}$ is
computable~\cite[Section~4]{Niwinski91}.

Rabinovich and Tiferet~\cite{RabinovichT21} have later shown, via a
characterisation using \enquote{heavy cycles}, that one
can decide in polynomial time whether the ambiguity of a Büchi automaton over
infinite trees is at most countable (see also~\cite{RabinovichT21parity} for
partial results on parity automata). Again, in this setting, uncountable necessarily
means of cardinality~$2^{\aleph_{0}}$, see~\cite[Theorem~1.2]{BaranyKR10}.

As mentioned in Remark~\ref{rem:reparam-ext}, Rabinovich has extended Theorem~\ref{thm:reparam-intro} (reparameterisation) to MSO \emph{finite set} interpretations from countable labelled linear orders~\cite{RabinovichWoLLIC,RabinovichICALP}. Such interpretations from $(\bbN,\leqslant)$ present automatic
structures~\cite[Proposition~3.1]{ColcombetL07}, and this was the
original motivation for introducing MSO set interpretations.
Colcombet et al.\ have shown that the MSO theory of any automatic $\omega$-word is decidable~\cite[Corollary~1.2]{StringMSOSI}, as a direct consequence of their recent results on string-to-string MSO set interpretations (cf.\ \S\ref{sec:transducers-intro} / \S\ref{sec:related-work-trans}).

The aforementioned
characterisation of regular string languages of polynomial growth by
\enquote{boundedness} (Definition~\ref{def:bounded-language}) has also been used by Bárány~\cite[Section~3.3.2]{BaranyPhD} to study the
subclass of \enquote{automatic structures of polynomial growth}, which led to
further works~\cite{GanardiK20,Zacek,blumensath2025simple}. For surveys on automatic
structures, see~\cite{Gradel20} or~\cite[Chapter~VII]{MorvanPhD}.

\subparagraph{Ambiguity of context-free grammars \& languages.}

A classical result of formal language theory is that the unambiguity of a
context-free grammar is undecidable, with three proofs appearing independently
in 1962~\cite{Cantor62,ChomskyMPS63,Floyd62}. Four decades later, Wich showed in
his PhD thesis~\cite{WichPhD} that exponential ambiguity is undecidable, even
with the promise that the input grammar is either unambiguous or exponentially
ambiguous~\cite[Theorem~7.43]{WichPhD} (this also implies for instance that
finite ambiguity is undecidable).

On the positive side, unambiguity is decidable
for context-free grammars that generate bounded languages in the sense of Definition~\ref{def:bounded-language}~\cite[Section~4]{CFLamb}. Also, the
ambiguity of any context-free grammar is either exponential or
$O(n^{\ell})$~\cite[Theorem~7.54]{WichPhD} for some $\ell$ that is computable
from the grammar (in fact, defined by a simple closed-form formula), and there
is a (non-effective) criterion for exponential
ambiguity~\cite[Theorem~7.22]{WichPhD} that is very similar to the
\enquote{heavy cycles} of Section~\ref{sec:exp-vs-poly}. But polynomially bounded does
not imply $\Theta(n^{k})$ here: for any computable unbounded and non-decreasing
function $f\colon\bbN\to\bbN$, e.g.\ $f=\log^{*}$, one can find a CFG
whose growth rate of ambiguity is below $f$, and yet tends to
$+\infty$~\cite[Theorem~6.18]{WichPhD}.

A closely related topic is the \emph{inherent ambiguity} of a context-free
language, quantifying over all grammars generating the language. It is
undecidable whether a context-free language is inherently
(un)ambiguous~\cite[Theorem~2.1]{CFLamb} (see also~\cite[Corollary on
p.~4]{Greibach1968}), but sophisticated techniques based on generating series
have been developed to prove that specific languages are inherently
ambiguous~\cite{Flajolet1987,Koechlin22}. It turns out that if a function
$\bbN\to\bbN$ is the growth rate of the ambiguity of some context-free grammar,
then it can also be realized as an inherent ambiguity function (in a suitable
sense) for some context-free language~\cite[Theorem~8.1]{WichPhD}. Therefore,
some context-free languages are inherently exponentially ambiguous (this had
been proved earlier by Naji~\cite[Satz~4.2.1]{NajiDiploma}) while some others
have an inherent ambiguity that grows very slowly. To wrap up this part, let us
mention that Wich has given a characterisation of the context-free languages
with polynomially bounded ambiguity~\cite[Theorem~7.39]{WichPhD}.

\subparagraph{More weighted automata \& matrix (semi)groups.}

A precursor from the 1960s to the works mentioned in
Section~\ref{sec:related-work-intro} is Schützenberger's characterisation of the
$\bbZ$-rational series $\Sigma^* \to \bbZ$ of polynomial growth,
with given degree~\cite{Schutz} --- cf.\ the modern
rephrasing in~\cite[Chapter~9]{BerstelReutenauer}. An effective version was
proved recently by Colcombet, Douéneau-Tabot and Lopez~\cite{Zpolyreg} (see
also~\cite[Chapter~5]{gaetanPhD}) whose motivations came from polyregular
functions. In both cases, the proofs are significantly more difficult than for
weights in $\bbN$.

The boundedness problem for $\bbN$-weighted automata ---
equivalently, the finite ambiguity problem for nondeterministic automata ---
admits two natural generalisations to rational weights. Namely, for a map
$\Sigma^{*}\to\bbQ$ computed by a $\bbQ$-weighted automaton:
\begin{itemize}
  \item Does it take bounded values? This is undecidable even for nonnegative
        weights, but some special cases are decidable,
        see~\cite{CzerwinskiLMPW22}.
  \item Does it take finitely many distinct values? A decision procedure was given by Mandel and
        Simon~\cite[Corollary~5.4]{MandelS77}, extending their work for weights
        in $\bbN$ (cf.~\S\ref{sec:related-work-intro}).
\end{itemize}
The latter problem is essentially equivalent (modulo minimization of weighted
automata, as discussed in~\cite[\S4.1]{BumpusHKST20}) to the following:
\emph{does a given finite set of matrices in $\bbQ^{d\times d}$ generate a
  finite multiplicative monoid?} In fact, Jacob has shown its decidability over
an arbitrary field, not just $\bbQ$~\cite[Théorème~2.9]{Jacob1977}. That said,
the problem was not known to be elementary recursive over $\bbQ$ before Bumpus,
Haase, Kiefer and Stoienescu~\cite[Theorem~13]{BumpusHKST20} put it in~$\mathsf{coNEXP^{NP}}$ four decades
later.
Speaking of computational complexity, in the case of \emph{groups} generated by a finite set of invertible matrices:
\begin{itemize}
  \item there is a polynomial-time algorithm by Babai, Beals and
        Rockmore~\cite[Theorem~1.4]{BabaiBR93} to decide finiteness when the
        coefficients are in a finite extension of $\bbQ$;
  \item see also Detinko and Flannery~\cite[\S3]{DetinkoF2019} for a practical
        algorithm that works in positive characteristic, and for a discussion of
        implementations in computer algebra systems.
\end{itemize}

\subparagraph{Min/max-automata \& regular cost functions.}

In our discussion of weighted automata throughout this paper, the semiring
structure was implicitly taken to be the usual addition and multiplication of
numbers.
One can also study the growth rate of weighted automata
over semirings that do not embed into fields, e.g.\ tropical semirings (for which the notion of joint spectral radius from \S\ref{sec:open-problems} can be adapted, cf.\ e.g.~\cite{JSRmaxplus}).
The equivalence between weighted automata and linear/affine cost register automata~\cite{AlurDDRY13,Benalioua2024}, mentioned over $\bbN$ in \S\ref{sec:related-work-trans}, extends to arbitrary semirings. 

For instance, Simon
showed~\cite[Theorem~13]{Melanges} that when $\Sigma$ has at least
two letters, these classes form a strict hierarchy for $k\in\bbN\setminus\{0\}$ covering all unbounded \((\bbN,\min,+)\)-automata:
\[ \left\{ f\colon\Sigma^{*}\to\bbN \;\middle|\; f\ \text{recognised by some
    \((\bbN,\min,+)\)-automaton},\; \growth[f](n) = \Omega\big(\sqrt[k]{n}\big) \right\} \]
As another example, Colcombet, Daviaud and Zuleger proved that
\((\bbN,\max,+)\)-automata exhibit an asymptotic behavior with fractional
degree~\cite{ColcombetDZ14,DaviaudPhD}, namely: if such a weighted automaton
recognises a function $f\colon\Sigma^{*}\to\bbN$, then
\[ \min_{|w|\geqslant n} f(w) = \Theta(n^{\alpha})\ \text{for some}\ \alpha\in(\bbQ\cap[0,1])\cup\{-\infty\}\ \text{computable from the automaton.} \]
See~\cite[Examples~4.1.1 and~4.1.2]{DaviaudPhD} for concrete max-plus
automata that realize the degrees $\alpha=1/2$ and $\alpha=2/3$ respectively.

In Colcombet's theory of regular cost functions~\cite{CostFunctions}, min-plus
automata and max-plus automata are both generalised by \enquote{B-automata}
(which are no longer weighted automata). Cost functions are the equivalence
classes of functions $\Sigma^{*}\to\bbN\cup\{+\infty\}$ for a suitable
equivalence relation, and those that have a representative computed by some
B-automaton are said to be regular. The point of this theory is that many
\enquote{quantitative} computational problems on formal languages and automata
can be solved by deciding the boundedness of regular cost functions; cf.\ the
survey~\cite{CostFunctionsLICS}. While the present paper also attempts to take a
unified look at several quantitative questions (of asymptotic growth), there is
no clear relationship at the moment between our work and regular cost functions.

\subparagraph{Combinations of topics.} Weighted automata with restricted ambiguity
(e.g.\ finite or polynomial) are a common object of study. The purpose of this
restriction can be to control their expressive power
(e.g.~\cite{ErikPaulDiploma,WeightedFO,MalettiNSU23}) or to make decision
problems easier (e.g.~\cite{KirstenLombardy,Paul24}).

There are also weighted tree transducers which generalise both weighted tree automata and some nondeterministic tree transducer models; they compute relations where each pair of trees is assigned a weight, and have been applied to natural language processing. Many references on the topic are given in~\cite[p.~6]{WTAbook}. It is possible that Lemma~\ref{lem:total-det-top-down} could be obtained as a consequence of closure properties of weighted tree transducers.

\printbibliography

\newpage

\appendix

\section{Further details for Section~\ref{sec:ambiguity}}

\subsection{Proof of Lemma~\ref{lem:acc-lin}}%
\label{app:acc-lin}

\emph{Horn satisfiability} is the special case of CNF-SAT where each clause contains at most one positive occurrence of a literal (it is then called a \emph{Horn clause}). It is well known that Horn-SAT can be solved in linear time~\cite{HornLin} (see also~\cite{TryAlgo} for a concise Python implementation); what is more, these algorithms actually compute a \emph{minimum satisfying assignment} (for the pointwise partial ordering with $\mathtt{false < true}$). 

Given a tree automaton, consider the following set of clauses: 
\[ \lnot q_{1} \lor \dots \lor \lnot q_{m} \lor q \quad\text{for each transition}\ (q_{1},\dots,q_{m},a,q) \]
whose atoms are the states. By definition, the set of atoms set to $\mathtt{true}$ in the minimum satisfying assignment is the smallest set of states such that for each transition $(q_{1},\dots,q_{m},a,q)$, if all the $q_i$ are in that set, then so is $q$. This is none other than the set of accessible states.

\begin{remark}
  The linear-time decision procedure for language emptiness in \textit{Tree Automata Techniques and Applications}~\cite[Theorem~1.7.4]{TATA} also proceeds by reduction to Horn-SAT.\@ The reduction takes the above clauses and adds $\lnot q$ for each accepting state $q$. This results in a Horn-SAT instance that is unsatisfiable if and only if some accepting state is accessible, i.e.\ the recognised language is nonempty.
\end{remark}

\begin{remark}
  Just as accessibility in word automata amounts to reachability in directed graphs, accessibility in tree automata amounts to reachability in \emph{directed hypergraphs}, cf.~\cite{DirectedHypergraphs} for a survey. It is well known that directed hypergraph reachability is very closely related to Horn-SAT, as explained in~\cite[Section~3]{DirectedHypergraphs}; see also the Dagstuhl meeting~\cite{DagRep.4.5.1} gathering researchers interested in Horn formulas and directed hypergraphs. The aforementioned linear-time algorithms for Horn-SAT can be seen as clever traversals of the hypergraph associated to a set of Horn clauses following~\cite[Section~2]{Ausiello1991}.  
\end{remark}

\subsection{Proof of Proposition~\ref{prop:center-ambiguous-cycles}}\label{app:center-ambiguous-cycles}

Given a tree automaton $\cA$ we build the product automaton $\cA \times \cA$.
A center-ambiguous cycle of $\cA$ is two runs $q \to_C q$ which differ on some node on the path 
from the hole of $C$ to the root (for some context $C$), which corresponds to a run of $\cA\times\cA$
on $C$ from $(q,q)$ to $(q,q)$ which goes through some $(q_1,q_2)$ with $q_1 \neq q_2$ along the 
path from the hole of $C$ to the root. The existence of a center-ambiguous cycle in $\cA$ is 
therefore equivalent to the existence in $\cA\times\cA$ of two runs $(q,q) \to_{C_{1}} (q_1,q_2)$ and
$(q_1,q_2) \to_{C_{2}} (q,q)$ with $q_1 \neq q_2$. 
This is equivalent to the existence, in the shallow digraph, of two vertices $(q,q)$
and $(q_1,q_2)$ with $q_1 \neq q_2$ in the same strongly connected component. We can build the 
digraph, compute its strongly connected components and check the property in time linear in the
size of $\cA\times\cA$, so in quadratic in the size of $\cA$.

\subsection{Final step of the proof of Proposition~\ref{prop:side-amb}}\label{app:side-amb}

The missing part in the main text was:

\begin{lemma}\label{lem:ambiguous-states}
  The ambiguous states of a tree automaton --- i.e.\ the states that appear at the root in multiple runs of a same tree --- are computable in quadratic time.
\end{lemma}
\begin{proof}
Noting $Q$ the set of states and $\Delta$ the set of transitions of a tree automaton $\cA$, we recall the 
construction by Seidl~\cite[proof of Proposition~1.2]{Seidl} of the tree automaton $\cA'$ in his decision procedure for unambiguity: \\
the set of states is $Q' = Q \times Q \cup Q \times \{\#\}$, and the set of transitions is: 
\begin{align*}
	\Delta' & = \{((q_1,p_1), \dots, (q_k,p_k), a, (q,p)) \mid (q_1, \dots, q_k, a, q), (p_1, \dots, p_k, a, p) \in \Delta\} \hspace{15mm} (1) \\
	& \cup \{((q_1,p_1) ,\dots, (q_k,p_k), a, (q,\#)) \mid \\
	& ~\hspace{23mm}~ (q_1, \dots, q_k, a, q) \text{ and } (p_1, \dots, p_k, a, q) \text{ are distinct transitions in } \Delta \} ~ (2) \\
	& \cup \{((q_1,q_1) ,\dots, (q_{i-1},q_{i-1}),(q_i,\#),(q_{i+1},q_{i+1}) , \dots, (q_k,q_k), a, (q,\#)) \mid \\
	& ~\hspace{76mm}~ (q_1, \dots, q_k, a, q) \in \Delta, 1 \leq i \leq k \} ~ (3)
\end{align*}
This automaton is of quadratic size in the size of $\cA$. We do not mention accepting states 
because they are not relevant to the problem of computing ambiguous states. 

Note that in any run of $\cA'$ on a tree $t$, the set of nodes with a label in 
$Q \times \{\#\}$ is either empty or forms a path in $t$ from a node $u$ to the root. In the 
latter case a rule from the subset $(2)$ of $\Delta'$ is used exactly once in this run: at node $u$. 

The first thing to show is that a tree $t$ has a run of $\cA'$ with $(q,p)$ at the root if and 
only if it has a run of $\cA$ with $q$ at the root, and a run with $p$ at the root (if $q=p$ then there is one run with $q$ at the root). 
This is proven by induction over $t$, using in $\cA'$ exclusively the transitions from the first 
part $(1)$ of $\Delta'$, which is exactly the set of transitions of the 
product-automaton $\cA \times \cA$. 

Next we prove that a tree $t$ has a run of $\cA'$ with $(q,\#)$ at the root if and only if 
there are two distinct runs of $\cA$ on $t$ with $q$ at the root. 

If $(q,\#)$ is accessible then the set of nodes with a label in $Q \times \{\#\}$ forms a path 
in $t$ from a node $u$ to the root. At this node $u$ a transition from part $(2)$ of $\Delta'$ 
is used, giving two distinct transitions from $\Delta$ and therefore two runs of $\cA$ on the 
subtree of $t$ rooted at node $u$. We complete each of these two runs with transitions from 
$\Delta$ given by the transitions from $\Delta'$ on other nodes of $t$: either from $(3)$ for 
nodes along the path from $u$ to the root, or from $(1)$ for other nodes. This gives two 
distinct runs of $\cA$ on $t$ with $q$ at the root.

If there are two distinct runs of $\cA$ on $t$ with $q$ at the root, then we note $S$ the 
(non-empty)set of nodes of $t$ where these two runs do not use the same transition. Let $u$ be 
a minimal element of $S$ for the ancestor order, i.e.\ the two runs use the same transitions 
along the path from $u$ to the root, but not at node $u$. Let $q'$ the label of node $u$ in both 
runs. Then we can use the transition from $(2)$ at path $u$ with state $(q',\#)$ at path $u$ 
given by the two runs of $\cA$, and transitions from $(1)$ at all the nodes 
below $u$ in $t$. We then complete this run using rules from $(3)$ along the path from $u$ to 
the root (this works because both runs use the same transitions along this path), and we use 
rules from $(1)$ outside this path. We get a run of $\cA'$ on $t$ with $(q,\#)$ at the root. 

Finally, a state $(q,\#)$ is accessible in $\cA'$ if and only if $q$ is ambiguous in $\cA$. 
To compute ambiguous states we therefore only need to apply Lemma~\ref{lem:acc-lin} to automaton
$\cA'$. This algorithm is linear in the size of $\cA'$, so quadratic in the size of $\cA$. 
\end{proof}

\begin{remark}
  Similarly, a tree automaton $\cA$ is unambiguous if and only if all states in $\cA\times\cA$ that are both accessible and co-accessible are of the form $(q,q)$. (For co-accessibility we define the accepting states of $\cA\times\cA$ to be pairs of accepting states of $\cA$.) This gives a decision procedure for unambiguity in quadratic time, by reduction to linear-time trimming (\S\ref{sec:trim}), that is slightly different from Seidl's proof of~\cite[Proposition~1.2]{Seidl} since it uses $\cA\times\cA$ without extra states.
\end{remark}

\subsection{Proof of the upper bound in Claim~\ref{clm:pump-lower-bound}}\label{app:pump-lower-bound}
We prove $|C[\pump(n,\Pi)]|=O(n)$ (ignoring $C$ because it has bounded size) by induction on $\Pi$.
There are two cases to this induction:
\begin{align*}
	|\pump(n,a(\Pi_{1},\dots,\Pi_{k}))| &= |a(\pump(n,\Pi_{1}),\dots,\pump(n,\Pi_{k}))| = 1+ k. O(n) = O(n)\\
	|\pump(n,\pumpnode{C'}(\Pi_1))| &= |C'^{n}[\pump(n,\Pi_1)| = |C'|.n + O(n) = O(n)
\end{align*}

\subsection{Proof of Lemma~\ref{lem:deg-abstract}}\label{app:deg-abstract}
We recall the definition of the fixed-point operator here:
\[ f \mapsto \left[ q \mapsto \max\left(\bigl\{f(q)\bigr\}\cup\left\{\sum_{i=1}^{k} f(q_i) \;\middle|\; (q_1,\dots,q_k,a,q)\ \text{trans.}\right\} \cup \Bigl\{f\bigl(q'\bigl)+1 \;\Big|\; q' \Rrightarrow q\Bigr\}\right) \right]. \]
Note that it increases the value of $f(q)$ in two cases, 
each of which corresponds to a different inductive case of Definition~\ref{def:pumping}.

Noting $f_0$ the function mapping each state $q$ to $-\infty$ and, for each $n \in \bbN$, $f_{n+1}$ the function obtained from $f_n$ by applying the fixed point operator as above, we can prove by induction on $n$ that $f_n(q)$ is the maximum degree of a pumping pattern of height $\leq n-1$ 
for $q$ (counting the height of $\pumpnode{C}(\Pi)$ as $1+$ the height of $\Pi$) for all integer $n$ and state $q$.

The base case ($n=0$) is trivial.
The induction goes two ways, first we show the existence of the pumping pattern $\Pi$ of 
degree $f_{n+1}(q)$, and conversely we show that any pumping pattern for $q$ of height $n$ 
has degree at most $f_{n+1}(q)$. The induction step has three cases:
\begin{itemize}
\item if $f_{n+1}(q) = f_n(q)$ then the induction hypothesis gives a pumping pattern 
of degree $f_n(q)$ and height $\leq n-1$. 
Conversely, if there is a pumping pattern of degree $f_n(q)$ and height $\leq n-1$, 
then its degree is $\leq f_{n+1}(q)$. 
\item if $f_{n+1}(q) = \sum_{i=1}^{k} f_n(q_i)$ for some transition $(q_1,\dots,q_k,a,q)$ 
then, noting $\Pi_{i}$ the pumping pattern of height $\leq n-1$ and degree $f_n(q_i)$ 
obtained from the induction hypothesis on $f_n(q_i)$ for each $i \leq k$, we have the 
pumping pattern $\Pi = a(\Pi_1, \dots, \Pi_k)$ of height $n$ and degree 
$\deg(\Pi) = \sum_{i=1}^{k} \deg(\Pi_i) = \sum_{i=1}^{k} f_n(q_i) = f_{n+1}(q)$. 
Conversely, if there is a pumping pattern of the form $\Pi = a(\Pi_1, \dots, \Pi_k)$ 
of height $\leq n$ then there is a transition $(q_1,\dots,q_k,a,q)$ and 
$f_{n+1}(q) = \sum_{i=1}^{k} f_n(q_i) \geq \sum_{i=1}^{k} \deg(\Pi_i) = \deg(\Pi)$. 
\item if $f_{n+1}(q) = f_n(q') + 1$ with $q' \Rrightarrow_C q$ for some state $q'$ and 
context $C$, then there is a pumping pattern $\Pi'$ for $q'$ of height $\leq n-1$ and 
degree $f_n(q')$, and so pumping pattern $\Pi = \pumpnode{C}(\Pi')$ is of height $\leq n$ 
and degree $f_n(q') + 1 = f_{n+1}(q)$. 
Conversely, if there is a pumping pattern $\Pi = \pumpnode{C}(\Pi')$ of height $n$ with 
$q' \Rrightarrow_C q$ then $\Pi'$ has height $n-1$ and therefore degree $\leq f_n(q')$. 
So $\deg(Pi) = \deg(\Pi') + 1 \leq f_n(q') + 1 \leq f_{n+1}(q)$. 
\end{itemize}

\section{Further details for the proof of Theorem~\ref{thm:reparam-details}}\label{app:reparam-details}

We detail the construction of the MSO formula $H(X_{1},\ldots,X_{\ell},Y)$ that relates each $\vec{P} \in \wn F(t)$ to the set of critical nodes for the accepting run of $\cB_{F}$ over $t$ corresponding to $\vec{P}$.
\begin{claim}[{cf.~e.g.~\cite[proof of Theorem~2.7]{HandbookAutomataTrees}}]\label{clm:runs-in-mso}
  Let $\cA$ be a tree automaton with the set of states $Q=\{q_1,\dots,q_N\}$. One can build an MSO formula $\mathsf{Run}_{\cA}$ such that
  \[ \wn \mathsf{Run}_{\cA}(t') = \{\widehat\rho = (\rho^{-1}(\{q_1\}),\dots,\rho^{-1}(\{q_{N}\})) \mid \rho\ \text{is an accepting run of}\ \cA\ \text{on}\ t'\} \]
\end{claim}
We apply this fact to the tree automaton $\cA_F$ over the ranked alphabet $\Sigma \times \{0,1\}^\ell$.
Using Proposition~\ref{prop:mso-marked}, we get a formula $\mathsf{nuR}(X_{1},\ldots,X_{\ell},Z_1,\dots,Z_N)$ over $\tree\Sigma$ such that
\[ t \models \mathsf{nuR}\bigl(\vec{P},\vec{R}\bigr) \iff \marked\bigl(t,\vec{P}\bigr) \models \mathsf{Run}_{\cA_F}\bigl(\vec{R}\bigr) \]
By definition of $\cA_F$, for any $t$ and $\vec P$ there exists such a $\vec{R}$ if and only if $t \models F(\vec{P})$, and then $\vec{R} = \widehat\rho_{t,\vec{P}}$ where $\rho_{t,\vec{P}}$ is the unique accepting run of $\cA_F$ on $\marked(t,\vec{P})$ by unambiguity. Via the bijection of Claim~\ref{clm:query-vs-ambiguity}, the run of $\cB_F$ on $t$ corresponding to $\vec{P}$ is then
\[ \phi_{t,\vec{P}} \colon v \in \nodes(t) \mapsto (\rho(v),(\text{1 if \(v \in P_i\), else 0})_{1 \leqslant i \leqslant \ell})\]
Therefore, we have $\phi_{t,\vec{P}}(v) = (q_i,\vec{b})$ if and only if $t \models \Phi_{(q_i,\vec{b})}(\vec{P},\widehat\rho_{t,\vec{P}},v)$, where
\[ \Phi_{(q_i,\vec{b})}\bigl(\vec{X},\vec{Z},x\bigr) = Z_i(x) \land \bigwedge_{b_j = 0} \lnot X_j(x) \land \bigwedge_{b_j = 1} X_j(x)\]
From this, for any transition $\tau = ((q_{i_1},\vec{b}_{1}),\dots,(q_{i_m},\vec{b}_{m}),a,(q,\vec{b}))$ in $\cB_F$ we can define
\[ \Psi_\tau\bigl(\vec{X},\vec{Z},x\bigr) = a(x) \land \Phi_{(q,\vec{b})}\bigl(\vec{X},\vec{Z},x\bigr) \land \bigwedge_{\mathclap{j \leqslant \rank(a)}} \exists y.\; \child_j(x,y) \land \Phi_{(q_{i_j},\vec{b}_j)}\bigl(\vec{X},\vec{Z},y\bigr) \]
so that the run $\phi_{t,\vec{P}}$ takes the transition $\tau$ at the node $v$ if and only if $t \models \Psi_\tau(\vec{P},\widehat\rho_{t,\vec{P}},v)$.

Recalling from Definition~\ref{def:critical} that critical nodes are just those at which critical transitions are taken, our construction of $H$ can now be finished:
\[ H\bigl(\vec{X},Y\bigr) = \exists \vec{Z}.\;  \mathsf{nuR}\bigl(\vec{X},\vec{Z}\bigr) \land \left[\forall y.\; Y(y) \Leftrightarrow \bigvee_{\mathclap{\text{\(\tau\) critical}}} \Psi_\tau\bigl(\vec{X},\vec{Z},y\bigr) \right] \]

\section{Proof of Claim~\ref{clm:joint-spectral}}\label{app:joint-spectral}

First, note that $F_S$ formally depends on a choice of norm on the space $\bbR^{Q\times Q}$ of $Q$-indexed square matrices ($Q$ is the set of states of $\cA$). However, the assertions \enquote{$f(n) = O(F_S(n))$} and \enquote{$F_S(n) = O(f(n))$} are independent of this choice, thanks to the equivalence of norms on finite-dimensional $\bbR$-vector spaces.

Let $M_a$ be the transition matrix of $\cA$ for $a \in \Sigma$. Recall (from e.g.~\cite[Chapter~1, Section~6]{BerstelReutenauer}) that for some row vector $I$ and column vector $T$,
\[ \forall w = w[1] \dots w[n] \in \Sigma^*,\quad \cA(w) = I \times M_{w[1]} \times \dots \times M_{w[n]} \times T \]
The linear operator $I \times (-) \times T \colon \bbR^{Q\times Q} \to \bbR$ has a finite-dimensional domain so it is Lipschitz:
\[ \cA(w) = O\bigl(\|M_{w[1]} \times \dots \times M_{w[n]}\|\bigr) = O\bigl(F_S(n)\bigr) \]

For the converse, we use the fact --- see e.g.~\cite[proof of Proposition~6.1 in Chapter~1]{BerstelReutenauer} --- that the $(i,j)$ coefficient of $M_{w[1]} \times \dots \times M_{w[n]}$ is the sum of weights of paths from $i$ to $j$ labeled by the word $w[1] \dots w[n]$ in $\cA$. Therefore, if $u \in \Sigma^*$ labels some path from an initial state to $i$ and $v \in \Sigma^*$ labels some path from $j$ to a final state, then
\[ 0 \leqslant \bigl(M_{w[1]} \times \dots \times M_{w[n]}\bigr)_{i,j} \leqslant \cA(uwv) \]
Since we have assumed that $\cA$ is trim, there always exist such $u,v$; and by pumping, we may take them to have length at most $|Q|$. By choosing the $\ell_\infty$ norm $\|A\|_\infty = \displaystyle\max_{i,j} |A_{i,j}|$ in $F_S$, we get:
\[ F_S(m) = \max_{|w|=n\leqslant m} \max_{i,j}\; \bigl(M_{w[1]} \times \dots \times M_{w[n]}\bigr)_{i,j} \leqslant \growth[\cA](m+2|Q|) \]
To conclude, note that $F_S(n) = O\bigl(F_S(n-2|Q|)\bigr)$ by, again, a Lipschitz continuity argument.

\end{document}